\documentclass[aip,jcp,reprint]{revtex4-1}
\usepackage[T1]{fontenc}       
\usepackage{verbatim}
\usepackage{graphicx}
\usepackage{amsmath}
\usepackage{xcolor}
\usepackage{amssymb}
\usepackage{bm}
\usepackage{textcomp}
\usepackage{longtable}
\usepackage{multirow}
\usepackage{url}
\usepackage{setspace}
\usepackage{array}
\usepackage{supertabular}
\usepackage{numprint}
\usepackage{float}
\usepackage{ragged2e}

\newcommand{\cm}{\ensuremath{\textrm{cm}^{-1}}}
\newcommand{\km}{\ensuremath{\textrm{km}\cdot\textrm{mol}^{-1}}}

\begin{document}

\title{Anharmonic vibrational spectroscopy of Polycyclic Aromatic Hydrocarbons (PAHs)}
\author{Giacomo Mulas}
\affiliation{IRAP, Universit\'e de Toulouse, CNRS, UPS, CNES, 9 Av. du Colonel Roche, 31028 Toulouse Cedex 4, France}
\affiliation{Istituto Nazionale di Astrofisica (INAF), Osservatorio Astronomico di Cagliari, 09047 Selargius (CA), Italy}
\email{gmulas@oa-cagliari.inaf.it}
\author{Cyril Falvo}
\affiliation{Institut des Sciences Mol\'eculaires d'Orsay (ISMO), CNRS, Univ. Paris-Sud, Universit\'e Paris-Saclay, 91405 Orsay, France}
\affiliation{LIPhy, Univ. Grenoble Alpes, CNRS, 38000 Grenoble, France}
\author{Patrick Cassam-Chena\"i}
\affiliation{Universit\'e C\^ote d'Azur, CNRS, LJAD, UMR 7351, 06100 Nice, France}
%
\author{Christine Joblin}
\affiliation{IRAP, Universit\'e de Toulouse, CNRS, UPS, CNES, 9 Av. du Colonel Roche, 31028 Toulouse Cedex 4, France}
\email{christine.joblin@irap.omp.eu}
\begin{abstract}
While powerful techniques exist to accurately account for anharmonicity in vibrational molecular spectroscopy, they are computationally very expensive and cannot be routinely employed for large species and/or at non-zero vibrational temperatures. Motivated by the study of Polycyclic Aromatic Hydrocarbon (PAH) emission in space, we developed a new code, which takes into account all modes and can describe all IR transitions including bands becoming active due to resonances as well as overtones, combination and difference bands. In this article, we describe the methodology that was implemented and discuss how the main difficulties were overcome, so as to keep the problem tractable. Benchmarking with high-level calculations was performed on a small molecule. We carried out specific convergence tests on two prototypical PAHs, pyrene (C$_{16}$H$_{10}$) and coronene (C$_{24}$H$_{12}$), aiming at optimising tunable parameters to achieve both acceptable accuracy and computational costs for this class of molecules. We then report the results obtained at 0\,K for pyrene and coronene, comparing the calculated spectra with available experimental data. The theoretical band positions were found to be significantly improved compared to harmonic Density Functional Theory (DFT) calculations. The band intensities are in reasonable agreement with experiments, the main limitation being the accuracy of the underlying calculations of the quartic force field. This is a first step towards calculating moderately high-temperature spectra of PAHs and other similarly rigid molecules using Monte Carlo sampling.
\end{abstract}


\maketitle

\section{Introduction}
Polycyclic Aromatic Hydrocarbons (PAHs) are a family of organic compounds composed of aromatic rings containing carbon atoms, and whose peripheral bonds are saturated by hydrogen atoms. PAHs (or closely related species) are thought to be the carriers of the so-called Aromatic Infrared Bands (AIBs) at $\sim$3.3, 6.7, 7.7, 8.6, 11.3 $\mu$m.\cite{Leger1984,Allamandola1985} These features are among the strongest emission features observed in the interstellar medium.\cite{Tielens2008} Therefore, the infrared (IR) spectroscopy of PAHs is of paramount importance in astrophysics. 
PAHs absorb starlight in the visible and ultraviolet domain via electronic transitions and re-emit most of this energy in the mid-IR in a vibrational de-excitation cascade. In such model PAHs should emit most of the flux in the AIBs from highly vibrationally excited states corresponding to a thermal temperature of the order of $\sim$1000~K.\cite{Draine2011} 
Each AIB results from the superposition of the emission of molecules at different temperatures, each of which being the superposition of a large number of individual vibrational transitions from a statistical distribution of excited states. 
Due to anharmonicity, each of these states is expected to have a shifted position with respect to the corresponding 1-0 fundamental transition.
Temperature-dependent effects of anharmonicity in modelling astronomical PAH spectra received only relatively sparse attention.\cite{Pech2002, Verstraete2001} These works followed measurements on a few neutral gas-phase PAHs, which reported the overall band shift and broadening of the most intense fundamental bands with temperature up to $\sim$900\,K.\cite{Joblin1995} Similar data became also available from theory. \cite{Basire2009, Joalland2010, Parneix2013}
In most AIB modelling studies though, theoretical calculations are performed at the harmonic approximation level.
The harmonic frequencies and first derivatives of the dipole moment are easily obtained using commonly available quantum chemistry computer codes that implement many different levels of theory for the electronic states of molecules such as the Hartree-Fock, Density Functional Theory (DFT) or Coupled-Cluster levels. Databases of such harmonic computed spectra are available\cite{Bauschlicher2010,Malloci2007} and include PAHs with up to a few hundreds C atoms. Anharmonicity in the Hamiltonian is then typically taken into account by some overall frequency scaling factor chosen in order to obtain  a better agreement with laboratory data measured in rare-gas matrices at low temperature. 
Note that laboratory data are themselves affected by matrix effects which can be modeled as anharmonicity effects.\cite{Cassam2008}
Scaling procedures can be refined in order to take into account the nature of the vibrational modes.\cite{Pauzat1995,Pauzat1997}
In any case, this completely neglects the anharmonic shifts of hot bands with respect to the fundamental, just correcting (empirically) for the effect of anharmonicity on the absolute position of the 1-0 transitions.
In order to perform simulations that can be compared with the AIB spectrum, temperature effects are then only simply included by adding some ad-hoc band redshift and broadening, see e.~g. Ref.~\citenum{Boersma2010} and the discussion of synthetic PAH spectra by Boersma \emph{et al.}\cite{Boersma2010} and Pauzat.\cite{Pauzat2011} 

In order to progress in the analysis of the astrophysical spectra, one has clearly to go beyond this simplified treatment. Efforts should be dedicated to describe connected states at non-zero temperatures. For this, the approximation of the electric dipole moment by its Taylor expansion truncated to the linear terms is not sufficient. Indeed this approximation only describes vibrational transitions connecting harmonic states which differ by only one quantum in only one (IR-active) vibrational mode. Any other vibrational transitions, including e.g. overtones, combination and difference bands, are thus completely neglected at this level of treatment.
The reason for using such seemingly crude approximations is that performing actual anharmonic calculations for anything but the smallest PAHs quickly becomes prohibitively expensive from a computational point of view. Complete variational calculations of rovibrational levels with  \textit{ab-initio} potential energy and electric dipole surfaces are only feasible for very small molecules with up to four or five atoms. \cite{Bowman2003,Tennyson2012,Pavlyuchko2015} However, the field is making constant progress and variational vibrational calculations can now be achieved for systems with up 11 atoms.\cite{Thomas2017} Vibrational Self Consistent Field (VSCF) methods, based on the \emph{ansatz} that the vibrational wave-function can be represented by a product of functions depending on a single coordinate,\cite{Bowman2008,Roy2013} can treat systems as large as a couple of hundreds of atoms, including PAHs. \cite{Hansen2010} However, they neglect all correlation between different degrees of freedom and are generally not accurate enough for our purposes. Techniques like the Vibrational Mean Field Configuration Interaction (VMFCI) \cite{Cassam2006,Cassam2012} improve on this by contracting several modes into one to describe correlated wave-functions, and can scale up to molecules as big as naphthalene (C$_{10}$H$_{8}$), provided that the modes involved in important resonances form disjoint subsets of limited sizes. For much larger molecules, perturbation theory up to second order (VPT2) or beyond,\cite{Mills1972} is the method of choice to treat inter-modes correlation starting from VSCF solutions, or both intra-mode anharmonicity and inter-modes correlation starting from the harmonic approximation. However, this treatment breaks down in the case of near degeneracy, be it accidental or due to symmetry.  While analytical VPT2 equations have existed for long,  symmetric and spherical tops deserve special care due to symmetry degeneracies of modes \cite{Gaw1991}, furthermore, accidental near-degeneracies of energy levels have to be detected and treated in a specific manner. Since in PAHs vibrational modes of the same type cluster at given frequencies, near-degeneracies are unavoidable for these molecules, in particular in the spectral region of C-H stretches and C-H bends. Resonances must therefore be taken out of the perturbation treatment, and explicitly accounted for, without omitting any interacting state nor resonant term of the vibrational potential, using for example the generalized second-order vibrational perturbation theory (GVPT2).\cite{Gaw1991,Martin1995,Barone2005,Piccardo2015}

This GVPT2 approach requires to first define some appropriate thresholds as a function of the desired target accuracy to determine which terms are to be considered ``resonances'' and which can safely be treated using perturbation theory. Then, one defines polyads of interacting harmonic vibrational states connected by resonant terms and solves the corresponding variational problems, while perturbative corrections for non-resonant terms can be added either before or after. In principle, for vanishingly small thresholds all terms are considered resonances and one goes back to the limit of a full variational calculation in a basis of harmonic vibrational states. In general, in addition to thresholds for discriminating between resonant and non-resonant terms, one also needs to implement some truncation scheme to keep the size of the polyads and of the associated variational problems small enough to be tractable without degrading the accuracy of the results. This kind of calculation has been implemented and successfully applied to compute the vibrational spectrum of PAHs involving transitions from the ground 
vibrational state and possibly some of the lowest lying ones in energy. \cite{Miani2000,Pirali2009,Mackie2015} Solving for all vibrational states up to higher energies quickly becomes prohibitive, since the number of states involved explodes, behaving as a multi-factorial with energy. However, if one is interested in reproducing the overall envelope of the spectral profile resulting from the superposition of hot bands as a function of vibrational excitation (or temperature), a possible approach is to give up a complete solution in favor of a Monte Carlo sampling of vibrational states.\cite{Parneix2013,Basire2009,Basire2011,Falvo:2012vn,Feraud:2014yg} We developed and implemented a code suitable for this approach,\cite{anharmonicaos} and in this paper we test its behaviour benchmarking its calculations of the 0~K spectrum against both experimental measurements for some small-medium PAHs and very high-level VMFCI calculations for a smaller species, namely ethylene oxyde.
In Section~\ref{computational} we describe the Van Vleck perturbation approach and we give computational details. Then in Section~\ref{results} we proceed to compare the outcomes of our calculations with reference calculations, which we use to define a strategy for choosing tuning parameters that provide both acceptable accuracies and computational costs. We also compare with experimental results, assessing the accuracy that can be expected from these calculations for fundamental bands, bands becoming weakly active due to resonances, and also overtone/combination/difference bands stemming from transitions mediated by the second order terms in the expansion of the electric dipole moment. In Section~\ref{discussion} we discuss our results, assessing the applicability of the method and its estimated computational costs for future calculations of spectra of PAHs and similarly rigid molecules considering all significantly populated levels at moderately high T using Monte Carlo sampling.
 
\section{Computational formalism} \label{computational}
\subsection{Van Vleck perturbation theory}
\label{vanvleck}
In this section, we give a brief review of the Van Vleck approach to perturbation theory applied to molecular vibrations.\cite{Sibert:1988kx,Sibert:1988aa,McCoy1992,Sibert:2013aa}
The anharmonic vibrational Hamiltonian $H$ is written as
\begin{equation}
H = H^{(0)}+\lambda H^{(1)}+\lambda^2 H^{(2)},
\label{eq:anharmH}
\end{equation}
where $\lambda$ is the perturbation parameter. The zero-order Hamiltonian is given as the harmonic normal-mode Hamiltonian
\begin{equation}
{H}^{(0)} =  H^{\text{harm}} = \sum_i \frac{\hbar\omega_i}{2}\left(p_i^2 + q_i^2 \right),
\label{eq:anharmH0}
\end{equation}
and the first- and second-order Hamiltonian are given by the cubic and quartic expansion of the vibrational potential
\begin{align}
H^{(1)} &= \frac{1}{3!} \sum_{ijk} \left(\frac{\partial^3 V}{\partial q_i \partial q_j \partial q_k}\right)_0 q_i q_j q_k, \label{eq:anharmH1}\\
{H}^{(2)} &= \frac{1}{4!} \sum_{ijkl}{\left(\frac{\partial^4 V}{\partial q_i \partial q_j \partial q_k \partial q_l}\right)_0 q_i q_j q_k q_l}.
\label{eq:anharmH2}
\end{align}
Higher-order terms in the potential are neglected.
The Van Vleck procedure relies on an infinitesimal contact transformation which will lead to an effective Hamiltonian. The contact transformation is defined by \mbox{$T=\exp\left(S\right)$}  where the operator $S$ is anti-hermitian \mbox{$S^\dagger = -S$}. This operator can be expanded up to the second-order in the perturbation parameter 
\begin{equation} \label{sdefinition}
{S} = \lambda {S}^{(1)}+\lambda^2 {S}^{(2)}+\ldots
\end{equation}
The transformed Hamiltonian $\widetilde{H}=T^\dagger H T$ up to the second-order in the pertubation parameter is then written as
\begin{equation}
\widetilde{{H}} = \widetilde{{H}}^{(0)}+\lambda \widetilde{{H}}^{(1)}+\lambda^2 \widetilde{{H}}^{(2)}+\ldots,
\label{eq:transfo}
\end{equation}
where 
\begin{align}
\widetilde{{H}}^{(0)} &= {H}^{(0)},\\
\widetilde{{H}}^{(1)} &= {H}^{(1)}-\left[{S}^{(1)},{H}^{(0)}\right],\\
\widetilde{{H}}^{(2)} &= {H}^{(2)} -
\left[{S}^{(2)},{H}^{(0)}\right] -
\left[{S}^{(1)},{H}^{(1)}\right] \label{eq:H2} \\
& + \frac{1}{2}\left[{S}^{(1)},
\left[{S}^{(1)},{H}^{(0)}\right]\right].\nonumber
\end{align}
Each term of the contact transformation ${S}^{(i)}$ is chosen such that it cancels out all non-diagonal terms of $\widetilde{{H}}^{(i)}$. This is in general possible as long as no resonance occurs. For example, in the case of the anharmonic vibrational Hamiltonian of Eqs.~(\ref{eq:anharmH}),  (\ref{eq:anharmH0}), (\ref{eq:anharmH1}) and (\ref{eq:anharmH2}), one can find in Ref.~\citenum{Herman:1948ys} the following expression for ${S}^{(1)}$
\begin{multline}
{S}^{(1)} = \sum_{ijk}  \frac{\text{i}\omega_k}{3!\Delta_{ijk}} \left(\frac{\partial^3 V}{\partial q_i \partial q_j \partial q_k}\right)_0 \Big( 2\omega_i\omega_j p_i p_j p_k \\ + \left(\omega_i^2 + \omega_j^2 - \omega_k^2\right) \left( q_i q_j p_k + q_i p_k q_j + p_k q_i q_j \right) \Big),
\label{eq:S1}
\end{multline}
where $\Delta_{ijk}$ is defined by
\begin{multline}
\Delta_{ijk} = \left(\omega_i - \omega_j - \omega_k\right)\left(\omega_i + \omega_j - \omega_k\right) \\ \times \left(\omega_i - \omega_j + \omega_k\right)\left(\omega_i + \omega_j + \omega_k\right)
\end{multline}
This expression can then be used in Eq.~(\ref{eq:H2}) and $S^{(2)}$ can be determined. In fact, at this point, it is not necessary to obtain an explicit expression for $S^{(2)}$. One just have to assume that the second order transformation exists and is not divergent, which is the case if no resonances occur.  Then, the eigenvalues of the effective Hamiltonian  are given by the usual Dunham expansion
\begin{multline}
E_{\mathbf{n}} = \chi_0 + \sum_i \omega_i \left(n_i + 1/2\right) \\ + \sum_{i\leqslant j} \chi_{ij} \left(n_i + 1/2\right) \left(n_j + 1/2\right)
\end{multline}
where the expression for the anharmonic coefficients $\chi_0$, and $\chi_{ij}$ can be found easily in the literature.\cite{Mills1972,Barone2005,Herman:1948ys}\par
In the case of exact or near resonances, the expression  for the first order contact transformation Eq.~(\ref{eq:S1}) cannot be used  as it is.  Indeed, Fermi resonances defined by the occurrences of triplets $i$,$j$ and $k$ such that  $\pm \omega_i \pm \omega_j \pm \omega_k \simeq 0$ would result in a diverging first-order contact transformation. This implies that the contact transformation should be determined to infinite order. In the case of quasi-degenerate perturbation theory these resonant terms are simply excluded from the definition of the contact transformation. As a result the first-order transformed Hamiltonian $\widetilde{{H}}^{(1)}$ is not diagonal.
Similarly the occurrence of Darling-Dennison\cite{Darling:1940fk} (DD) resonances arises by diverging terms in the second-order  contact transformation $S^{(2)}$ and occurring when $\pm \omega_i \pm \omega_j \pm \omega_k\pm \omega_l \simeq 0$. These terms should also be excluded, therefore preventing for a full cancellation of the non-diagonal terms in the second-order transformed Hamiltonian $\widetilde{{H}}^{(2)}$.  Note that DD resonances can occur even if there are no corresponding quartic derivatives since quartic coupling terms appear due to the application of the first order contact transformation. After using the contact transformation, we obtain an effective and non-diagonal Hamiltonian $\widetilde{{H}}$ whose eigenvalues are identical to the Hamiltonian $H$ at the second order of the perturbation. This effective Hamiltonian can be diagonalized using a finite variational basis.\par
%
After being determined, the infinitesimal contact transformation can be applied to any operators such as the dipole operator $\bm{\mu}$ resulting in a transformed operator accounting for both mechanical and electric anharmonicity up to the second order in the perturbation parameter. Similarly as for the Hamiltonian, the dipole operator is expanded as
\begin{equation}
\bm{\mu} = \bm{\mu}^{(0)} + \lambda \bm{\mu}^{(1)} + \lambda^2 \bm{\mu}^{(2)} + \ldots ,
\end{equation}
where $\bm{\mu}^{(0)}$ is the permanent dipole, and where $\bm{\mu}^{(1)}$ and $\bm{\mu}^{(2)}$ are respectively the linear and quadratic terms of the expansion of the dipole as a function of the normal mode coordinates about the equilibrium position, they are written as
\begin{align}
&\bm{\mu}^{(1)} =  \sum_i \bm{\mu}_i^{(1)} q_i =
\sum_i \left(\frac{\partial \bm{\mu}}
{\partial q_i} \right)_0 q_i,\\
&\bm{\mu}^{(2)} = 
\frac{1}{2} \sum_{ij} \bm{\mu}_{ij}^{(2)} q_i q_j =
\frac{1}{2} \sum_{ij} \left(\frac{\partial^2 
\bm{\mu}} {\partial q_i \partial q_j} \right)_0 
q_i q_j.
\end{align}
Applying the perturbative transformation, the expression for the transformed dipole operator up to the second-order in the perturbation parameter is given by
\begin{equation} \label{smutransform}
\widetilde{\bm{\mu}} =  \bm{\mu}^{(0)} + 
\lambda \bm{\mu}^{(1)} + 
\lambda^2 \left(\bm{\mu}^{(2)}  -
\left[{S}^{(1)}, \bm{\mu}^{(1)} \right] 
\right) + \ldots,
\end{equation}
where one should remember that resonant terms are omitted from the definition of ${S}^{(1)}$, as explained before.
The intensities of the transitions are then calculated by computing the matrix elements of the dipole operator between two eigenstates of the effective Hamiltonian $\widetilde{H}$. The intensity of a vibrational transition,  neglecting rotational degrees of freedom, is then written as
\begin{equation}
I_{ab} = \frac{2\pi}{3\hbar c \epsilon_0} \omega_{ab} \langle \left| \psi_a | \widetilde{\bm{\mu}} | \psi_b \rangle \right|^2,
\label{eq:lineint}
\end{equation}
where $\omega_{ab} = \omega_b - \omega_a$ are the transitions between eigen-frequencies and where $| \psi_a \rangle$ are the eigenstates of $\tilde{H}$. $I_{ab}$ is  comparable to the experimental band intensity integrated over its rotational substructure.
%
%
\subsection{Polyad construction and variational problem}
In this section we give the computational details of our implementation of the Van Vleck perturbation theory described in Sec.~\ref{vanvleck}. The computational protocol described in this section was implemented in our code: AnharmoniCaOs (the Cagliari-Orsay model for anharmonic molecular spectra in 2nd order perturbation theory).
\par
The first step is to identify resonances. For this purpose, starting from an initial harmonic state $\bm{n} = (n_1,\dots,n_N)$, we first identify all harmonic states $\bm{n}'\neq \bm{n}$ directly connected to $\bm{n}$ through the cubic couplings. For these states, the ratio of the coupling term $V^{(1)}_{\bm{n}\bm{n}'}=\langle \bm{n} | H^{(1)} | \bm{n}' \rangle$ and the harmonic energy difference between these states is used to define a resonance. A Fermi resonance occurs then if 
\begin{equation}
\left| \frac{V^{(1)}_{\bm{n}\bm{n}'}}{E^{(0)}_{\bm{n}}-E^{(0)}_{\bm{n}'}}\right| \geqslant r_3
\label{eq:resonfermi}
\end{equation}
where $r_3$ is a small threshold parameter. A list of Fermi resonances is thus built. From this list the first-order contact transformation ${S}^{(1)}$ can be defined by excluding all terms corresponding to a resonance as in Eq.~(\ref{eq:resonfermi}). Applying the first-order contact transformation ${S}^{(1)}$ onto the Hamiltonian $H$ we obtain the second-order part of the transformed Hamiltonian
\begin{equation}
{H}^{(2)}-\left[{S}^{(1)},{H}^{(1)}\right]+\frac{1}{2}\left[{S}^{(1)},\left[{S}^{(1)},{H}^{(0)}\right]\right].
\label{eq:op2nd}
\end{equation}
This operator is then used to identify harmonic states $\bm{n}''\neq \bm{n}$ directly connected to the initial state $\bm{n}$ through the quartic couplings which include the effect of the first-order transformation. Darling-Dennison (DD) resonances are then defined by
\begin{equation}
\left| \frac{V^{(2)}_{\bm{n}\bm{n}''}}{E^{(0)}_{\bm{n}}-E^{(0)}_{\bm{n}''}}\right| \geqslant r_4,
\label{eq:resondd}
\end{equation}
where $V^{(2)}_{\bm{n}\bm{n}''}$ is non-diagonal matrix element of operator~(\ref{eq:op2nd}). The list of DD resonances having been built, the second-order contact transformation can be defined implicitly. This transformation is used to define the  transformed Hamiltonian $\widetilde{{H}}$ (Eq.~\ref{eq:transfo}).\par
\begin{figure*}
\includegraphics[width=0.9\textwidth]{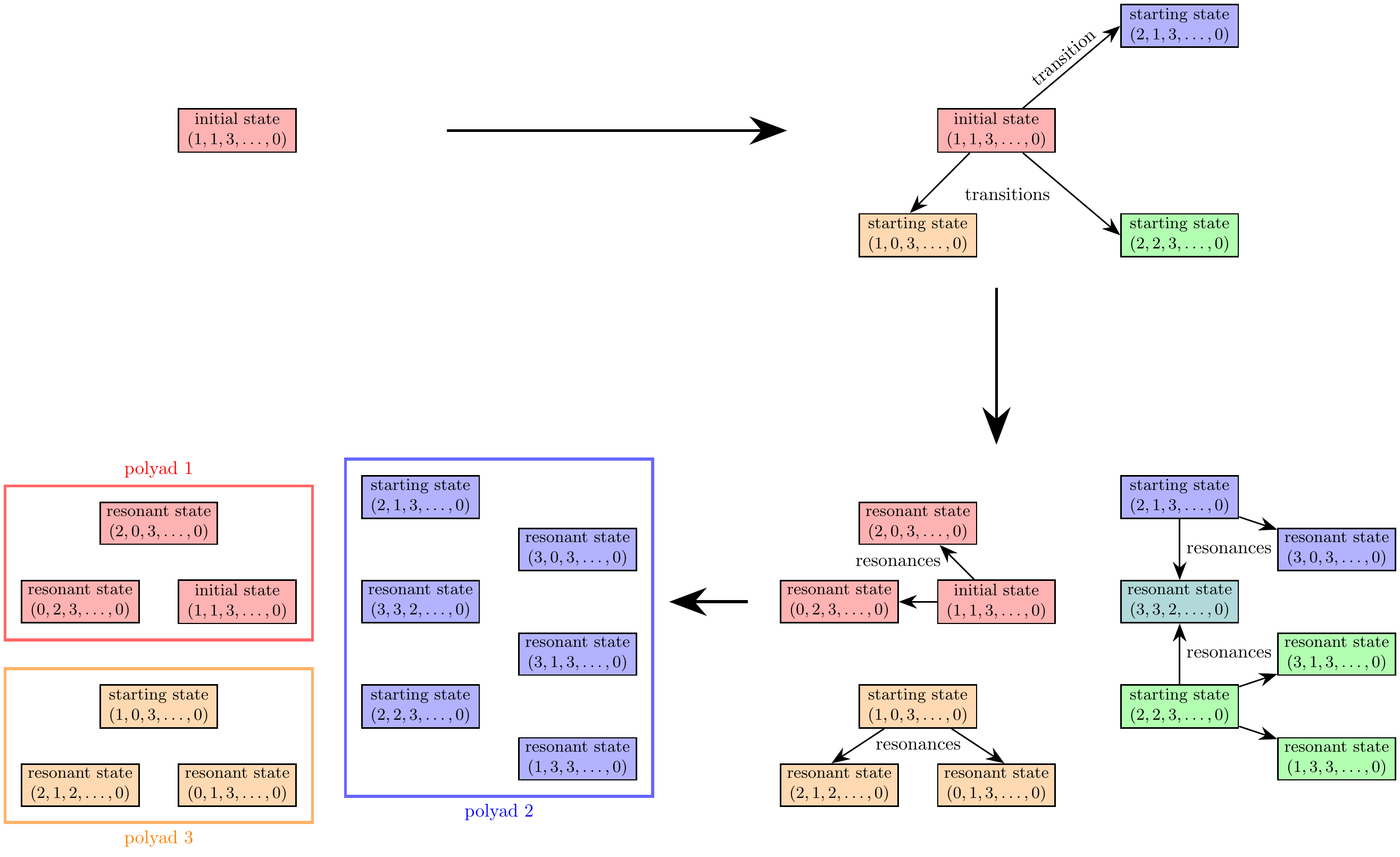}
\caption{Schematic algorithm of polyad definition}\label{polyadfigure}
\end{figure*}

After this preliminary work, polyads are iteratively constructed as resumed in Fig.~\ref{polyadfigure}. A list of starting states is built by including the  initial harmonic state $\bm{n}$ and the final harmonic states $\widehat{\bm{n}}$ 
reached by a dipolar transition through the transformed dipole operator $\widetilde{\bm{\mu}}$. Each starting state will constitute the start of a polyad. Running trough the resonances the list of states in each polyad is increased iteratively. When two polyads share a common state, they are merged.
Symmetry can be taken advantage of by considering only states of a given symmetry type at a time.  This way, one could think of separating  explicitly the problem into a number of subproblems, However, in practice, given the way in which polyads are recursively built by following resonances, resonant terms can only be non-zero between harmonic states of the same symmetry. Hence symmetry separation is automatically enforced and there would not be any significant performance gain by implementing the explicit separation, which would make the code more complicated. It is only necessary to include a cutoff parameter to remove the numerical noise, introduced by cubic and quartic, non-symmetry-adapted Hamiltonian terms, in case the separate code used to obtain the potential does not explicitly use symmetry itself and thus produces very small non-zero spurious terms. \par
This procedure, depending on how resonances combine, can lead to polyads of very large, in principle even infinite, size. To keep the problem tractable, a truncation mechanism is therefore necessary. To obtain finite size polyads we used a ``cost model''.  The very initial harmonic states, i.~e. the initial one and the ones connected to it by permitted transitions, are assigned a ``budget'' initialized to 1. Additional states newly added to a polyad ``inherit'' the budget of the state they are connected to, minus a ``cost'' which is inversely proportional to the ``strength'' of the resonance (the ratio between the non-diagonal element and the difference of the diagonal elements of the connected harmonic states) divided by a tuning parameter $h$. Therefore, a very strong resonance will add many states to a polyad, whereas a weak one will add few, possibly only one. Larger values of $h$ will produce larger polyads, reducing truncation errors at the price of an increased computational cost.

To each distinct polyad corresponds a variational problem, for which an effective Hamiltonian is defined. What we want to obtain is an accurate description of the anharmonic states containing a non-negligible component of the initial harmonic starting states (the very first harmonic state and the ones connected to it by allowed transitions). Depending on how strong resonances are, and on chosen tuning parameter values, the eigensolutions we need may be a small fraction of the size of a given polyad. If the polyad is relatively small, complete solution using a standard direct method (e.g. Divide and Conquer \cite{Cuppen1981} or Relatively Robust Representations \cite{Dhillon1997}) is more efficient. If, instead, the polyad is large (e.~g. tens of thousands of harmonic states or more) and a small fraction of the eigensolutions is needed (e.~g. less than $\sim$10\%), then an iterative method like Jacobi-Davidson,\cite{Sleijpen1996,Sleijpen2000} explicitly tuned to select the eigenvalues with the largest component in the space of the initial starting states, may become more competitive. After the eigensolutions are found, the line intensities between anharmonic states are computed using Eq.~(\ref{eq:lineint}). 
The final spectrum is then built from individual transitions weighted by the square of the component of the starting state it contains. In this way, when the same polyad is obtained more than once from different starting states, each transition will eventually converge to its exact value for complete coverage of the starting states.

%
%
\par
\subsection{Electronic structure calculations \label{elecstructure}}
The Van-Vleck method described in the previous sections relies on the use of a quartic force field corresponding to harmonic frequencies, cubic and quartic derivatives of the potential energy surface as well as the first and second derivatives of the dipole moment. These parameters are easily obtained from  quantum chemical calculations. All electronic structure calculations were performed using the Gaussian09 suite of programs.\cite{Frisch2009} Geometry optimizations and frequency calculations were performed using density functional theory (DFT) with the hybrid functional B97-1,\cite{Hamprecht1998} and with the TZ2P \cite{VanLenthe2003} and 6-31G* \cite{Ditchfield1971,Hehre1972} basis sets. We chose this exchange-correlation functional over some more commonly used ones as e.g. B3-LYP \cite{Becke1993,Lee1988} because the former was found to provide more accurate band position and intensities in particular for aromatic molecules.\cite{Miani2000,Cane2007,Pirali2009} As in Ref.~\citenum{Cane2007} we used a (150,770) grid for the Kohn-Sham (KS) integration and a (75,194) grid for the Coupled Perturbed Kohn-Sham (CPKS) steps.\par
For each molecule, geometry optimization was first performed, then harmonic frequencies and normal modes were computed. Cubic and quartic normal mode derivatives of the potential around the equilibrium position were obtained using numerical differentiation of the analytical Hessian  matrix, using a displacement step of \mbox{0.01 \AA$\cdot$amu$^{1/2}$}. The calculation of all quartic derivatives  requires $36N^2$ Hessian calculations, where $N$ is the number of atoms. For large systems like pyrene or coronene such a calculation could only be performed with semi-empirical force-field and not with \emph{ab-initio} or DFT methods. Note that only the semi-diagonal involving at most two different indices are needed for perturbation theory, the other terms just contribute to DD resonances. Therefore, we only computed the quartic derivatives with two identical indices $\displaystyle \left(\frac{\partial^4 V}{\partial q_i \partial q_i \partial q_j \partial q_k}\right)_0$. Similarly, second derivatives of the dipole moment were obtained by numerical differentiation of the analytical dipole first derivatives. Invariance of derivatives on the order of differentiation was used to check the numerical stability.
\section{Results} \label{results}

We now compare the results we obtained with AnharmoniCaOs with some reference calculations and experimental results. Comparison with reference high-level vibrational calculations performed with the VMFCI method using exactly the same quartic force field will enable us to benchmark the effect of the different user-defined thresholds. In this way we can separate the effect of using the van Vleck method, and different levels of resonance detection and truncation of the size of variational calculations, from the effect of using a quartic force field and, in turn, from the level of theory used to obtain it. On the other hand, comparison with available experimental data for PAHs enables us to gauge the accuracy of this method \emph{including also} the effect of truncating the nuclear potential to the quartic expansion and of the level of theory used to compute it. In addition, we perform some convergence tests on pyrene and coronene, to study the accuracy vs. computational cost as a function of the tunable parameters of AnharmoniCaOs specifically for the family of PAHs.

\subsection{Comparison with VMFCI calculations for Ethylene oxyde}\label{sec:ethox}

A new VMFCI calculation of all vibrational energy levels of ethylene oxyde up to about $3600$ cm$^{-1}$  has been performed (see supplementary material). The results compare favorably with previously  published ones,\cite{begue2007comparison,Thomas2015,Thomas2017,Odunlami2017} using the same quartic force field for the nuclear potential. We  used the very same force field with AnharmoniCaOs, and compared the results with the VMFCI ones for different values of the tuning parameters $r_3$, $r_4$, and $h$. Table~\ref{tab:c2h4o_1} shows how the accuracy of AnharmoniCaOs changes for decreasing values of $r=r_3=r_4$, while keeping fixed the parameter $h=4$. Decreasing values of $r$ cause more and more terms to be taken out of the perturbative treatment, and instead considered as resonances, in setting up and solving polyads in a variational way. For comparison, we also list in the same Table~\ref{tab:c2h4o_1} the VSCF results, which we obtained as an intermediate step of our reference VMFCI calculation. Clearly, the VSCF discrepancies are much larger than those of our GVPT2 results, for all sets of tuning parameters used.
\begin{table*}
\begin{center}
\begin{tabular*}{\hsize}{@{\extracolsep{\fill}}cccccccccc}
\hline
mode & VMFCI & VSCF & $r$=1.0 & $r$=0.5 & $r$=0.3 & $r$=0.2 & $r$=0.1 & $r$=0.05 & $r$=0.01\\
\hline
1	&2920$^a$& 3015  &2955	&2955	&2918	&2918	&2918	&2920	&2930\\
2	&1496   & 1525	&1502	&1502	&1502	&1503	&1502	&1503	&1503\\
3	&1271	&1286	&1271	&1271	&1271	&1271	&1271	&1272	&1272\\
4	&1122	&1160	&1128	&1128	&1128	&1128	&1129	&1127	&1132\\
5	&879	&891	&878	&878	&878	&878	&879	&878	&879\\
6	&3029	&3108	&3043	&3043	&3043	&3043	&3060	&3065	&3097\\
7	&1148	&1172	&1154	&1154	&1154	&1154	&1156	&1156	&1166\\
8	&1018	&1050	&1025	&1025	&1025	&1025	&1027	&1028	&1037\\
9	&2910	&3032	&2920	&2920	&2920	&2921	&2921	&2920	&2949\\
10	&1468	&1487	&1474	&1474	&1474	&1474	&1475	&1475	&1482\\
11	&1124	&1158	&1131	&1131	&1131	&1131	&1132	&1132	&1133\\
12	&822	&842	&822	&822	&822	&822	&822	&822	&826\\
13	&3041	&3125	&3058	&3058	&3058	&3058	&3055	&3054	&3085\\
14	&1146	&1169	&1151	&1151	&1151	&1151	&1152	&1152	&1152\\
15	&793	&837	&802	&802	&802	&802	&802	&803	&803\\
\hline
\multicolumn{2}{c}{root mean square error} &55.3& 12.1	&12.1	&8.0	&8.0	&10.8	&11.8	&25.2\\
\hline
\end{tabular*}
\end{center}
\begin{justify} \footnotesize $^a$ VMFCI assignments of step $n$ are in terms of eigenstates of step $n-1$ and not in terms of the initial harmonic oscillator (HO) basis functions. So, we use Thomas \emph{et al.} \cite{Thomas2017} assignments to relate the VMFCI frequencies to those assigned in this work on the basis of the dominant HO basis function. In fact, VMFCI assignments coincides with those of Ref.~\citenum{Thomas2017} except for $\nu_1$ see supplementary material.
\end{justify}
\caption{Fundamental frequencies for ethylene oxyde as a function of the threshold $r$ and for $h=4$.} \label{tab:c2h4o_1}
\end{table*}
Conversely, Table~\ref{tab:c2h4o_2} compares the results of AnharmoniCaOs for fixed $r=r_3=r_4=0.05$ and varying $h$.
\begin{table}
\begin{tabular*}{\hsize}{@{\extracolsep{\fill}}cccccc}
\hline
mode & VMFCI & $h$=4 & $h$=8 & $h$=16 & $h$=18\\
\hline
1	&2920 \footnote{Same remark as in Tab.~\ref{tab:c2h4o_1}.}	&2920	&2920	&2920	&2918\\
2	&1496	&1503	&1503	&1503	&1503\\
3	&1271	&1272	&1272	&1271	&1271\\ 
4	&1122	&1127	&1127	&1127	&1127\\ 
5	&879	&878	&878	&878	&878\\  
6	&3029	&3065	&3065	&3053	&3043\\ 
7	&1148	&1156	&1156	&1155	&1155\\
8	&1018	&1028	&1028	&1027	&1026\\ 
9	&2910	&2920	&2919	&2918	&2917\\ 
10	&1468	&1475	&1474	&1474	&1474\\ 
11	&1124	&1132	&1132	&1132	&1132\\ 
12	&822	&822	&822	&822	&822\\  
13	&3041	&3054	&3054	&3051	&3050\\
14	&1146	&1152	&1152	&1152	&1152\\ 
15	&793	&803	&803	&803	&802\\ 
\hline
\multicolumn{2}{c}{root mean square error} & 11.8	&11.8	&8.9	&7.0\\
\hline
\end{tabular*}
\caption{Fundamental frequencies for ethylene oxyde as a function of the parameter $h$ and for $r=0.05$.} \label{tab:c2h4o_2}
\end{table}
Interestingly enough, the accuracy of AnharmoniCaOs steadily improves for larger and larger $h$ (at the price of a considerable increase in computational cost), while this is not the case for $r$. Indeed, we see in Table~\ref{tab:c2h4o_1} that the best accuracy, for $h=4$, is obtained with $r\approx0.3$, and it gets \emph{worse}, not better, if $r$ is reduced while keeping $h$ fixed. This is due to the way in which polyads are built, and their truncation mechanism: no matter how weak it is, a resonance \emph{always} adds at least one basis state to the starting ones. The idea behind this is that it is pointless to add a term to the list of resonances and then just neglect it because it is too weak. Conversely, if one decreases too much $r$ without in parallel increasing $h$ one gets an unbalanced truncation, with some weakly coupled basis states included (namely the ones connected to starting states via very weak resonances) while others, comparatively more strongly coupled, are not.
Indeed, Table~\ref{tab:c2h4o_2} shows that with $r=0.05$ one does obtain more accurate results than with $r=0.3$ \emph{provided} that one then uses a large enough value of $h$ to allow for a balanced truncation of polyads. All in all, it appears that $r\approx0.3$ and $h\approx4$ provide a good compromise in terms of accuracy versus computational cost.

\subsection{Convergence tests on pyrene}\label{sec:convpyrene}

The quartic force field and first and second derivatives of the electric dipole moment were obtained using DFT with the \mbox{B-971} \cite{Hamprecht1998} exchange-correlation functional and the \mbox{TZ2P} \cite{VanLenthe2003} Gaussian basis set, as described in Sect.~\ref{elecstructure}.
We then used the AnharmoniCaOs code with the resulting quartic force field and second order Taylor expansion of the electric dipole, using the harmonic ground vibrational state as a starting state. We started with $r = 0.3$ and $h = 4$, based on the results of Sect.~\ref{sec:ethox}, as a reasonable compromise between speed and accuracy, bearing in mind that our final goal will eventually be to perform (at least) tens or hundreds of thousands of such individual calculations to obtain a good enough Monte Carlo sampling of the energy dependence of vibrational bands. Still, in the case of pyrene, we conducted exploratory calculations where these parameters were pushed further, to check convergence with respect to the number of resonances and the number of states being included in the polyads to be explicitly diagonalised. For the smallest value of the threshold ($r=0.05$) we obtained very large polyads. The size of the largest polyad, keeping fixed $h = 8$, ramped up from $\sim 650$ with $r=0.3$, to $\sim 2700$ with $r=0.2$, to $\sim 15000$ with $r=0.1$, and to $\sim 36000$ with $r=0.05$ harmonic states. In particular, the polyads containing the states involved in the fundamental transitions in the C\textendash H region include states spanning an energy range increasing from $\sim$3015-5120~cm$^{-1}$ for $r=0.3$, to $\sim$2540-5120~cm$^{-1}$ for $r=0.2$, $\sim$500-9800~cm$^{-1}$ for $r=0.1$, and $\sim$450-10000~cm$^{-1}$ for $r=0.05$. 

\begin{figure}
\includegraphics[width=\hsize]{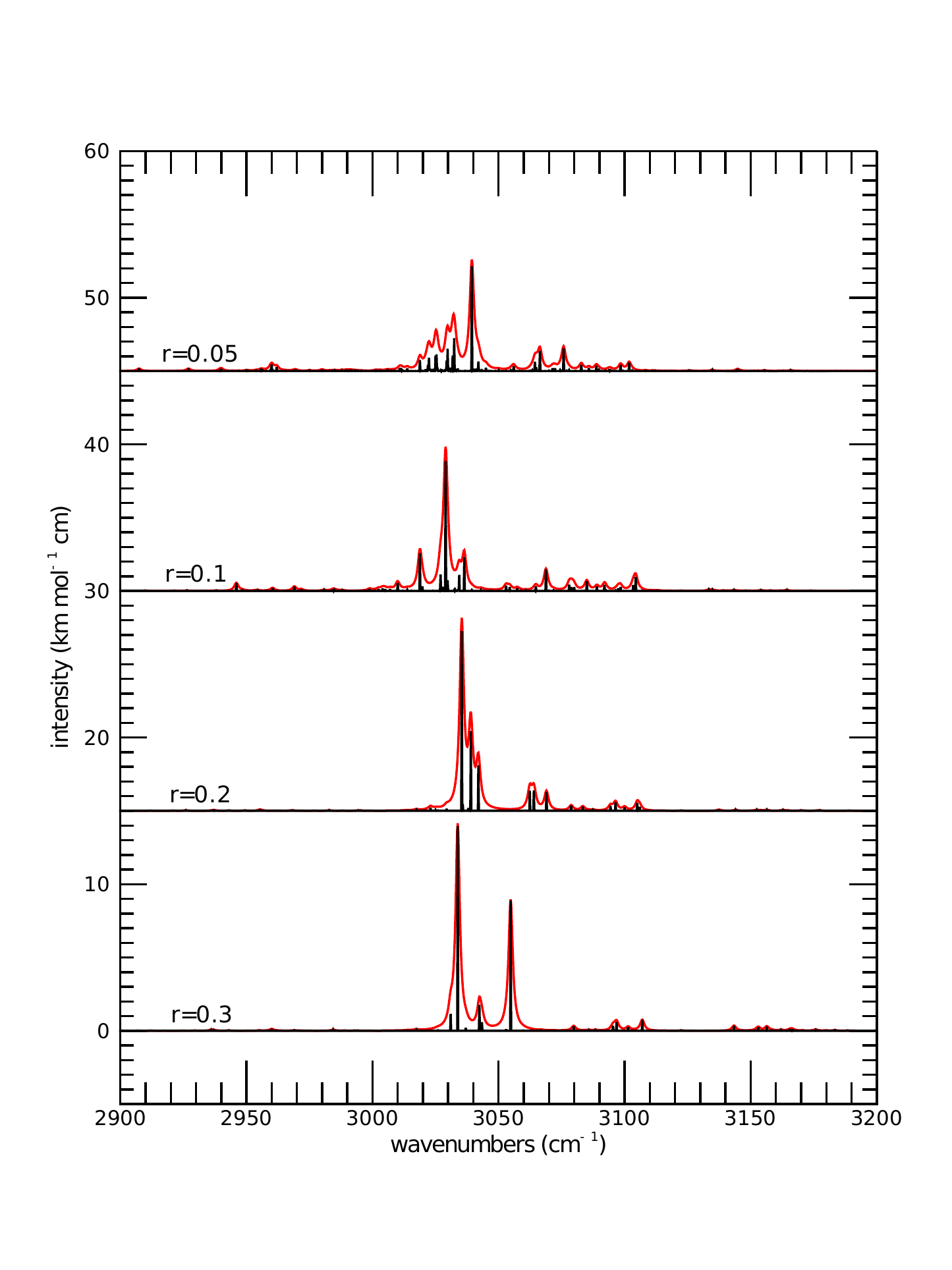}
\caption{Anharmonic spectrum of pyrene in the C\textendash H stretch region for various values of the threshold $r$, keeping fixed $h=8$. Black bars indicate the precise position of individual bands, whereas red envelopes are convolved with Lorentzian profiles with a 1 cm$^{-1}$ width. Spectra computed with different values of $r$ are shifted for clarity by multiples of 15~km~mol$^{-1}$~cm with respect to the previous one.}\label{plotpyrthres}
\end{figure}

\begin{figure}
\includegraphics[width=\hsize]{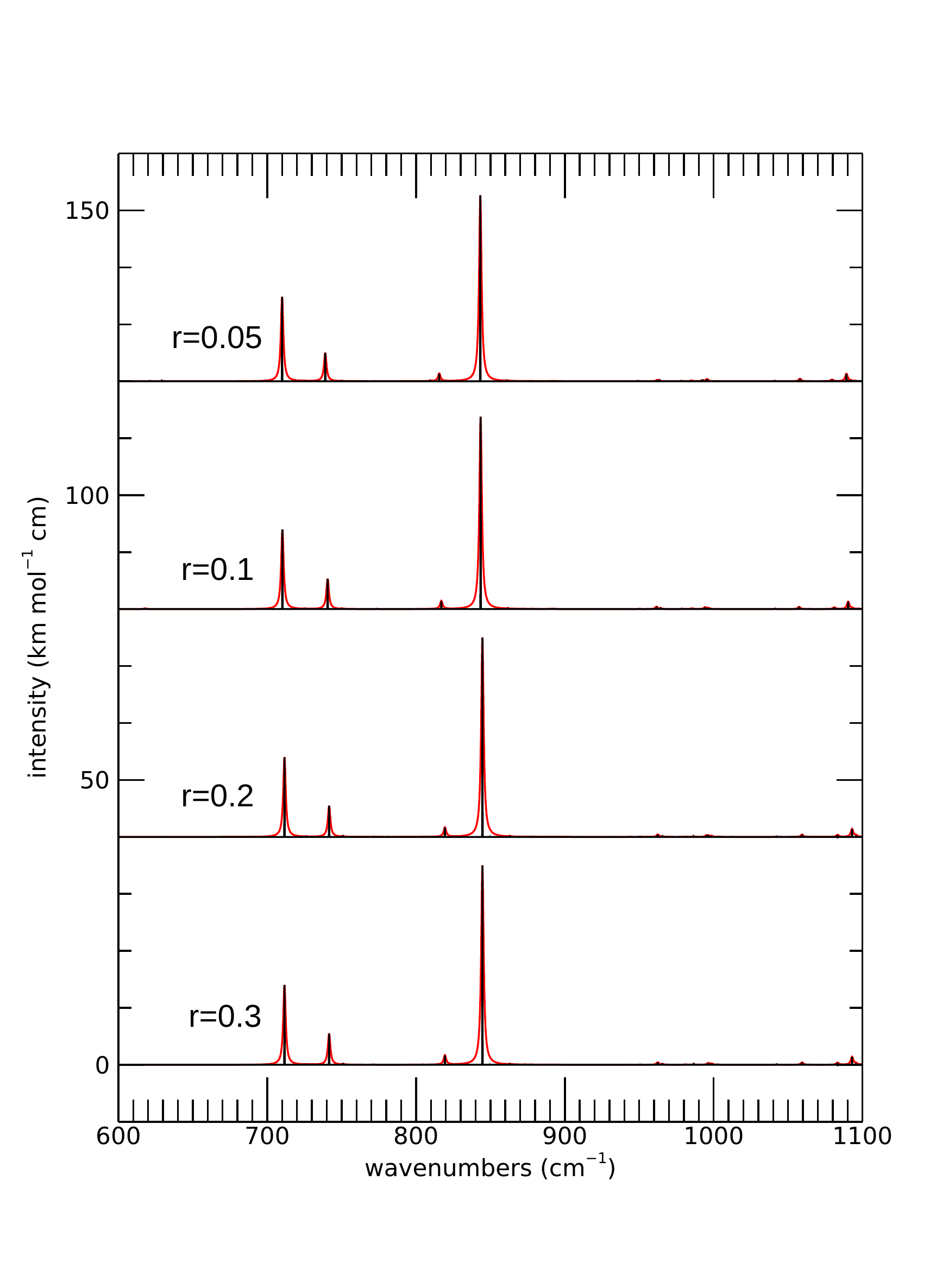}
\caption{Anharmonic spectrum of pyrene  from 600 to 1100~cm$^{-1}$ computed with various values of the threshold $r$, keeping fixed $h=8$. Black bars indicate the precise position of individual bands, whereas red envelopes are convolved with Lorentzian profiles with a 1 cm$^{-1}$ width. Spectra computed with different values of $r$ are shifted for clarity by multiples of 40~km~mol$^{-1}$~cm.}\label{plotpyrthres2}
\end{figure}

\begin{figure}
\includegraphics[width=\hsize]{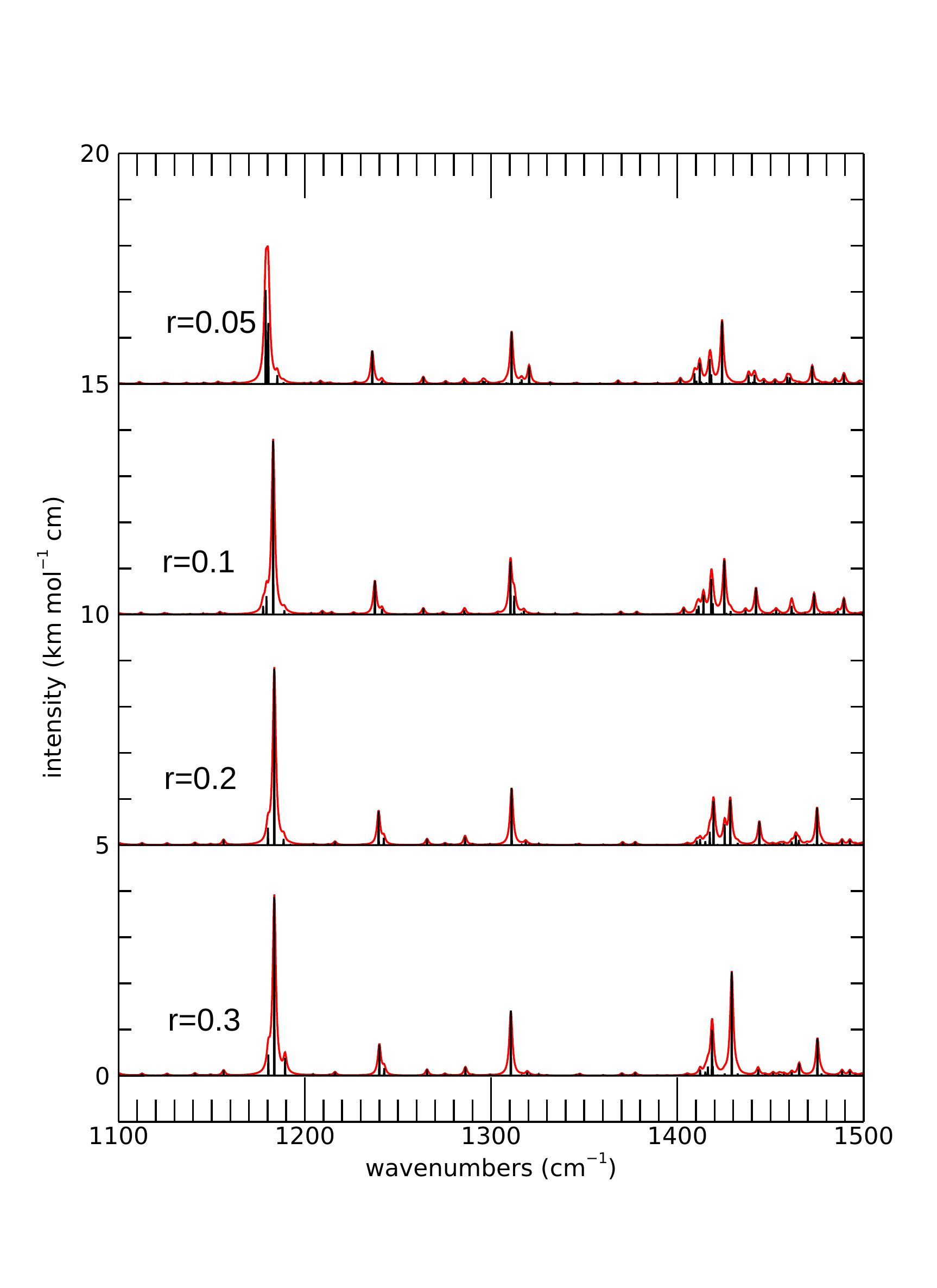}
\caption{Anharmonic spectrum of pyrene  from 1100 to 1500~cm$^{-1}$ computed with various values of the threshold $r$, keeping fixed $h=8$. Black bars indicate the precise position of individual bands, whereas red envelopes are convolved with Lorentzian profiles with a 1 cm$^{-1}$ width. Spectra computed with different values of $r$ are shifted for clarity by multiples of 5~km~mol$^{-1}$~cm.}\label{plotpyrthres3}
\end{figure}

\begin{figure}
\includegraphics[width=\hsize]{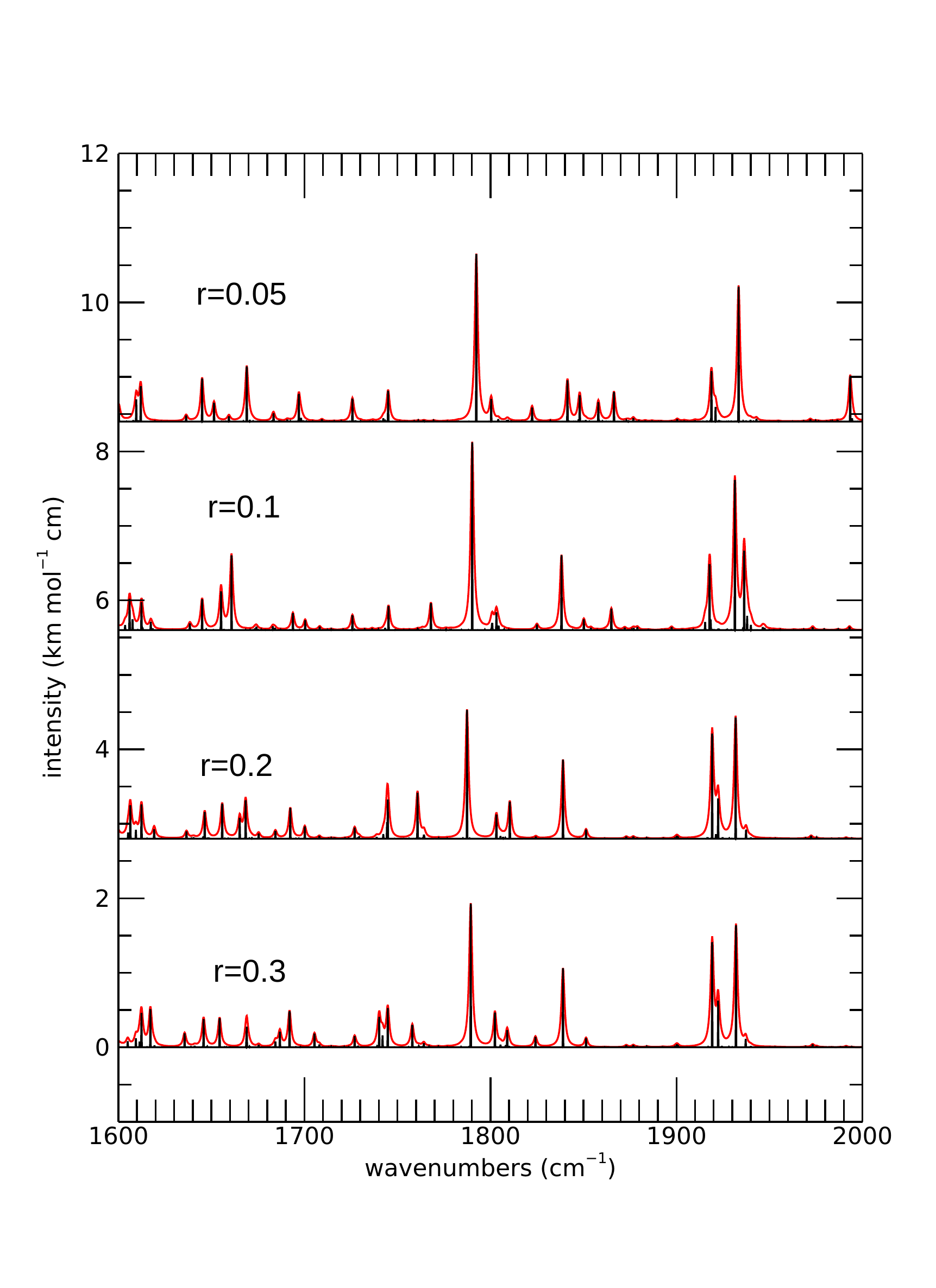}
\caption{Anharmonic spectrum of pyrene  from 1600 to 2000~cm$^{-1}$ computed with various values of the threshold $r$, keeping fixed $h=8$. Black bars indicate the precise position of individual bands, whereas red envelopes are convolved with Lorentzian profiles with a 1 cm$^{-1}$ width. Spectra computed with different values of $r$ are shifted for clarity by multiples of 2.8~km~mol$^{-1}$~cm.}\label{plotpyrthres4}
\end{figure}

\begin{figure}
\includegraphics[width=\hsize]{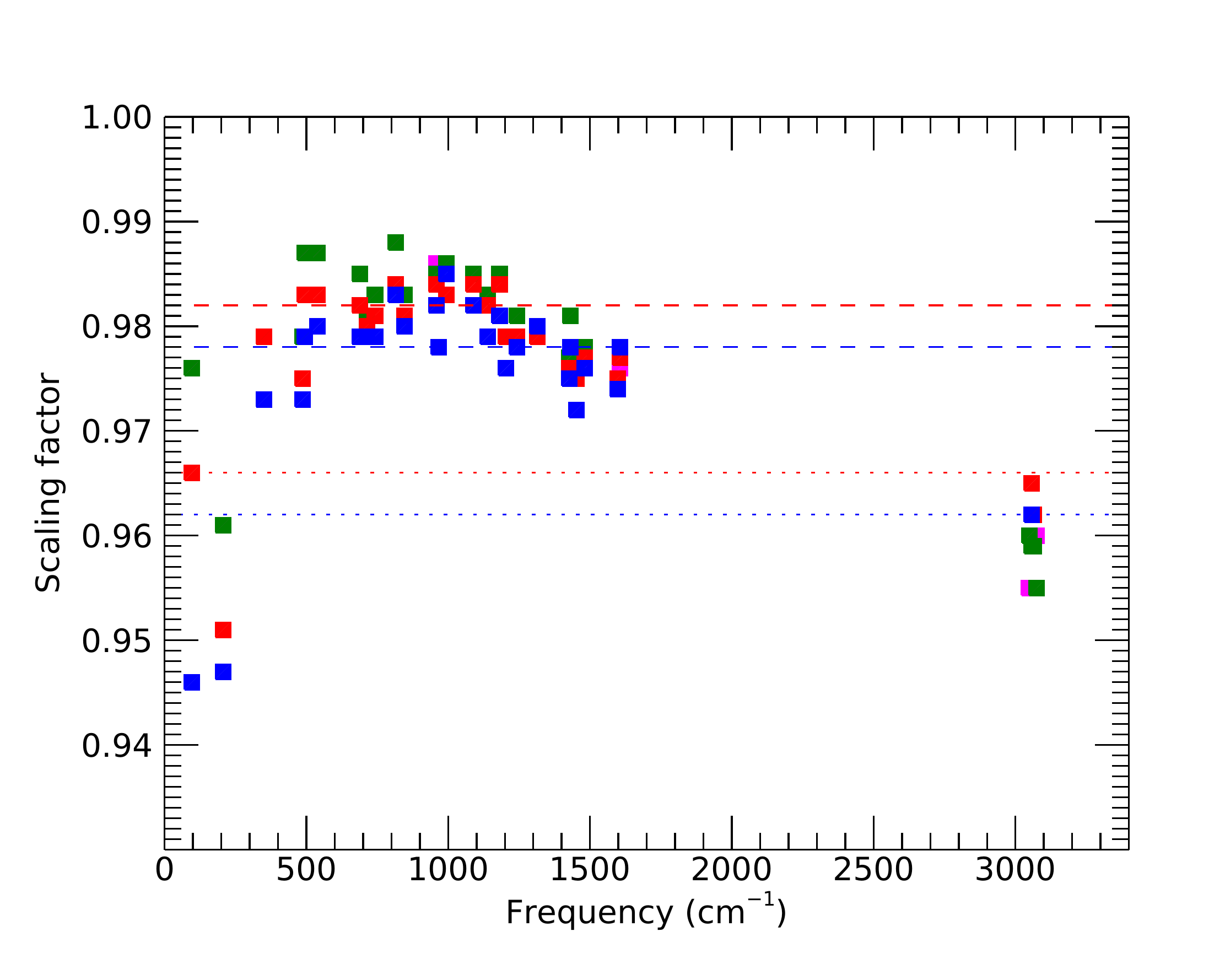}
\caption{Ratio between the computed anharmonic frequencies and the harmonic frequencies of unambiguously identified fundamentals of pyrene, as a function of the harmonic frequencies. Magenta squares are for $r=0.3$, green ones for $r=0.2$, red ones for $r=0.1$, blue ones for $r=0.05$. The overlaid lines represent the scaling factors for C-H stretches (dotted) and for all other bands (dashed). Red lines mark scaling factors from Ref.~\citenum{Cane2007}, namely 0.966 for C-H stretches and 0.982 for all the other bands. Blue lines mark scaling factors obtained from our best anharmonic calculation for pyrene, namely 0.962 for C-H stretches and 0.978 for all other bands except the two lowest modes, for which it is 0.946}\label{plotpyrscalings}
\end{figure}

Some spectra for various values of $r$ are shown in Fig.~\ref{plotpyrthres} for the region of C\textendash H stretches and in Figs.~\ref{plotpyrthres2}, \ref{plotpyrthres3},  and \ref{plotpyrthres4} for three other representative spectral ranges. 
We studied how the positions of specific bands change with different $r$ values. A large set of bands as they result from calculations with different $r$ values is given in Table~V in Supplementary Material. In some cases, most notably in the C-H stretch spectral region, with decreasing $r$ values some bands split in several ones due to resonances. This happens, for example, for the bands which are computed at 3034, 3037, and 3042 at the $r=0.3$ level, which split into a multitude of bands at smaller $r$ values. Some of these results are summarised in Fig.~\ref{plotpyrscalings}, which shows the ratio between anharmonic and harmonic frequencies of unambiguously identified fundamentals, for the different $r$ values.
In general, this exploration shows that $r = 0.3$ (and  $h = 8$) already provides an acceptable level of accuracy for most bands, when compared with harmonic calculations. In almost all cases in which the same fundamental band can be unambiguously identified for all $r$ values, the calculation for $r=0.3$ already provides $\geq$90\% of the anharmonic correction to frequencies. Going from $r = 0.3$ to $r = 0.05$ only changes the positions of fairly strong bands (i.~e. with peak intensities larger than $\sim$1~km~mol$^{-1}$~cm when convolved with a 1~cm$^{-1}$ wide Lorentzian) by less than $\sim$3-5~cm$^{-1}$, usually (but not always) slightly redwards. Only in very few particular cases, e.~g. around 1430~cm$^{-1}$, when some resonance is taken into account due to decreasing $r$, a significant band may split in two close ones, but the overall spectral structure does not change much anyway except for the C-H stretch region. Very weak bands (i.~e. bands with peak intensities $<1$~km~mol$^{-1}$~cm when convolved with a 1~cm$^{-1}$ wide Lorentzian) are more sensitive. However, in most cases this is because some of them are ``peripheric'' states in big polyads, ``borrowing'' just very little intensity from fundamentals to which they are very indirectly connected by resonances. These states are indeed expected to be less well described, due to basis truncation errors. 

Bands in the C\textendash H stretches region, in contrast to other strong ones, appear very sensitive to thresholds and more difficult to get to converge. This is clearly due to the crowding of a much larger number of resonating states than the ones involved in transitions in other spectral regions. When more and more resonances are included, and larger polyads considered, the fundamental bands split in a few fairly strong and a multitude of weak bands. In some cases, this makes it actually impossible to ``identify'' a given fundamental in this spectral region with any specific anharmonic transition in the $r=0.05$ calculation. Upon examining the spectra in Fig.~\ref{plotpyrthres}, a side effect of this is also apparent: the total, integrated band intensity somewhat decreases when more and more resonances are treated explicitly. This happens due to the multitude of very weak Fermi resonances. Fermi resonances cause some transitions, which would be IR-inactive in the double harmonic approximation, to ``borrow'' some intensity from a fundamental transition. When such a Fermi resonance is treated explicitly, the total intensity is conserved, i.~e. the ``borrowed'' intensity that appears from the band becoming active simultaneously disappears from the resonating fundamental. In contrast, when the same Fermi resonance is treated via a perturbative treatment, the ``borrowed'' intensity from a fundamental by a combination/difference band appears as a second order contribution, whereas the intensity decrease from the resonating fundamental appears as a third order one. Consequently, the latter is \emph{not} included in the second order perturbative treatment. Some caution is therefore in order when considering band intensities obtained by these kinds of calculations, when a large number of weak resonances can add up and reach a total ``borrowed'' intensity which amounts to a sizeable fraction of the fundamental transition intensities. We actually implemented in our code an optional intensity correction which enforces the conservation of total intensity when computing the contribution of mechanical anharmonicity to the van Vleck transformed second derivatives of the electric dipole moment. This correction is not really well-balanced, since it approximately includes only one selected 3$^{\rm rd}$ order correction to the transformed dipole moment operator, and not for example the one, at the same order, transferring intensity in the opposite direction from an IR-active combination/difference band (produced by a sizeable second derivative of the dipole moment itself) to a fundamental resonating with it. This new functionality should be more fully tested under conditions in which other sources of error are negligible, hence it is disabled by default in AnharmoniCaOs, we just use it in the next section to get an estimate of how large its effect might be.

We carried out similar convergence tests for coronene. Its quartic force field and derivatives of the dipole moment were obtained using DFT with the \mbox{B97-1} \cite{Hamprecht1998} exchange-correlation functional. However, for coronene we used the \mbox{6-31G*} Gaussian basis set to perform all the numerical derivatives required to obtain the quartic force field, since using the TZ2P \cite{VanLenthe2003} basis-set would be computationally too expensive. For AnharmoniCaOs we used $r=0.3$, $r=0.2$, and $r=0.1$, keeping fixed $h=6$. Calculations with $r=0.05$ and $h=8$ were computationally too expensive. The results of this exploration for coronene were by and large the same as for pyrene, with the C-H stretch region being the only fairly sensitive one to thresholds and relatively difficult to converge.

\subsection{Comparison with experimental results and previous theoretical calculations}\label{sec:experiment}

Benchmarking our calculations on PAHs at 0\,K would ideally require spectra recorded in gas-phase at very low temperatures. Such data are now becoming available for the CH stretch region.\cite{Maltseva2015,Maltseva2016} 
We therefore took the band positions reported for pyrene in ref.~\citenum{Maltseva2016}. Since this data is limited to the CH stretch range and does not report absolute intensities, we also used 
experimental data from Joblin \emph{et al.},\cite{Joblin1994} both in Ne matrix at 4~K and in gas phase at relatively high temperatures ($\sim$ 300-500 \textdegree C) due to the low vapour pressure of PAHs.
Since the matrix effect on the band position is expected to be weak for neon \cite{Cassam2008}, we used the measured positions in Ne as the best available experimental data to be compared with our calculated positions.
For most of the spectrum, the bands in Ne are very sharp and well resolved. We remark though that the band structure in the CH stretch range is less resolved in Ne matrices compared to gas-phase at low temperature, which can be observed from the values reporded in Table\,\ref{tab:pyrene1}. Again this indicates a strong role of anharmonicity in this range. Ne matrix spectra lack an absolute calibration of band intensities. In order to compare the experimental integrated band intensities with calculated band strengths, we therefore used the gas-phase spectra from Joblin \emph{et al.}\cite{Joblin1994}
Error bars are not easy to derive but we proceeded as follows for the gas-phase data that are available to us.\cite{Joblin1992} A key point in deriving absolute intensities is to estimate the PAH column density. For gas-phase experiments in which all the PAH sample was evaporated at the temperature of interest,  we were able to derive the column density of molecules in the gas-phase. If the evaporation is not total then this density is controlled by the vapour pressure, which is less precisely known. For coronene at 723\,K we could use two spectra corresponding to total evaporation with different column densities.
For pyrene at 523\,K, we had only one such case and used a second case by applying an average global scaling on all bands. We deemed this useful to get an estimate of the relative errors from one experiment to the other but the absolute values might be less accurate in this case compared to coronene. In hot gas phase spectra it is often difficult to separate overlapping bands.
After substraction of a continuum from the spectra, we defined intervals to calculate band intensities. We then identified the corresponding bands in the theoretical spectra and summed them for comparison. The slight differences that can be found between the new gas-phase intensities and the ones previously reported \cite{Joblin1994} is attributed to changes in the considered intervals and to the way the continuum level is substracted.

\subsubsection{Pyrene}\label{sec:pyrene}

The quartic force field and first and second derivatives of the electric dipole moment were obtained as described in Sections~\ref{elecstructure}  and \ref{sec:convpyrene}.
We then used the AnharmoniCaOs code with the resulting quartic force field and second order Taylor expansion of the electric dipole, using the harmonic ground vibrational state as a starting state. 

The results of our calculations can be compared with those obtained by Mackie \emph{et al.} \cite{Mackie2016} and with the experimental data as described above. The comparison is shown in Table~\ref{tab:pyrene1} for band positions and Table~\ref{tab:pyrene2} for band intensities. We remark that the calculated spectrum shown in Ref.~\citenum{Mackie2016} is nearly identical to the one we obtained with $r = 0.1$ and $h = 8$, shown in Fig.~\ref{plotpyrthres}, which is not completely converged yet. We tentatively suppose that the smallest thresholds they tested for convergence, necessarily limited by computational constraints (the run including the largest polyads we tested required about 100 Gbytes of RAM) were smaller than the ones we arrived at. Since the computed spectrum changes abruptly every time the decrease of the threshold causes a significantly larger number of states to be included in the polyads, they probably did not reach threshold values low enough to obtain the top spectrum we show in Fig.~\ref{plotpyrthres}.

Table~\ref{tab:pyrene2} also includes the results of AnharmoniCaOs including the partial correction enforcing total band intensity conservation in the perturbative calculation (see previous section), to get an estimate of how large this effect is.
\setlength{\LTcapwidth}{2.0\hsize}
\begin{longtable*}{@{\extracolsep{\fill}}ccllll}
\caption{List of the main band positions for pyrene. Theoretical data are from (a) this work and (b) Ref.~\citenum{Mackie2016}. The experimental band positions are those of the main bands in the Ne matrix spectrum recorded at 4\,K taken from Ref.~\citenum{Joblin1994}. For the CH stretch range we also report the values recorded in gas-phase at low-temperature from Ref.~\citenum{Maltseva2016}.}\label{tab:pyrene1}\\[2mm]

\hline\\[.5mm]
Gross pos. ($\mu$m) & \multicolumn{5}{c}{Position (\cm)} \\[.5mm]
\cline{2-6} \\[.5mm]
 & Ne Matrix 4\,K & \multicolumn{4}{c}{Theory 0\,K} \\[.5mm]
 & *Gas-phase (low T) & \multicolumn{4}{c}{} \\[.5mm]
  \cline{3-6} \\[.5mm]
  &     &  harm. scaled & \multicolumn{3}{c}{anharmonic} \\[.5mm]  
  \cline{4-6}\\[.5mm]
   &     & (a) & (a) & & (b) \\ [.5mm]
\hline\\[.5mm]
\endfirsthead

\multicolumn{6}{c}{\tablename\ \thetable{} -- continued from previous page} \\[2mm]
\hline\\[.5mm]
Gross pos. ($\mu$m) & \multicolumn{5}{c}{Position (\cm)} \\[.5mm]
\cline{2-6}\\[.5mm]
  & Ne Matrix 4\,K & \multicolumn{4}{c}{Theory 0\,K} \\[.5mm]
 & *Gas-phase (low T) & \multicolumn{4}{c}{} \\[.5mm]
  \cline{3-6} \\[.5mm]
  &      &  harm. scaled & \multicolumn{3}{c}{anharmonic} \\  [.5mm]
  \cline{4-6}\\[.5mm]
   &      & (a) & (a) & & (b) \\ [.5mm]
\hline\\[.5mm]
\endhead

\\[.5mm] \hline \\[1mm] \multicolumn{6}{r}{{Continued on next page}} \\[.5mm] \hline
\endfoot

\\[.5mm] \hline
\endlastfoot

\multirow{29}{*}{3.3}	&  *3118.7	& & & & \\
	& *3108.9	& & 3102, 3098 & $\nu_{10}+\nu_ {52}+\nu_{57}$& \\
& *3096.0 & & 3095, 3092 & $\nu_{30}+\nu_{44}+\nu_{53}$, $\nu_{21}+\nu_{34}+\nu_{40}+\nu_{57}$ &\\
& *3087.8 & & 3089, 3086 & $\nu_{4}+\nu_{45}$, $\nu_{30}+\nu_{46}+\nu_{64}$ & \\
& *3071.9 & & 3084, 3083 &  $\nu_{45}+\nu_{62}+\nu_{64}$, $\nu_{9}+\nu_{13}+\nu_{27}$ & \\
& ``3066'', *3064.9 & 3074 & 3076, 3073, 3071 & $\nu_{23}$,  $\nu_{26}+\nu_{57}$, $\nu_{4}+\nu_{36}+\nu_{70}$	& 3072 \\
& *3063.3 &  & 3066, 3065 &  $\nu_{29}+\nu_{45}+\nu_{53}$, $\nu_{28}+\nu_{32}$ & \\
& *3059.5  & & 3062 &  $\nu_{26}+\nu_{34}+\nu_{51}$ & \\
& *3057.5 & & 3056, 3050 &  $\nu_{4}+\nu_{28}$, $\nu_{10}+\nu_{28}+\nu_{65}$,  $\nu_{4}+\nu_{10}+\nu_{53}$ & \\
& *3055.6 & & 3045 & $\nu_{35}+\nu_{56}+\nu_{70}$ & 3043 \\
& *3052.9 & & 3042 &  $\nu_{33}+\nu_{49}+\nu_{53}+\nu_{63}$, $\nu_{11}+\nu_{51}+\nu_{63}+\nu_{65}$ & \\
&  3053,*3049.8 & 3057 & 3040 & $\nu_{42}$ & 3048 \\
& *3044.0 & & 3034 &  $\nu_{12}+\nu_{31}+\nu_{32}+\nu_{52}$ & \\
& & & 3033 &  $\nu_{17}+\nu_{44}+\nu_{66}$, $\nu_{21}+\nu_{26}+\nu_{37}$ & \\
& & & 3032, 3031 &  $\nu_{12}+\nu_{29}+\nu_{32}+\nu_{53}$, $\nu_{21}+\nu_{26}+\nu_{37}$ & \\
& & & 3030 &  $\nu_{52}+\nu_{57}+2\nu_{70}$, $\nu_{5}+\nu_{27}$,$\nu_{26}+\nu_{30}+\nu_{53}$ & \\
& & & 3026 &  $\nu_{19}+\nu_{39}+\nu_{45}$, $\nu_{35}+\nu_{48}+\nu_{53}$ & \\
& & & 3025 &  $\nu_{12}+\nu_{37}+\nu_{66}$ & \\
& & & 3024 &  $\nu_{17}+\nu_{34}+\nu_{62}+\nu_{66}$ & \\
& & & 3023 &  $\nu_{13}+\nu_{46}+\nu_{62}$, $\nu_{33}+\nu_{56}+\nu_{63}$, $\nu_{5}+\nu_{45}$ & \\
& & & 3022 &  $\nu_{21}+\nu_{36}+\nu_{51}+\nu_{64}$ & \\
& & & 3019, 3018 &  $\nu_{5}+\nu_{27}$, $\nu_{30}+\nu_{58}+\nu_{64}$& \\
& & & 3014 &  $\nu_{5}+\nu_{35}+\nu_{70}$ & \\
& & & 3011 & $\nu_{12}+\nu_{32}+\nu_{56}$ & \\
& & & 3002 & $\nu_{28}+\nu_{56} $ & \\
& & & 2960 &  $\nu_{15}+\nu_{40}+\nu_{57}$, $\nu_{45}+\nu_{57}$ & \\
& & & 2940 &  $\nu_{35}+\nu_{39}+\nu_{47}$, $\nu_{26}+\nu_{59}$, $\nu_{7}+\nu_{30}+\nu_{64}$ & \\
& & & 2927 & $\nu_{4}+\nu_{47}$ & \\
& & & 2908 & $\nu_{21}+\nu_{59}+\nu_{66}$ & \\[1mm]
\multirow{4}{*}{5.2}	& \multirow{2}{*}{1935} 		& & 1937 & $\nu_{14}+\nu_{35}$ & 1945\\
& & & 1934 & $\nu_{35}+\nu_{66}$ & 1937  \\
& 1923 & & 1921 & $\nu_{14}+\nu_{36}$ & 1932 \\
	& 1918 	&	& 1919 & $\nu_{36}+\nu_{66}$ & \\[1mm]
\multirow{2}{*}{5.3}	& 1870, 1867, 1863 	&	& 1866 & $\nu_{14}+\nu_{19}$ & 	 1860 \\
 	&  1851		& & 1858, 1848, 1842 & $\nu_{10}+\nu_{69}$, $\nu_{15}+\nu_{35}$ $\nu_{15}+\nu_{36}$	& \\[1mm]
\multirow{2}{*}{5.6} & 1800  &	& 1801 & $\nu_{37}+\nu_{66}$ & 1804 \\
& 1796 & & 1792 & $\nu_{15}+\nu_{19}$ & 1799 \\[1mm]
\multirow{3}{*}{5.7} & 1761	&	& 1766 & $\nu_{14}+\nu_{20}$   &  1774 \\
& 1749 & & 1745	& $\nu_{19}+\nu_{67}$ & \\
& & & 1725 &  $\nu_{15}+\nu_{37}$ & \\[1mm]
5.9	& 1695 	& & 1697 & $\nu_{35}+\nu_{68}$  & \\[1mm]
\multirow{2}{*}{6.0} & 1670, 1665  &	& 1669 & $\nu_{36}+\nu_{69}$ & \\
    & 1648  &	& 1651, 1645 & $\nu_{16}+\nu_{35}$, $\nu_{20}+\nu_{67}$  &  \\[1mm]
\multirow{2}{*}{6.2} & 1617 &  & 1617, 1612, 1610 &$\nu_{19}+\nu_{68}$, $\nu_{38}+\nu_{67}$, $\nu_{19}+\nu_{69}$ & \\
& 1605 & 1606 & 1600, 1596 & $\nu_{44}$, $\nu_{12}+\nu_{34}$ & \\[1mm]
\multirow{3}{*}{6.5} & & 1597 & 1585, 1575 & $\nu_{26}$, $\nu_{12}+\nu_{41}+\nu_{68}$, $\nu_{16}+\nu_{19}$ & 1588  \\
& 1545 &  & 1537, 1544	& $\nu_{39}+\nu_{66}$, $\nu_{20}+\nu_{68}$ & \\
& 1517 &  & 1521, 1505	&  $\nu_{16}+\nu_{37}$, $\nu_{38}+\nu_{68}$  & \\[1mm]
\multirow{2}{*}{6.8} & \multirow{2}{*}{1496} & \multirow{2}{*}{1481} & 1473, 1489, 1485  & $\nu_{45}$, $\nu_{14}+\nu_{21}$, $\nu_{21}+\nu_{66}$  & 1479 \\
  & & & 1460, 1459 &  $\nu_{15}+\nu_{39}$, $\nu_{51}+\nu_{64}$  & \\[1mm]
\multirow{3}{*}{7.0}	& \multirow{3}{*}{1438, 1436} 	& 1442 & 1442, 1438  &$\nu_{36}+\nu_{70}$, $\nu_{27}$ & 1444  \\
& & 1430 & 1424 & $\nu_{28}$ 	&  1431\\
& & 1427 & 1418,1417,1412 & $\nu_{34}+\nu_{63}$, $\nu_{46}$, $\nu_{10}+\nu_{53}$& \\[1mm]
7.7	& 1311 		& 1314 & 1311, 1320, 1317 & $\nu_{47}$, $\nu_{13}+\nu_{67}$, $\nu_{32}+\nu_{64}$ & 1319  \\[1mm]
8.0	& 1244 		& 1242 & 1241, 1236 & $\nu_{40}+\nu_{68}$, $\nu_{29}$  & 1243  \\[1mm]
\multirow{2}{*}{8.4}	& \multirow{2}{*}{1185, 1182} 		& \multirow{2}{*}{1181} & \multirow{2}{*}{1180} & $\nu_{49}$, $\nu_{16}+\nu_{40}$,  &  \multirow{2}{*}{1188}   \\
& & & & $\nu_{39}+\nu_{41}+\nu_{53}$, $\nu_{39}+\nu_{41}+\nu_{53}$ & \\[1mm]
9.1	& 1097, 1087 		& 1089 & 1089,1081,1080 & $\nu_{30}$, $\nu_{12}+\nu_{34}$, $\nu_{53}+\nu_{63}$	& 1081, 1095  \\[1mm]
9.4 & 1063 &  & 1058	& $\nu_{39}+\nu_{70}$  \\[1mm]
10.0 & 1004 & 993 & 996,993	& $\nu_{31}$, $\nu_{52}+\nu_{65}$  \\[1mm]
10.4 & 966  &   966   & 964,963 & $\nu_{66}$, $\nu_{41}+\nu_{69}$ & 969 
\\[1mm]
\multirow{2}{*}{11.8} & 844		& 845 & 843	& $\nu_{67}$ & 861\\
     & 822      & 815 & 816	& $\nu_{32}$ & 822 \\[1mm]
13.4 & 745		& 742 & 739	& $\nu_{68}$ & 750 \\[1mm]
14.0 & 713		& 713 & 710	& $\nu_{69}$ & 741  \\[1mm]
18.5 & 542		& 539 & 538	& $\nu_{52}$ \\[1mm]
20.0 & 499		& 495 & 493	& $\nu_{34}$   \\[1mm]
20.5 & 488		& 487 & 483 & $\nu_{70}$  \\[1mm]
29.0 & & 350 & 347,342	& $\nu_{53}$, $\nu_{22}+\nu_{72}$  \\[1mm]
50 & & 206 & 198	& $\nu_{71}$  \\[1mm]
\end{longtable*}

%
\begin{table*}
\begin{tabular*}{\hsize}{@{\extracolsep{\fill}}ccccccc}
\hline
Gross pos.($\mu$m)  & Integration range (\cm)  & \multicolumn{5}{c}{Intensities (\km)}  \\
\cline{3-7}
  & & Gas 570\,K  &  \multicolumn{4}{c}{Theory 0\,K} \\
 \cline{4-7}
  & &  & harm. & \multicolumn{3}{c}{anharm.}\\ 
 \cline{5-7}
& &  &  (a) & \multicolumn{2}{c}{(a)} & (b)\\ 
\cline{5-6}
& &   &  & nc & c & \\   
\hline
3.3	 & [2850-3250] & 140($\pm$10)     &  95 & 99 & 74 &\\
5.2  & [1900-1950] & 9.5 ($\pm$0.5)   &  & 9.0 & 9.3 & 18.2 \\
5.3	 & [1830-1880] & 3.8 ($\pm$0.1)   & & 5.5 & 5.7 & 2.0 \\
5.6  & [1780-1830] & 7.8 ($\pm$0.7)   &  & 8.8 & 9.1 & 14.1 \\
5.7  & [1715-1760] & 5.8 ($\pm$0.2)   &  & 2.7 & 2.9 & \\
5.9  & [1677-1704]    & --		      & & 1.9 & 2.0 & \\
6.0	 & [1655-1677] & 2.5 ($\pm$0.2)   & & 2.7 & 2.7 & \\
6.1	 & [1630-1655] & --   & & 2.8 & 2.9 & \\
6.2  & [1560-1620] & 11.4 ($\pm$1.2)  &  13.2 & 17.3 & 16.4 & 7.1 \\
6.6  & [1531-1552] & --		     &  & 2.6 & 2.7 &  \\
6.8  & [1467-1525] & --		     & & 3.7 & 3.5 & 3.0 \\
7.0  & [1405-1430] & 11.4 ($\pm$0.1)   & 11.2 & 8.8 & 7.8 & 6.1 \\
7.7	 & [1300-1327] & 2.4 ($\pm$0.1)  & 4.9 & 5.2 & 4.5 & 6.1 \\
8.0	 & [1224-1245] & 1.9 ($\pm$0.3)    & 2.3 & 2.6 & 2.4 & 2.0 \\
8.4	 & [1165-1200] & 10.5 ($\pm$0.1)   & 14.0 & 13.7 & 12.1 & 14.1 \\
9.1	 & [1074-1100] & 5.4 ($\pm$1.2)	  & 7.0 & 5.4 & 4.7 & 8.1 \\
10.0 & [988-1003]  & 2.2 ($\pm$0.6)      & & 1.7 & 1.6  &  \\
11.8 & [824-863]   & 100 ($\pm$6) 	  & 112 & 101 & 93 & 101 \\
13.4 & [730-750]   & 20.8 ($\pm$1.4)  	  & 17.5 & 15.2 & 14.0 & 24.2\\
14.0 & [695-725]   & 46 ($\pm$1) 	  & 44.8 & 45.3 &  42.9 & 41.4 \\
18.5 & [526-550]   &              & 2.5 & 2.3 & 2.1 &  \\
20.0 & [487-501]   &              & & 2.7	& 2.4 & \\
20.5 & [474-487]   &	 	 	  & 2.1 & 1.9 & 1.7 & \\
29.0 & [335-354]   &	 	 	  &  & 1.4 & 1.3 & \\
50 & [185-212]   &	 	 	  &  & 9.7 & 9.0 & \\
\hline
\end{tabular*}
\caption{List of the integrated intensities for pyrene. Theoretical data are (a) this work and (b) from Ref.~\citenum{Mackie2016}. Since only relative intensities are reported in this latter study, we have scaled the values to the theoretical anharmonic intensity of the main CH out-of-plane bend mode. For our anharmonic results we list two columns, the ``nc'' one is standard GPVT2, ``c'' includes our correction borrowing some third order perturbation theory terms to enforce conservation of the intensity of transitions in resonances treated with perturbation theory. Experimental values have been derived from measurements in gas-phase at 570~K from two independent spectra taken from Refs.~\citenum{Joblin1994} and~\citenum{Joblin1992}. The mean values are listed and the values in brackets provide the scatter around these values.
Since in these spectra band overlapping is unavoidable, we selected integration ranges and integrated the corresponding theoretical bands. The listed ranges are the ones considered for the theoretical spectra and have been slightly shifted to cover the band envelopes in gas-phase spectra. }\label{tab:pyrene2}
\end{table*}
To assess the overall accuracy of our calculations, we compare in Tables~\ref{tab:pyrene1} and~\ref{tab:pyrene2} our results with the laboratory data that have been described above, as well as with the scaled harmonic calculation (see Supplementary material for all harmonic results and scaling factors). We also compared with the previously published anharmonic calculations by Mackie \emph{et al.}\cite{Mackie2016} Since in this latter paper only relative intensities are given, we estimated absolute intensities for all possible bands by multiplying them by the absolute intensity we calculated for the band the authors chose for reference. This excludes C-H stretches that the authors studied separately from the rest of the spectrum with an unknown scaling factor. Concerning band positions, we listed all states which carry a significant intensity and only one position was listed when these states are too close to be resolved. Assigning a given theoretical band to an experimental one can be tricky. We did our best on the basis of band proximity. This can be disputable especially in the case of the CH stretch region for which the calculations face some difficulties, as discussed above and in Sect.~\ref{discussion}.
For anharmonic calculations we label bands by the leading harmonic base state in the expansion of the upper state of the transition. The harmonic modes are given in tables in the supplementary material. While in some cases anharmonic states are close enough to a single, well-defined harmonic state, in others they happen to be linear combinations of many harmonic states with coefficients of about the same magnitude, which can be as small as 10-20\% for the leading harmonic state (which is nonetheless chosen to label the transition). Clear examples of both cases occur in the C-H stretch region. The $\nu_{42}$ harmonic state remains fairly well defined in the anharmonic calculation, with a large fraction of its intensity therefore in a single band at a computed position of 3040 cm$^{-1}$. The corresponding anharmonic state is a linear combination of the harmonic states $\nu_{42}$ (38\%), $\nu_{45}+\nu_{48}+\nu_{53}$ (19\%), $\nu_{4}+\nu_{45}$ (4\%), plus many other harmonic states contributing each less than 3\%.
Conversely, the $\nu_{43}$ and $\nu_{23}$ harmonic states appear to distribute over a large number of anharmonic states via strong resonances. For example, the anharmonic state corresponding to the band at 3076~cm$^{-1}$, which we labelled $\nu_{23}$, is a linear combination of the harmonic states $\nu_{23}$ (the ``leading'' one with 12\%), $\nu_{4}+\nu_{27}$ (11\%), $\nu_{24}$ (9\%), plus a large number of other harmonic states contributing less. The harmonic state $\nu_{43}$, while contributing to many states, is not the leading one in any.

When compared to values in Ne matrices and excluding the CH stretch range, band positions appear to be accurate on average to better than 0.8\%. The worst case is the band at 20.0\,$\mu$m for which the shift reaches 1.2\%. There is a very slight tendency to underestimate band positions, with an average ratio between theoretical and experimental positions of 0.998, which shows that systematic error is minimal. Of course Ne matrix positions will also be affected by some matrix shift effect, even if it is expected to be small. Concerning integrated band intensities, we listed the integration limits we adopted to allow comparison with the gas phase spectra. We give here for pyrene the results both using our higher order correction enforcing conservation of band intensity and without it. The difference between the two theoretical sets of values are much smaller than the differences with experimental values, with the uncorrected ones appearing, if any, slightly better on average. For band intensities, the largest differences between calculated and experimental values appear in some of the weakest bands, the one at 5.7~$\mu$m being lower by a factor of two in calculations, while the one at 7.7~$\mu$m is larger by the same factor. The C-H stretch at 3.3\,$\mu$m, typically one of the most difficult bands to compute accurately with DFT, is lower in calculations by $\sim$40\%, the remaining bands are within 25\%.  The fractional differences have both signs and average to about zero.

Our results are generally consistent with those published by Mackie \emph{et al.}\cite{Mackie2016} for pyrene, even if a detailed one-to-one comparison depends on the possibility of unambiguously matching them based on what is given in Table~1 of their supplementary material. For this reason, not all theoretical band positions reported by Mackie \emph{et al.}\cite{Mackie2016} are listed in last column of Table~\ref{tab:pyrene1}, but only those whose identification could be matched with a corresponding band in our calculations. In particular, in the C-H stretch region we only included the three bands with the largest component of the three C-H stretch IR-active fundamentals. Differences are non-negligible nonetheless, as could be expected from using two different implementation of Van-Vleck theory (see Sect.~\ref{discussion}) and could also originate from small differences in the initial quartic force field. We remark that the best agreement with the low-temperature gas-phase data in the CH stretch range is found with our not completely converged calculation with $r=0.1$, extremely similar to the one shown in Ref.~\citenum{Mackie2016}. This hints that the accuracy we can achieve with this kind of calculation is, in this case, limited by the accuracy of the underlying DFT calculations yielding the quartic force field. The apparent slightly better accuracy of the $r=0.1$ calculation with respect to the nominally better one with $r=0.05$ is likely due to a partial accidental cancellation of errors.
The complete, anharmonic spectrum of pyrene, computed with $r=0.05$ and $h=8$, is shown in the supplementary material, and available in tabulated form from the online database by Malloci, Joblin, and Mulas.\cite{Malloci2007b}


\subsubsection{Coronene}\label{sec:coronene}
As for pyrene, the quartic force field and first and second derivatives of the electric dipole moment were obtained as described in Sections~\ref{elecstructure}  and \ref{sec:convpyrene}. We then used the results obtained by our AnharmoniCaOs code with $r = 0.1$ and $h = 6$, leading to the largest polyads we could computationally handle, to obtain positions and intensities of the permitted transitions. We did not use our correction to enforce intensity conservation, since we saw in the case of pyrene that its effect is small compared to the errors arising from the underlying DFT calculations.

\setlength{\LTcapwidth}{2.0\hsize}
\begin{longtable*}{@{\extracolsep{\fill}}ccccl}
\caption{List of the main band positions for coronene. Theoretical data are from this work. The experimental band positions are those of the main bands in the Ne matrix spectrum recorded at 4\,K taken from Ref.~\citenum{Joblin1994}.}\label{tab:coronene1}\\[2mm]

\hline\\[.5mm]
Gross pos. ($\mu$m) & \multicolumn{4}{c}{Position (\cm)} \\[.5mm]
\cline{2-5} \\[.5mm]
 & Ne Matrix 4\,K& \multicolumn{3}{c}{Theory 0\,K} \\[.5mm]
  \cline{3-5} \\[.5mm]
  &      &  harm. scaled & \multicolumn{2}{c}{anharmonic} \\[.5mm]  
\hline\\[.5mm]
\endfirsthead

\multicolumn{5}{c}{\tablename\ \thetable{} -- continued from previous page} \\[2mm]
\hline\\[.5mm]
Gross pos. ($\mu$m) & \multicolumn{4}{c}{Position (\cm)} \\[.5mm]
\cline{2-5}\\[.5mm]
 & Ne Matrix 4\,K& \multicolumn{3}{c}{Theory 0\,K} \\[.5mm]
  \cline{3-5} \\[.5mm]
  &      &  harm. scaled & \multicolumn{2}{c}{anharmonic} \\  [.5mm]
\hline\\[.5mm]
\endhead

\\[.5mm] \hline \\[1mm] \multicolumn{5}{r}{{Continued on next page}} \\[.5mm] \hline
\endfoot

\\[.5mm] \hline
\endlastfoot

\multirow{9}{*}{3.3} & & & 3121 & $\nu_{30}+2\nu_{61}$, $\nu_{30}+\nu_{61}+\nu_{62}$, $\nu_{30}+2\nu_{62}$ \\
		& & & 3120 & $\nu_{2}+\nu_{51}$, $\nu_{2}+\nu_{52}$ \\
		& & & 3112 & $\nu_{30}+\nu_{71}$, $\nu_{30}+\nu_{71}$ \\
		& 3070 & 3066 & 3071 & $\nu_{45}$, $\nu_{46}$ \\
		& & & 3056 & $\nu_{10}+\nu_{51}$, $\nu_{10}+\nu_{52}$, $\nu_{21}+\nu_{71}$, $\nu_{21}+\nu_{72}$ \\
		& & & 3049 & $\nu_{21}+2\nu_{61}$, $\nu_{21}+\nu_{61}+\nu_{62}$, $\nu_{21}+2\nu_{62}$ \\
		& & & 3050 & $\nu_{49}+\nu_{75}$, $\nu_{49}+\nu_{76}$, $\nu_{50}+\nu_{75}$, $\nu_{50}+\nu_{76}$ \\
		& & & 3038 & $\nu_{10}+\nu_{51}$ $\nu_{10}+\nu_{52}$ \\
		& ``3035'' & 3048 & 3025 & $\nu_{47}$, $\nu_{48}$ \\[1mm]
\multirow{3}{*}{5.3} & \multirow{2}{*}{1926, 1913, 1898} & & 1918, 1920 & $\nu_{35}+\nu_{91}$,$\nu_{35}+\nu_{92}$, $\nu_{36}+\nu_{91}$,$\nu_{36}+\nu_{92}$\\
		& & & 1889, 1892, 1895 & $\nu_{7}+\nu_{35}$, $\nu_{7}+\nu_{36}$ \\
		& & & 1886 & $\nu_{6}+\nu_{14}$ \\[1mm]
5.6		& 1809, 1801, 1786 & & 1783 & $\nu_{7}+\nu_{37}$, $\nu_{7}+\nu_{38}$ \\[1mm]
\multirow{3}{*}{5.9}		& 1721 & & 1708 & $\nu_{14}+\nu_{37}$, $\nu_{14}+\nu_{38}$ \\
		& 1697 & & 1692 & $\nu_{17}+\nu_{91}$, $\nu_{17}+\nu_{92}$ \\
		& & & 1681 & $\nu_{37}+\nu_{93}$, $\nu_{37}+\nu_{94}$, $\nu_{38}+\nu_{94}$, $\nu_{38}+\nu_{94}$ \\[1mm]
6.2 & 1621 & 1614 & 1619 & $\nu_{49}$,$\nu_{50}$ \\[1mm]
\multirow{2}{*}{7.6} & 1317 & 1309 & 1315 & $\nu_{55}$,$\nu_{56}$ \\
 & & & 1310 & $\nu_{14}+\nu_{41}$, $\nu_{14}+\nu_{42}$ \\[1mm]
8.8 & 1139 & 1134 & 1143 & $\nu_{59}$,$\nu_{60}$  \\[1mm]
11.7 & 857 & 857 & 862 & $\nu_{14}$ \\[1mm]
13.0 & 772 & 767 & 767 & $\nu_{63}$,$\nu_{64}$ \\[1mm]
18.2 & 549 & 548 & 553 & $\nu_{15}$ \\[1mm]
26.5   & & 376 & 379 & $\nu_{65}$,$\nu_{66}$\\[1mm]
81  	& & 121 & 126 & $\nu_{16}$ \\[1mm]
\end{longtable*}
\begin{table*}
\begin{tabular*}{\hsize}{@{\extracolsep{\fill}}ccccc}
\hline
Gross pos. ($\mu$m) & Exp. Range (\cm) & \multicolumn{3}{c}{Intensities (\km)}  \\
\cline{3-5}
&  & Gas 570\,K & \multicolumn{2}{c}{Theory 0\,K} \\
\cline{4-5}
&  & & harm. & anharm.\\
\hline
3.3 	& [2995-3110]	& 190 (1)	 & 134.7 & 177\\
5.3 	& [1870-1906]	& 23 (3)	  & & 31.5\\
5.6 	& [1770-1825]	& 18 (4)	 & & 5.9\\
5.9 	&[1670-1720]	& 20 (3)	 & & 13.9 \\
6.2 & [1600-1629] & 19 (2)  & 15.2 & 24.9 \\
7.6 & [1300-1330] & 37 (2)  & 42.3 & 22.7 \\
8.8 & [1128-1158] & 22 (4)  & 17.3 & 14.4 \\
11.7 & [845-880] & 138 (s)  & 167.9 & 147 \\
13.0 & [745-785]	& 32 (1) & 13.6 & 13.2 \\
18.2 &  [530-565] & 44 (2)  & 42.7 & 23.0 \\
26.5 &  [360-395] &  &   7.7 & 5.9 \\
81   &  [110-140] &  &   6.9 & 4.6 \\
\hline
\end{tabular*}
\caption{List of the integrated intensities for coronene. Theoretical data are from this work. Experimental values have been derived from measurements in gas-phase at 570~K from two independent spectra taken from Refs.~\citenum{Joblin1994} and~\citenum{Joblin1992}. The mean values are listed and the values in brackets provide the scatter around these values. The strongest band at 11.7\,$\mu$m (annotation (s)) was saturated in one of the two experimental spectra so only a single value could be reported.
Since in these spectra band overlapping is unavoidable, we selected integration ranges and integrated the corresponding theoretical bands. The listed ranges are the ones considered for the theoretical spectra and have been slightly shifted to cover the band envelopes in gas-phase spectra.}\label{tab:coronene2}
\end{table*}

In Table~\ref{tab:coronene1} we compare our computed results for the band positions of coronene with laboratory data that were described above and with the scaled harmonic calculation, using the same approach as that described for pyrene. The band positions appear to be accurate to better than 0.7\% except for the weak band at $\sim$5.3\,$\mu$m for which the normalized shift between theory and experiment is 1.1\%. As for pyrene, there is a balance between negative and positive values for this shift, since the average ratio between theoretical and Ne matrix positions is $\sim$0.998.

Theoretical band intensities appear somewhat less accurate than the ones for pyrene. The extreme cases are the bands at 5.3~$\mu$m and 13.0~$\mu$m, which are both lower by a factor $\sim$3 in the calculations compared to the experiment. The average difference is about 35\%, with errors on both the positive and negative sides, with a prevalence of theoretical underestimation of experimental values. However, the accuracy is rather good for the most intense bands at 3.3 and 11.7\,$\mu$m. This is consistent with what we found in the case of pyrene. The slightly worse agreement with experimental data relative to the case of pyrene, is likely due to the smaller basis set used to obtain the quartic force field for coronene.
The complete, anharmonic spectrum of coronene is shown in the supplementary material, and available in tabulated form from the online database by Malloci, Joblin, and Mulas.\cite{Malloci2007b}


%
\section{Discussion and Conclusions}\label{discussion}

We performed anharmonic calculations of the vibrational spectra of neutral pyrene and coronene at 0~K using our AnharmoniCaOs code, and compared them to the best available experimental data. The results are overall fairly satisfactory, yielding a significant improvement over the conventional double harmonic DFT calculations normally used for molecules of this size. As we already remarked at the beginning of Sect.~\ref{sec:experiment}, the most meaningful comparison should be with high resolution, low temperature gas phase measurements, which are  only available in the C-H stretch region for a small set of PAHs.\cite{Maltseva2015,Maltseva2016} For the present work, we mostly relied on Ne matrix spectra for band positions and high temperature gas-phase spectra for band intensities.\cite{Joblin1994,Joblin1992} Our conclusions might slightly change if/when full high resolution, low temperature gas phase spectra of such species become available.

The accuracy of our anharmonic calculations is clearly related to several different independent limiting factors, respectively: 
\begin{enumerate}
\item The representation of the true adiabatic potential energy surface as a quartic force field, and the dipole moment as a Taylor expansion truncated to second order, both in terms of cartesian normal coordinates \label{tayloritem}; 
\item The accuracy of the determination of the quartic force field and quadratic dipole parameters via quantum chemistry calculations\label{dftitem}; 
\item The accuracy of the GVPT2 method itself\label{gvpt2item}; 
\item The accuracy of our AnharmoniCaOs implementation (as a function of its parameters)\label{anharmonicaositem}. 
\end{enumerate}
Point~\ref{tayloritem} is an acceptable approximation when oscillations have a relatively small amplitude. Therefore, it is suitable for semirigid molecules at not too high vibrational energies. So, it should be appropriate for PAHs, unless they have side groups that give rise to internal rotations, that are poorly represented by a truncated Taylor expansion in normal coordinates.\cite{Friha:2013uq} We notice that it would be desirable in some cases, to opt for non-cartesian coordinates so that the associated quartic force field would have a better asymptotic behaviour, and the potential would be reasonably described even along large amplitude motions.\cite{Dateo1994,Fortenberry2013} Such coordinates, however, are not directly applicable in our case: the Van Vleck approach used here for GVPT2 requires a description of the Hamiltonian in terms of normal coordinates, since it exploits their commutation properties to obtain a formally simple representation of the infinitesimal contact transformations and of the transformed Hamiltonian and dipole moment operators.

Point~\ref{dftitem} is basically a matter of trade-off between computational cost and accuracy. Band positions can be fairly accurate for PAHs, as shown also by our results, even using moderate size basis sets (such as the triple zeta ones we used) with DFT using hybrid exchange-correlation functionals.\cite{Candian2016} This can be further improved, approaching spectroscopic accuracy, by using more accurate (and computationally way more expensive) methods with larger basis sets, either only for the harmonic frequencies or also for some of the higher derivatives of the potential, and for the first and/or second derivatives of the dipole moment (see Ref.~\citenum{Bloino2012} and references therein). Absolute intensities are more difficult to get with the same accuracy as for positions using DFT calculations, but they also become fairly precise when using high (computationally expensive) levels of theory.\cite{Begue2006} 

As to point~\ref{gvpt2item}, GVPT2 has been used successfully for many years (see e.g. Ref. \citenum{Martin1995}), and it has been reviewed by Bloino \emph{et al.} in Ref. \citenum{Bloino2012}. In general, GVPT2 should be appropriate as long as point~\ref{tayloritem} is valid.

Finally point~\ref{anharmonicaositem} has been assessed in our comparisons with higher level vibrational calculations, in Sect.~\ref{sec:ethox}. Our code has the limitation that rotational degrees of freedom, and their interaction terms with vibrational ones (i.e. terms due to the Coriolis and centrifugal pseudo forces) are not included in our approximated Hamiltonian. This is not expected to be an important limit for relatively large, semi-rigid, asymmetric top molecules
such as e.~g. pyrene. However, it is always possible that a small number of individual states undergo accidental Coriolis resonances. In contrast, strong Coriolis interactions are bound to occur for vibrational states which are degenerate due to symmetry reasons, and this will be more likely for symmetric species like coronene. However, we remark that the agreement between theoretical and experimental data for coronene does not seem significantly worse than that of pyrene, hinting that Coriolis coupling is not crucial.

Summing up, we find that for the cases of pyrene and coronene presented here the main limitation appears to be point~\ref{dftitem}, since we clearly reached a point at which increasing the accuracy of our GPVT2 calculation did not improve the agreement with experimental data. Indeed, it looks like there is a ``sweet spot'' in the accuracy of the GPVT2 calculations where some accidental cancellation of errors with the underlying DFT-based quartic force field produces the best results, even if this may not be general.

The calculations presented here are similar to those performed by Mackie \emph{et al.} in Ref.~\citenum{Mackie2016}. The latter authors reported an agreement of calculated band positions with experimental data (in rare gas matrices) of $0.4 \pm 0.6\%$, to be compared with our value of better than 0.8\%. Since both studies include different sets of laboratory data, it is difficult to conclude which one provides the best results. Still, it can be noticed that some individual bands positions differ by more than numerical noise, when comparing both theoretical studies (cf. Table~III). The implementation of GVTP2 in Ref.~\citenum{Mackie2016} is different from ours. In particular, a different strategy is used to define which cubic and quartic terms of the potential should be included in the treatment of resonances rather than in the perturbation expansion, and subsequently the construction (and truncation) of polyads  differ. In spite of these differences, the overall agreement can be seen as a validation of both approaches. With the parameters used, the results reported here appear slightly more accurate for the region of combination bands, whereas the results in Ref.~\citenum{Mackie2016} appear somewhat better in the C-H stretches region. We remark once more that we actually did obtain very nearly the same results as Mackie \emph{et al.}\cite{Mackie2016} for the C-H stretches of pyrene with our not completely converged anharmonic calculation, as discussed in Sect.\ref{sec:pyrene} and recalled in the previous paragraph, where we hinted that this can be due to an accidental cancellation of errors. The availability of two independent codes performing similar calculations is of course important, as this can be used to test both. Indeed, the close similarity of the results, at least with some choice of the tunable parameters, validates both codes and the correctness of their results. In addition we remark that AnharmoniCaOs is freely distributed under an open source license via SourceForge, guaranteeing it will remain available and easy to find and download for the foreseeable future.

Scaled harmonic calculations of fundamentals appear to be not much less accurate than our anharmonic ones. However, they depend on the use of empirically calibrated parameters. Moreover, harmonic calculations are limited only to fundamental transitions and completely neglect combination, difference and overtone bands, some of which may be of sizeable intensity especially when strong Fermi resonances occur, as shown also in Refs.~\citenum{Mackie2015}, \citenum{Mackie2016} and \citenum{Maltseva2016}. Indeed, Fig.~\ref{plotpyrthres}, and Table~V in Supplementary Material, clearly show that some bands, most notably (but not only) in the C-H stretch region, split from their simplistic harmonic structure into several, sometimes a multitude of, close ones, due to resonances. Such a complex structure was confirmed for the C-H stretch region by the low temperature, gas-phase measurements of Refs.~\citenum{Mackie2015}, \citenum{Mackie2016} and \citenum{Maltseva2016}. Only anharmonic calculations can account for this. We also note that as a by-product we can also use this kind of anharmonic calculation to estimate purely theoretical frequency scaling factors. Fig.~\ref{plotpyrscalings} shows the empirical scaling factors for the level of theory B97-1/TZ2P,\cite{Cane2007,Maltseva2015,Maltseva2016} as well as the ones we can derive from our calculations on pyrene, which are consistent with them.

The main limiting factor to this kind of anharmonic calculations, at least for species of up to a few tens of atoms in size, is related to the computational cost of computing the cubic and quartic force field constants, as well as the second derivatives of the electric dipole moment, which must be obtained via numerical differentiation since currently there is no readily available code that can compute them analytically. However, some general schemes have been proposed \cite{Ringholm2014} to obtain such higher order derivatives in the framework of the Density Functional Theory, even if their computational implementation is still a work in progress and is still private. When such codes will become available and well-tested, anharmonic calculations, at least of the ground vibrational state and of the states connected to it by permitted transitions, are likely to become commonplace. The AnharmoniCaOs code, besides being usable to process quartic force fields and first and second order electric dipole moment derivatives to obtain spectra from the vibrational ground state, can be used for vibrationally excited states as well, and is currently being tested to obtain moderately high temperature spectra of the same PAHs studied here.

\section*{Supplementary material}

See supplementary material for:
\begin{itemize}
\item A description of the different ethylene oxide force fields that have been used for benchmarking purposes.
\item A table with the comparison of the VMFCI reference calculation with HI-RRBPM calculation D of Ref.~\citenum{Thomas2017} for the force field adimensionned with DFT quadratic force constants;
\item A table with the comparison of the VMFCI reference calculation with a previous VMFCI calculation using a different contraction-truncation scheme for the force field adimensionned with coupled-cluster quadratic force constants;
\item A table with the list of harmonic vibrational normal modes of pyrene;
\item A table with the list of harmonic vibrational normal modes of coronene;
\item A table listing the (significantly IR-active) anharmonic states of pyrene as a function of $r$ values, for fixed $h=8$;
\item Figures of the complete anharmonic spectra of pyrene and coronene.
\end{itemize}

\begin{acknowledgments}
C.~ Joblin and G.~Mulas acknowledge support from the European Research Council under the European Union's Seventh Framework Programme ERC-2013-SyG, Grant Agreement n.~610256 NANOCOSMOS. 
C.~Falvo gratefully acknowledges financial support by the Agence Nationale de la Recherche (ANR) grant ANR-16-CE29-0025 as well as the use of the computing center M\'esoLUM of the LUMAT research federation (FR LUMAT 2764). P.~Cassam-Chenai acknowledges the SIGAMM Mesocentre for hosting the CONVIV code project. This work was granted access to the HPC and vizualization resources of the  ``Centre de Calcul Interactif'' hosted by University Nice Sophia Antipolis. We acknowledge the ``Accordo Quadro INAF-CINECA (2017)'' for the availability of high performance computing resources and support.
\end{acknowledgments}

\bibliography{biblio}
\end{document}


\title{Anharmonic vibrational spectroscopy of Polycyclic Aromatic Hydrocarbons (PAHs)}
\author{Giacomo Mulas}
\affiliation{IRAP, Universit\'e de Toulouse, CNRS, UPS, CNES, 9 Av. du Colonel Roche, 31028 Toulouse Cedex 4, France}
\affiliation{Istituto Nazionale di Astrofisica (INAF), Osservatorio Astronomico di Cagliari, 09047 Selargius (CA), Italy}
\email{gmulas@oa-cagliari.inaf.it}
\author{Cyril Falvo}
\affiliation{Institut des Sciences Mol\'eculaires d'Orsay (ISMO), CNRS, Univ. Paris-Sud, Universit\'e Paris-Saclay, 91405 Orsay, France}
\affiliation{Univ. Grenoble Alpes, CNRS, LIPhy, 38000 Grenoble, France}
\author{Patrick Cassam-Chena\"i}
\affiliation{Universit\'e C\^ote d'Azur, CNRS, LJAD, UMR 7351, 06100 Nice, France}
%
\author{Christine Joblin}
\affiliation{IRAP, Universit\'e de Toulouse, CNRS, UPS, CNES, 9 Av. du Colonel Roche, 31028 Toulouse Cedex 4, France}
\email{christine.joblin@irap.omp.eu}
\begin{abstract}
This supplementary material contains:
\begin{itemize}
\item A description of the different ethylene oxide force fields that have been used for benchmarking purposes
\item Table~\ref{tab:SM1}: Comparison of the VMFCI reference calculation with HI-RRBPM calculation D of Ref.~\citenum{Thomas2017} for the force field adimensionned with DFT quadratic force constants
\item Table~\ref{tab:SM2}: Comparison of the VMFCI reference calculation with a previous VMFCI calculation using a different contraction-truncation scheme for the force field adimensionned with coupled-cluster quadratic force constants
\item Table~\ref{tab:harmpyrene}: List of harmonic vibrational normal modes of pyrene.
\item Table~\ref{tab:harmcoronene}: List of harmonic vibrational normal modes of coronene
\item Table~\ref{tab:pyrtestthresnocor}: List of (significantly IR-active) anharmonic states of pyrene as a function of $r$ values, for fixed $h=8$
\item Figures~\ref{plotpyr1}--\ref{plotcor14}: Anharmonic spectra of pyrene and coronene 
\end{itemize}
\end{abstract}


\maketitle

\section{VMFCI calculations}\label{VMFCI}

\subsection{Different ethylene oxide force fields}\label{VMFCI_FFdesc}

The ethylene oxide quartic force field of B\'egu\'e \emph{et al.}\cite{begue2007comparison} has now become a classical benchmark for methods solving the anharmonic vibrational \sch equation. However, there are three different versions of this force field:
\begin{enumerate}
\item The version provided as supplementary material of Ref.~\citenum{begue2007comparison}, which was actually  only employed for the largest P-VMWCI calculation. It lacks one cubic and seven quartic force constants that were removed from the original DFT quartic force field. This was necessary to alleviate the computational effort required to perform the $\gamma=3$ P-VMWCI calculation of Ref.~\citenum{begue2007comparison}.
\item The version used in the main text of present work and in the VMFCI calculations of Refs.~\citenum{Thomas2017} and \citenum{begue2007comparison}. It is the combination of the CCSD(T)/cc-pVTZ quadratic constants and the full set of B3LYP/6-31+G(d,p) cubic and quartic force constants in mass-weighted, Cartesian, normal coordinates. In terms of adimensional coordinates, the force field obtained amounts to cubic and quartic force constants rescaled by using CCSD(T)/cc-pVTZ quadratic constants. It is this force field that has been used for benchmarking AnharmoniCaOs.
\item The version used in recent investigations.\cite{Thomas2015,Thomas2017,Odunlami2017} It differs from the previous one by the fact that the cubic and quartic force constants were adimensioned by using the B3LYP/6-31+G(d,p) quadratic constants. This results in a noticeably different force field: for instance, the zero point energy with this force field differs by $2$~\cm from that obtained with the previous force field, (compare the two tables below). 
\end{enumerate}

\subsection{Comparison of the new VMFCI reference calculation with  HI-RRBPM results}\label{VMFCI_calc1}
Two fairly large numerical calculations of the $200$ lowest eigenvalues of the vibrational Hamiltonian associated to force field 3 have appeared recently. One is based on the Adaptative Vibrational Configuration Interaction (A-VCI) algorithm,\cite{Odunlami2017} and the other on the Hierarchical Intertwined Reduced-Rank Block Power Method (HI-RRBPM).\cite{Thomas2017}  We postpone a discussion of the remarkable A-VCI results to a forthcoming publication. Here, we just validate a new Vibrational Mean Field Configuration Interaction (VMFCI) contraction-truncation scheme against the latest HI-RRBPM calculation (calculation ``D'' of Ref.~\citenum{Thomas2017}).\par
The new contraction-truncation scheme is the following:\\
- The calculation starts with the same modal basis set of $100$ harmonic function on each of the $15$ modes, then a VSCF calculation is converged to machine precision in seven VMFCI iterations. The common zero point energy (ZPE) of the 15 modes is then $12539.269841$~\cm.\\
- Then, we make $6$ contractions: the four C$-$H stretching modes $\nu_{1}-\nu_{6}-\nu_{9}-\nu_{13}$, the two H$-$C$-$H scissoring modes $\nu_{2}-\nu_{10}$, the three ring breathing and deformation modes $\nu_{3}-\nu_{5}-\nu_{12}$, the two wagging modes $\nu_{4}-\nu_{11}$,
 the two twisting modes $\nu_{7}-\nu_{14}$,  the two rocking modes $\nu_{8}-\nu_{15}$, with truncation thresholds on the sum of the energies of $34072$~\cm\ for the stretchings, $17000$~\cm\ for the ring deformations and $30000$~\cm\ for the four other contractions. This partition is iterated once and the 6 ZPE's decrease down to $12501.358\pm0.001$~\cm. All this is abbreviated by
 VSCFCI($\{\nu_{1},\nu_{6},\nu_{9},\nu_{13}\}$, $34072$; $\{\nu_{2},\nu_{10}\}$, $30000$; $\{\nu_{3},\nu_{5},\nu_{12}\}$, $17000$; $\{\nu_{4},\nu_{11}\}$, $30000$; $\{\nu_{7},\nu_{14}\}$ , $30000$; $\{\nu_{8},\nu_{15}\}$, $30000$).\\
- Next, we make a coarser partition of only two contractions and play on both individual energy and sum of energy  thresholds in a single VMFCI step:\\
 VMFCI({$\{\{\nu_{1},\nu_{6},\nu_{9},\nu_{13}\}_{17523},\{\nu_{2},\nu_{10}\}_{12272},\{\nu_{8},\nu_{15}\}_{12077}\}$, $17914$; \\
 $\{\{\nu_{3},\nu_{5},\nu_{12}\}_{11459}, \{\nu_{4},\nu_{11}\}_{11156}, \{\nu_{7},\nu_{14}\}_{9633} \}$, $15280$ ),\\ (the subscripts are the thresholds in \cm\ on the individual energies of the component of the new contractions). \\
- Finally, the two previous contraction are contracted in a VCI step with a truncation threshold of $6367.5~\cm$ on the second component and a threshold of $11956.6~\cm$ on the sum of the component energies. The size of this VCI is 430769 basis functions, it splits into 4 $C_{2V}$-symmetry blocks of dimensions 108147 for A1, 107421 for A2, 107740 for B1, 107461 for B2.\par
The whole calculation is denoted VMFCI(11956.6) in the table below and the wave numbers are compared to those of calculation D of Ref.~\citenum{Thomas2017}.
\begin{center}
\setlength\LTcapwidth{\hsize}
\begin{longtable}{@{\extracolsep{\fill}}cccc}
\caption{Lowest 200 vibrational energy levels (in  \cm ) of ethylene oxide for force field 3}\label{tab:SM1} \\[1mm]
\hline \\[.5mm]
irreps.&VMFCI(11956.6)&Ref.~\citenum{Thomas2017}&Assignment \cite{Thomas2017} \\[.5mm]
\hline \\[.5mm] \endfirsthead
\hline \\[.5mm]
irreps.&VMFCI(11956.6)&Ref.~\citenum{Thomas2017}&Assignment \cite{Thomas2017}\\[.5mm]
\hline \\[.5mm] \endhead
\\[.5mm] \hline \\[.5mm] \multicolumn{4}{l}{\small\sl continued on next page}\\[.5mm] \endfoot
\\ \endlastfoot
$A_1$ &   12461.57 &   12461.65  &    ZPE                    \\
$B_2$ &     792.80 &     793.10  & $\nu_{15}                $\\
$B_1$ &     821.98 &     822.00  & $\nu_{12}                  $\\
$A_1$ &     878.32 &     878.33  & $\nu_5                   $\\
$A_2$ &    1017.31 &    1017.49  & $\nu_8                   $\\
$A_1$ &    1121.64 &    1121.94  & $\nu_4                   $\\
$B_1$ &    1124.07 &    1124.37  & $\nu_{11}                  $\\
$B_2$ &    1146.00 &    1146.19  & $\nu_{14}                  $\\
$A_2$ &    1148.22 &    1148.40  & $\nu_7                   $\\
$A_1$ &    1270.95 &    1270.94  & $\nu_3                   $\\
$B_1$ &    1467.59 &    1467.72  & $\nu_{10}                  $\\
$A_1$ &    1495.49 &    1495.74  & $\nu_2                   $\\
$A_1$ &    1588.03 &    1589.09  & $2\nu_{15}                 $\\
$A_2$ &    1611.50 &    1611.63  & $\nu_{15} + \nu_{12}         $\\
$A_1$ &    1641.39 &    1641.21  & $2\nu_{12}                 $\\
$B_2$ &    1670.69 &    1670.78  & $\nu_{15} + \nu_5          $\\
$B_1$ &    1695.36 &    1695.13  & $\nu_5 + \nu_{12}          $\\
$A_1$ &    1755.07 &    1754.81  & $2\nu_5                  $\\
$B_1$ &    1806.39 &    1806.81  & $\nu_{15} + \nu_8          $\\
$B_2$ &    1832.96 &    1832.89  & $\nu_{12} + \nu_8          $\\
$A_2$ &    1889.73 &    1889.81  & $\nu_5 + \nu_8           $\\
$A_2$ &    1908.54 &    1908.90  & $\nu_{15} + \nu_{11}         $\\
$B_2$ &    1910.58 &    1910.80  & $\nu_{15} + \nu_4          $\\
$B_1$ &    1928.71 &    1929.15  & $\nu_{15} + \nu_7          $\\
$A_1$ &    1938.65 &    1938.92  & $\nu_{15} + \nu_{14}         $\\
$B_1$ &    1940.37 &    1940.58  & $\nu_{12} + \nu_4          $\\
$A_1$ &    1944.67 &    1944.71  & $\nu_{12} + \nu_{11}         $\\
$A_2$ &    1962.19 &    1962.04  & $\nu_{12} + \nu_{14}         $\\
$B_2$ &    1970.84 &    1970.60  & $\nu_{12} + \nu_7          $\\
$A_1$ &    1998.09 &    1997.73  & $\nu_5 + \nu_4           $\\
$B_1$ &    1999.74 &    1999.39  & $\nu_5 + \nu_{11}          $\\
$B_2$ &    2021.35 &    2021.12  & $\nu_5 + \nu_{14}          $\\
$A_2$ &    2024.24 &    2024.00  & $\nu_5 + \nu_7           $\\
$A_1$ &    2032.16 &    2032.78  & $2\nu_8                  $\\
$B_2$ &    2060.62 &    2060.74  & $\nu_{15} + \nu_3          $\\
$B_1$ &    2086.60 &    2086.38  & $\nu_{12} + \nu_3          $\\
$A_2$ &    2128.92 &    2128.77  & $\nu_8 + \nu_4           $\\
$B_2$ &    2135.29 &    2135.21  & $\nu_8 + \nu_{11}          $\\
$A_1$ &    2140.80 &    2140.60  & $\nu_5 + \nu_3           $\\
$B_1$ &    2152.76 &    2152.89  & $\nu_8 + \nu_{14}          $\\
$A_1$ &    2165.51 &    2165.40  & $\nu_8 + \nu_7           $\\
$A_1$ &    2236.23 &    2236.66  & $2\nu_4                  $\\
$B_1$ &    2242.75 &    2243.99  & $\nu_4 + \nu_{11}          $\\
$A_2$ &    2246.58 &    2247.45  & $\nu_{15} + \nu_{10}         $\\
$A_1$ &    2249.90 &    2250.81  & $2\nu_{11}                 $\\
$B_2$ &    2264.70 &    2266.13  & $\nu_7 + \nu_{11}          $\\
$A_2$ & \textit{2267.87} & \textit{2268.95}  & $\nu_{14} + \nu_{11}         $\\
$B_2$ & \textit{2268.07} & \textit{2268.67}  & $\nu_{14} + \nu_4          $\\
$A_2$ &    2275.16 &    2275.77  & $\nu_7 + \nu_4           $\\
$B_2$ &    2282.99 &    2284.58  & $\nu_{15} + \nu_2          $\\
$A_1$ &    2288.29 &    2288.50  & $\nu_{12} + \nu_{10}         $\\
$A_2$ &    2289.88 &    2290.67  & $\nu_3 + \nu_8           $\\
$A_1$ &    2290.79 &    2291.41  & $2\nu_{14}                 $\\
$B_1$ &    2293.56 &    2294.02  & $\nu_7 + \nu_{14}          $\\
$A_1$ &    2296.10 &    2296.40  & $2\nu_7                  $\\
$B_1$ &    2310.81 &    2311.26  & $\nu_{12} + \nu_2          $\\
$B_1$ &    2345.42 &    2345.49  & $\nu_5 + \nu_{10}          $\\
$A_1$ &    2370.35 &    2370.74  & $\nu_5 + \nu_2           $\\
$B_2$ &    2381.33 &    \textbf{2388.00}  & $3\nu_{15}                 $\\
$A_1$ &    2389.76 &    2389.51  & $\nu_3 + \nu_4           $\\
$B_1$ &    2391.37 &    2391.01  & $\nu_3 + \nu_{11}          $\\
$B_1$ &    2402.50 &    2405.27  & $2\nu_{15} + \nu_{12}        $\\
$B_2$ &    2415.20 &    2415.12  & $\nu_3 + \nu_{14}          $\\
$A_2$ &    2415.99 &    2415.73  & $\nu_3 + \nu_7           $\\
$B_2$ &    2428.33 &    2429.22  & $\nu_{15} + 2\nu_{12}        $\\
$B_1$ &    2458.05 &    2458.83  & $3\nu_{12}                 $\\
$A_1$ &    2465.15 &    2468.36  & $2\nu_{15} + \nu_5         $\\
$B_2$ &    2481.23 &    2482.05  & $\nu_{10} + \nu_8          $\\
$A_2$ &    2484.11 &    2484.74  & $\nu_{15} + \nu_5 + \nu_{12} $\\
$A_2$ &    2506.69 &    2507.98  & $\nu_2 + \nu_8           $\\
$A_1$ &    2510.12 &    2513.46  & $\nu_5 + 2\nu_{12}         $\\
$A_1$ &    2537.36 &    2537.41  & $2\nu_3                  $\\
$B_2$ &    2546.70 &    2547.55  & $\nu_{15} + 2\nu_5         $\\
$B_1$ &    2567.16 &    2567.33  & $2\nu_5 + \nu_{12}         $\\
$B_1$ &    2583.86 &    2584.32  & $\nu_{10} + \nu_4          $\\
$A_1$ &    2587.72 &    2588.24  & $\nu_{10} + \nu_{11}         $\\
$A_2$ &    2595.27 &    2597.09  & $2\nu_{15} + \nu_8         $\\
$A_2$ &    2599.54 &    2600.34  & $\nu_{10} + \nu_{14}         $\\
$B_2$ &    2601.60 &    2602.28  & $\nu_{10} + \nu_7          $\\
$A_1$ &    2612.35 &    2613.87  & $\nu_2 + \nu_4           $\\
$B_1$ &    2616.37 &    2617.02  & $\nu_2 + \nu_{11}          $\\
$A_1$ &    2616.89 &    2618.09  & $\nu_{15} + \nu_{12} + \nu_8 $\\
$A_1$ &    2629.75 &    2630.88  & $3\nu_5                  $\\
$A_2$ &    2640.84 &    2641.75  & $\nu_2 + \nu_7           $\\
$B_2$ &    2641.13 &    2641.91  & $\nu_2 + \nu_{14}          $\\
$A_2$ &    2646.04 &    2646.24  & $2\nu_{12} + \nu_8         $\\
$B_1$ &    2676.07 &    2677.43  & $\nu_{15} + \nu_5 + \nu_8  $\\
$B_1$ &    2693.79 &    2696.02  & $2\nu_{15} + \nu_{11}        $\\
$A_1$ & \textit{2699.42} &  \textit{2701.69}  & $2\nu_{15} + \nu_4         $\\
$B_2$ & \textit{2700.39} &  \textit{2701.38}  & $\nu_5 + \nu_{12} + \nu_8  $\\
$A_2$ &    2711.51 &    2713.94  & $2\nu_{15} + \nu_7         $\\
$B_2$ &    2721.54 &    2723.68  & $\nu_{15} + \nu_{12} + \nu_{11}$\\
$A_2$ &    2728.32 &    2730.67  & $\nu_{15} + \nu_{12} + \nu_4 $\\
$B_2$ &    2734.81 &    2736.72  & $2\nu_{15} + \nu_{14}        $\\
$B_1$ &    2736.84 &    2737.37  & $\nu_{10} + \nu_3          $\\
$A_1$ &    2743.92 &    2744.96  & $\nu_{15} + \nu_{12} + \nu_7 $\\
$B_1$ &    2751.45 &    2752.71  & $\nu_{15} + \nu_{12} + \nu_{14}$\\
$A_1$ &    2757.10 &    2758.29  & $2\nu_{12} + \nu_4         $\\
$A_2$ &    2760.48 &    2761.35  & $2\nu_5 + \nu_8          $\\
$B_1$ &    2760.82 &    2761.37  & $2\nu_{12} + \nu_{11}        $\\
$A_1$ &    2762.81 &    2763.88  & $\nu_2 + \nu_3           $\\
$B_2$ &    2775.18 &    2775.59  & $2\nu_{12} + \nu_{14}        $\\
$A_2$ &    2784.06 &    2787.72  & $\nu_{15} + \nu_5 + \nu_{11} $\\
$B_2$ &    2785.74 &    2788.41  & $\nu_{15} + \nu_5 + \nu_4  $\\
$A_2$ &    2789.89 &    2790.40  & $2\nu_{12} + \nu_7         $\\
$B_1$ &    2803.09 & \textbf{2809.98} & $\nu_{15} + \nu_5 + \nu_7  $\\
$A_1$ &    2811.45 &    2813.50  & $\nu_5 + \nu_{12} + \nu_{11} $\\
$A_1$ &    2812.63 &    2817.12  & $\nu_{15} + \nu_5 + \nu_{14} $\\
$B_2$ &    2815.48 &    2817.46  & $\nu_{15} + 2\nu_8         $\\
$B_1$ &    2816.02 &    2817.70  & $\nu_5 + \nu_{12} + \nu_4  $\\
$A_2$ &    2832.39 &    2834.71  & $\nu_5 + \nu_{12} + \nu_{14} $\\
$B_1$ & \textit{2840.51} & \textit{2842.50} & $\nu_{12} + 2\nu_8         $\\
$B_2$ & \textit{2841.30} & \textit{2842.15} & $\nu_5 + \nu_{12} + \nu_7  $\\
$A_1$ &    2851.94 &    2853.47  & $2\nu_{15} + \nu_3         $\\
$A_1$ &    2872.12 &    2873.05  & $2\nu_5 + \nu_4          $\\
$A_2$ & \textit{2872.23} & \textit{2874.54} & $\nu_{15} + \nu_{12} + \nu_3 $\\   
$B_1$ & \textit{2872.96} & \textit{2873.88} & $2\nu_5 + \nu_{11}         $\\   
$B_2$ &    2894.46 &    2895.89  & $2\nu_5 + \nu_{14}         $\\
$A_2$ &    2897.94 &    2898.62  & $2\nu_5 + \nu_7          $\\
$A_1$ &    2898.22 &    2901.60  & $\nu_5 + 2\nu_8          $\\
$A_1$ &    2900.68 &    2902.18  & $2\nu_{12} + \nu_3         $\\
$B_1$ &    2907.72 &    2911.04  & $\nu_9                   $\\
$A_1$ &    2913.03 &    2917.53  & $\nu_{15} + \nu_8 + \nu_{11}\ ^a$\\
$B_1$ &    2915.00 &    2917.90  & $\nu_{15} + \nu_8 + \nu_4  $\\
$A_1$ &    2918.79 &    2923.02  & $\nu_1                   $\\
$B_2$ &    2929.63 &    2932.84  & $\nu_{15} + \nu_5 + \nu_3  $\\
$B_2$ &    2934.76 &    2937.33  & $\nu_{12} + \nu_8 + \nu_4  $\\
$A_2$ &    2940.11 &    2942.52  & $\nu_{15} + \nu_8 + \nu_{14} $\\
$B_2$ &    2946.04 &    2949.06  & $\nu_{15} + \nu_8 + \nu_7  $\\
$A_2$ &    2949.03 &    2951.56  & $\nu_{12} + \nu_8 + \nu_{11} $\\
$B_1$ &    2952.56 &    2954.44  & $\nu_5 + \nu_{12} + \nu_3  $\\
$A_1$ &    2953.40 &    2958.32  & $2\nu_{10}                 $\\
$A_1$ &    2965.35 &    2967.45  & $\nu_{12} + \nu_8 + \nu_{14} $\\
$B_1$ &    2980.82 &    2981.94  & $\nu_{12} + \nu_8 + \nu_7  $\\
$B_1$ &    2990.72 &    2995.63  & $\nu_2 + \nu_{10}          $\\
$A_1$ & \textit{2995.54} & \textit{\textbf{3002.34}} & $2\nu_2                  $\\
$A_2$ & \textit{2999.17} & \textit{3000.44} & $\nu_5 + \nu_8 + \nu_4   $\\
$B_2$ &    3003.31 &    3006.30  & $\nu_5 + \nu_8 + \nu_{11}  $\\
$A_1$ &    3009.07 &    3009.47  & $2\nu_5 + \nu_3          $\\
$B_2$ &    3013.56 & \textbf{3020.05} & $\nu_{15} + 2\nu_{11}\ ^{c1}      $\\
$A_2$ & \textit{3021.24} & \textit{3025.44} & $\nu_{15} + \nu_4 + \nu_{11} $\\
$B_1$ & \textit{3023.14} & \textit{3024.79} & $\nu_5 + \nu_8 + \nu_{14}  $\\
$A_2$ &    3026.36 &    3027.95  & $\nu_6                   $\\
$B_2$ &    3030.94 &    3033.86  & $\nu_{15} + 2\nu_4\ ^{c1}      $\\
$B_1$ & \textit{3034.01} & \textit{3038.36}  & $2\nu_{15} + \nu_{10}        $\\
$A_1$ & \textit{3034.03} & \textit{3035.70}  & $\nu_5 + \nu_8 + \nu_7   $\\
$B_2$ &    3038.39 &    3040.21  & $\nu_{13}                  $\\
$A_2$ & \textit{3041.93} & \textit{\textbf{3048.87}} & $3\nu_8                  $\\
$A_1$ &   3042.71 & \textbf{3048.70} & $\nu_{15} + \nu_7 + \nu_{11} $\\
$B_1$ & \textit{3043.77} & \textit{3048.42} & $\nu_{12} + 2\nu_{11}\ ^b      $\\
$B_1$ &    3050.73 & \textbf{3057.21} & $\nu_{15} + \nu_{14} + \nu_{11}$\\
$A_1$ &    3053.21 & \textbf{3060.39} & $\nu_{12} + \nu_4 + \nu_{11} $\\
$A_1$ &    3056.21 & \textbf{3061.69} & $\nu_{15} + \nu_{14} + \nu_4 $\\
$B_1$ &    3060.53 & \textbf{3066.08} & $\nu_{15} + \nu_7 + \nu_4  $\\
$B_2$ &    3063.15 &    3066.94  & $\nu_{15} + \nu_{12} + \nu_{10}$\\
$A_1$ & \textit{3063.78} & \textit{\textbf{3071.18}}  & $2\nu_{15} + \nu_2         $\\
$B_1$ & \textit{3065.83} & \textit{3069.95}  & $\nu_{12} + 2\nu_4\ ^b       $\\
$A_2$ & \textit{3069.09} & \textit{\textbf{3074.73}} & $\nu_{12} + \nu_{14} + \nu_4\ ^d$\\
$B_2$ & \textit{3069.29} & \textit{3072.99} & $\nu_{15} + 2\nu_{14}\ ^{c2}      $\\   
$B_1$ &    3074.08 & \textbf{3079.51} & $\nu_{15} + \nu_3 + \nu_8  $\\
$A_2$ & \textit{3079.24} & \textit{\textbf{3085.65}} & $\nu_{15} + \nu_{12} + \nu_2 $\\
$B_2$ & \textit{3080.70} & \textit{3084.54} & $\nu_{12} + \nu_{14} + \nu_{11}$\\
$A_2$ &    3083.97 &    3088.11  & $\nu_{15} + \nu_7 + \nu_{14} $\\
$B_2$ &    3085.07 &    3088.58  & $\nu_{15} + 2\nu_7\ ^{c2}       $\\
$B_2$ &    3092.78 &    \textbf{3098.61}  & $\nu_{12} + \nu_7 + \nu_4  $\\
$A_2$ &    3099.06 &    3102.95  & $\nu_{12} + \nu_7 + \nu_{11} $\\
$B_2$ &    3100.52 &    3103.16  & $\nu_{12} + \nu_3 + \nu_8  $\\
$B_1$ &    3100.73 &    3104.30  & $\nu_{12} + 2\nu_{14}        $\\
$B_1$ &    3106.74 &    3109.15  & $2\nu_{12} + \nu_{10}        $\\
$A_1$ & \textit{3109.44} & \textit{\textbf{3115.58}}  & $\nu_{12} + \nu_7 + \nu_{14}\ ^e $\\
$A_1$ & \textit{3109.53} & \textit{\textbf{3125.67}}  & $\nu_5 + 2\nu_4\ ^e $\\
$B_1$ &    3114.87 &    3119.83  & $\nu_5 + \nu_4 + \nu_{11}  $\\
$B_1$ &    3118.20 &    3121.56  & $\nu_{12} + 2\nu_7         $\\
$A_1$ & \textit{3121.77} & \textit{\textbf{3112.80}}  & $\nu_5 + 2\nu_{11}\ ^e $\\
$A_2$ &    3122.74 &    3125.96  & $\nu_{15} + \nu_5 + \nu_{10} $\\
$A_1$ &    3125.17 &    3127.61  & $2\nu_{12} + \nu_2         $\\
$A_1$ &    3131.40 &    3134.62  & $2\nu_8 + \nu_4          $\\
$A_2$ & \textit{3137.83} &\textit{\textbf{3143.98}} & $\nu_5 + \nu_{14} + \nu_{11} $\\
$B_2$ & \textit{3138.43} &\textit{3141.97} & $\nu_5 + \nu_7 + \nu_{11}  $\\
$B_2$ & \textit{3140.71} &\textit{3142.77} & $\nu_5 + \nu_{14} + \nu_4  $\\
$B_1$ &    3141.72 &    3146.30  & $2\nu_8 + \nu_{11}         $\\
$A_2$ &    3148.76 & \textbf{3154.15}  & $\nu_5 + \nu_7 + \nu_4   $\\
$A_2$ &    3155.47 & \textbf{3160.67} & $\nu_5 + \nu_3 + \nu_8   $\\
$B_2$ &    3156.27 & \textbf{3162.47} & $\nu_{15} + \nu_5 + \nu_2  $\\
$B_2$ &    3156.67 & \textbf{3163.50} & $2\nu_8 + \nu_{14}         $\\
$A_1$ &    3160.98 &    3163.70  & $\nu_5 + \nu_{12} + \nu_{10} $\\
$A_1$ &    3162.51 &    3164.77  & $\nu_5 + 2\nu_{14}         $\\
$B_1$ &    3165.89 &    3169.44  & $\nu_5 + \nu_7 + \nu_{14}  $\\
$A_1$ &    3168.96 &    3171.54  & $\nu_5 + 2\nu_7          $\\
$A_1$ &    3170.54 &   ???    & $4\nu_{15}                 $\\
$A_2$ &    3171.59 &    3175.51  & $\nu_{15} + \nu_3 + \nu_{11} $\\
$B_2$ &    3175.16 &    3179.51  & $\nu_{15} + \nu_3 + \nu_4  $\\
$A_2$ &    3178.65 &    3182.09  & $2\nu_8 + \nu_7          $\\
$B_1$ &    3181.45 & \textbf{3189.11}  & $\nu_5 + \nu_{12} + \nu_2  $\\
$A_2$ &    3190.24 &   ???    & $3\nu_{15} + \nu_{12}        $\\
$B_1$ &    3192.71 & \textbf{3198.95}  & $\nu_{15} + \nu_3 + \nu_7  $\\
$A_1$ &    3201.63 &    3205.73  & $\nu_{12} + \nu_3 + \nu_{11} $\\
$B_1$ &    3201.86 &    3206.84  & $\nu_{12} + \nu_3 + \nu_4  $\\
$A_1$ &    3206.74 &    3209.19  & $\nu_{15} + \nu_3 + \nu_{14} $\\
$A_1$ &    3216.11 &   ???    &  $2\nu_5 + 2\nu_{12}         $\\        
$B_1$ &    3221.25 &    3225.03  & $2\nu_5 + \nu_{10}         $\\
$A_2$ &    3224.44 &    3227.09  & $\nu_{12} + \nu_3 + \nu_{14} $\\
$B_2$ &    3231.59 &    3235.46  & $\nu_{12} + \nu_3 + \nu_7  $\\
\hline 
\end{longtable}
\end{center}
Table I. Inverted or permuted wave numbers are italicized, the  HI-RRBPM numbers which deviate by more than $5$ \cm\ are in bold. Some wave numbers are missing in the HI-RRBPM list and are marked by the symbol $???$. This is probably  because they would appear higher in the spectrum. They are overtones or combinations bands of low frequency modes which does not seem to be well described in the HI-RRBPM calculation arguably due to too low basis truncation criteria. As a matter of fact, most of the bold number correspond to bands involving $\nu_5$, $\nu_{12}$ or $\nu_{15}$.  Some footnotes in this tables are related to those in Tab.~2 of Ref. ~\citenum{Thomas2017}.\\
 
\noindent
$^a$ main weight on $\nu_1$ in VMFCI calculation, as for the eigenstate corresponding to $2918.79$ \cm. \\
$^b$ both assignments to $\nu_{12} + 2\nu_{11}$ in the new calculation.\\
$^{c1,c2}$ inversion confirmed in the new calculation.\\
$^d$ still assigned to $\nu_{15} + \nu_7 + \nu_{14}$ in the new calculation.\\
$^e$ permutation of 3 levels in the new calculation instead of just 2.\\

\subsection{Comparison of the new VMFCI reference calculation with a previous VMFCI calculation using a different contraction-truncation scheme}\label{VMFCI_calc2}

Here, we display the 200 lowest eigenvalues of the new VMFCI calculation used as a reference for the AnharmoniCaOs calculations in the main text. It is compared to a previous VMFCI calculation, referred to as VMFCI(11300), which uses force field 2 but was compared to force field 3 HI-RRBPM calculations in Ref~\citenum{Thomas2017}.\par
The contraction-truncation schemes of the two calculations are different.
The contraction-truncation scheme of the new VMFCI(11956.6) calculation is the same as that of the previous section except that the 2-contraction step\\  VMFCI({$\{\{\nu_{1},\nu_{6},\nu_{9},\nu_{13}\}_{17523},\{\nu_{2},\nu_{10}\}_{12272},\{\nu_{8},\nu_{15}\}_{12077}\}$, $17914$;\\
 $\{\{\nu_{3},\nu_{5},\nu_{12}\}_{11459}, \{\nu_{4},\nu_{11}\}_{11156}, \{\nu_{7},\nu_{14}\}_{9633} \}$, $15280$ ), is iterated once, leading to a common ZPE for the two contractions of $12492.353\pm0.001$ \cm. This ZPE is to be compared with that of the 6-contraction VSCFCI step of $12502.883\pm0.001$ \cm and with the converged VSCF ZPE of $12540.280269$ \cm.\\
 The size of the final VCI is 429499 splitted into 4 $C_{2V}$-symmetry blocks of dimensions 107852 for A1, 107096 for A2, 107415 for B1, 107136 for B2.\par
 The contraction-truncation scheme of the old VMFCI(11300) calculation is essentially the same as that of Ref.\citenum{begue2007comparison}, with a larger final threshold: $11300~\cm$ instead of $10721~\cm$. It can be abbreviated as:\\
 VSCF/VSCFCI($\{\nu_{1},\nu_{6},\nu_{9},\nu_{13}\}$, $34072$; $\{\nu_{3},\nu_{5},\nu_{12}\}$, $17000$)/VSCFCI($\{\{\nu_{1},\nu_{6},\nu_{9},\nu_{13}\},\nu_{10},\nu_{15}\}$, $19464$)/VCI(11300)\\
 The size of the final VCI is larger than for the new one with 456201 basis functions splitted into 4 $C_{2V}$-symmetry blocks of dimensions 114852 for A1, 113489 for A2, 114278 for B1, 113582 for B2. However, the results are clearly less accurate (see Table S2 below).
               
\begin{center}
\setlength\LTcapwidth{\hsize}
\begin{longtable}{@{\extracolsep{\fill}}ccc}
\caption{Lowest 200 vibrational energy levels (in  $\cm$) of ethylene oxide for force field 2}\label{tab:SM2}\\[1mm]
\hline\\[.5mm]
irreps.&VMFCI(11956.6)&VMFCI(11300) \cite{Thomas2017}\\[.5mm]
\hline \endfirsthead
\hline \\[.5mm]
irreps.&VMFCI(11956.6)&VMFCI(11300) \cite{Thomas2017}\\[.5mm]
\hline \\[.5mm] \endhead
\\[.5mm] \hline \\[.5mm] \multicolumn{3}{l}{\small\sl continued on next page}\\[.5mm]  \endfoot
\\ \endlastfoot
$A_1$ &  12463.57  & 12463.65 \\
$B_2$ &    793.03  &   793.37 \\
$B_1$ &    822.48  &   822.75 \\
$A_1$ &    878.76  &   879.04 \\
$A_2$ &   1017.97  &  1018.96 \\
$A_1$ &   1122.14  &  1122.96 \\
$B_1$ &   1124.37  &  1125.18 \\
$B_2$ &   1146.45  &  1147.09 \\
$A_2$ &   1148.42  &  1149.02 \\
$A_1$ &   1271.11  &  1271.50 \\
$B_1$ &   1467.98  &  1468.34 \\
$A_1$ &   1496.00  &  1496.65 \\
$A_1$ &   1588.62  &  1589.38 \\
$A_2$ &   1612.27  &  1612.92 \\
$A_1$ &   1642.48  &  1642.86 \\
$B_2$ &   1671.38  &  1671.94 \\
$B_1$ &   1696.43  &  1696.71 \\
$A_1$ &   1755.98  &  1756.21 \\
$B_1$ &   1807.45  &  1808.96 \\
$B_2$ &   1834.31  &  1835.47 \\
$A_2$ &   1890.96  &  1891.96 \\
$A_2$ &   1909.35  &  1910.19 \\
$B_2$ &   1911.54  &  1912.38 \\
$B_1$ &   1929.48  &  1930.21 \\
$A_1$ &   1939.41  &  1940.01 \\
$B_1$ &   1941.36  &  1941.82 \\
$A_1$ &   1945.44  &  1945.80 \\
$A_2$ &   1963.29  &  1963.62 \\
$B_2$ &   1971.50  &  1971.83 \\
$A_1$ &   1999.08  &  1999.34 \\
$B_1$ &   2000.51  &  2000.76 \\
$B_2$ &   2022.30  &  2022.57 \\
$A_2$ &   2024.90  &  2025.19 \\
$A_1$ &   2033.60  &  2036.08 \\
$B_2$ &   2061.03  &  2061.51 \\
$B_1$ &   2087.26  &  2087.49 \\
$A_2$ &   2130.39  &  2132.05 \\
$B_2$ &   2136.49  &  2138.11 \\
$A_1$ &   2141.59  &  2141.80 \\
$B_1$ &   2154.23  &  2155.72 \\
$A_1$ &   2166.39  &  2167.73 \\
$A_1$ &   2237.29  &  2237.65 \\
$B_1$ &   2243.54  &  2243.80 \\
$A_2$ &   2247.41  &  2247.97 \\
$A_1$ &   2250.51  &  2250.86 \\
$B_2$ &   2265.43  &  2266.28 \\
$A_2$ &   2268.70  &  2269.27 \\
$B_2$ &   2269.00  &  2269.43 \\
$A_2$ &   2275.80  &  2276.17 \\
$B_2$ &   2283.65  &  2284.40 \\
$A_1$ &   2289.19  &  2289.57 \\
$A_2$ &   2290.60  &  2291.52 \\
$A_1$ &   2291.68  &  2292.21 \\
$B_1$ &   2294.23  &  2294.73 \\
$A_1$ &   2296.53  &  2297.09 \\
$B_1$ &   2311.89  &  2312.71 \\
$B_1$ &   2346.25  &  2346.58 \\
$A_1$ &   2371.38  &  2372.12 \\
$B_2$ &   2382.39  &  2382.86 \\
$A_1$ &   2390.42  &  2390.76 \\
$B_1$ &   2391.82  &  2392.13 \\
$B_1$ &   2403.66  &  2404.62 \\
$B_2$ &   2415.82  &  2416.22 \\
$A_2$ &   2416.33  &  2416.72 \\
$B_2$ &   2429.68  &  2430.50 \\
$B_1$ &   2459.82  &  2460.42 \\
$A_1$ &   2466.23  &  2467.01 \\
$B_2$ &   2482.36  &  2483.52 \\
$A_2$ &   2485.46  &  2486.14 \\
$A_2$ &   2508.00  &  2510.34 \\
$A_1$ &   2511.90  &  2512.47 \\
$A_1$ &   2537.63  &  2538.02 \\
$B_2$ &   2547.89  &  2548.58 \\
$B_1$ &   2568.81  &  2569.28 \\
$B_1$ &   2584.82  &  2585.75 \\
$A_1$ &   2588.50  &  2589.46 \\
$A_2$ &   2596.91  &  2599.54 \\
$A_2$ &   2600.64  &  2601.69 \\
$B_2$ &   2602.50  &  2603.53 \\
$A_1$ &   2613.54  &  2615.39 \\
$B_1$ &   2617.19  &  2618.85 \\
$A_1$ &   2618.61  &  2621.28 \\
$A_1$ &   2631.18  &  2631.55 \\
$A_2$ &   2641.56  &  2643.15 \\
$B_2$ &   2642.02  &  2643.48 \\
$A_2$ &   2648.15  &  2650.27 \\
$B_1$ &   2677.78  &  2680.29 \\
$B_1$ &   2695.22  &  2697.64 \\
$A_1$ &   2700.93  &  2703.48 \\
$B_2$ &   2702.44  &  2704.29 \\
$A_2$ &   2712.80  &  2715.21 \\
$B_2$ &   2722.99  &  2725.03 \\
$A_2$ &   2729.73  &  2731.66 \\
$B_2$ &   2735.91  &  2737.56 \\
$B_1$ &   2737.38  &  2737.96 \\
$A_1$ &   2745.45  &  2747.23 \\
$B_1$ &   2752.86  &  2754.35 \\
$A_1$ &   2758.55  &  2760.02 \\
$B_1$ &   2762.24  &  2763.28 \\
$A_2$ &   2762.32  &  2763.83 \\
$A_1$ &   2763.49  &  2764.81 \\
$B_2$ &   2777.03  &  2778.08 \\
$A_2$ &   2785.37  &  2786.85 \\
$B_2$ &   2787.21  &  2788.85 \\
$A_2$ &   2791.15  &  2792.07 \\
$B_1$ &   2804.45  &  2805.89 \\
$A_1$ &   2812.84  &  2813.86 \\
$B_1$ &   2814.12  &  2815.20 \\
$A_1$ &   2817.31  &  2818.25 \\
$B_2$ &   2817.53  &  \textbf{2822.93} \\
$A_2$ &   2834.11  &  2834.86 \\
$B_2$ &   2842.57  &  2843.35 \\
$B_1$ &   2842.87  &  2847.79 \\
$A_1$ &   2852.76  &  2854.33 \\
$A_2$ &   \textit{2873.20} &  \textit{2874.44} \\
$A_1$ &   \textit{2873.63} &  \textit{2874.30} \\
$B_1$ &   2874.26  &  2874.81 \\
$B_2$ &   2895.96  &  2896.62 \\
$A_2$ &   2899.13  &  2899.84 \\
$A_1$ &   2900.38  &  \textbf{2905.43} \\
$A_1$ &   2901.94  &  2902.67 \\
$B_1$ &   2909.55  &  2910.86 \\
$A_1$ &   2914.73  &  2916.69 \\
$B_1$ &   2916.95  &  \textbf{2922.14} \\
$A_1$ &   2920.29  &  2924.52 \\
$B_2$ &   2930.71  &  2931.82 \\
$B_2$ &   2936.85  &  2941.73 \\
$A_2$ &   2942.08  &  2946.78 \\
$B_2$ &   2947.75  &  2951.93 \\
$A_2$ &   2950.91  &  2954.90 \\
$B_1$ &   2953.94  &  2954.55 \\
$A_1$ &   2955.01  &  2956.38 \\
$A_1$ &   2967.42  &  2971.25 \\
$B_1$ &   2982.39  &  2985.81 \\
$B_1$ &   2992.22  &  2993.97 \\
$A_1$ &   2997.03  &  3000.08 \\
$A_2$ &   3001.27  &  3005.26 \\
$B_2$ &   3005.13  &  3009.46 \\
$A_1$ &   3010.51  &  3011.16 \\
$B_2$ &   3015.27  &  \textbf{3020.67} \\
$A_2$ &   3022.83  &  3027.65 \\
$B_1$ &   3025.19  &  3028.81 \\
$A_2$ &   3028.63  &  3029.75 \\
$B_2$ &   3032.42  &  3035.82 \\
$B_1$ &   3035.32  &  3038.05 \\
$A_1$ &   3035.53  &  3039.04 \\
$B_2$ &   3040.60  &  3041.99 \\
$A_1$ &   3043.97  &  3047.88 \\
$A_2$ &   \textit{3044.37} &  \textit{\textbf{3055.15}} \\
$B_1$ &   \textit{3045.56} &  \textit{3049.21} \\
$B_1$ &   3052.13  &  3055.92 \\
$A_1$ &   3054.74  &  3057.68 \\
$A_1$ &   3057.67  &  3061.06 \\
$B_1$ &   3061.83  &  3064.98 \\
$B_2$ &   3064.57  &  3066.76 \\
$A_1$ &   3065.11  &  3068.03 \\
$B_1$ &   3066.99  &  3069.49 \\
$B_2$ &   3070.40  &  3073.86 \\
$A_2$ &   3070.63  &  3073.86 \\
$B_1$ &   3075.31  &  3078.94 \\
$A_2$ &   3080.74  &  3084.02 \\
$B_2$ &   3082.16  &  3084.92 \\
$A_2$ &   3085.30  &  3087.97 \\
$B_2$ &   3086.36  &  3089.06 \\
$B_2$ &   3093.99  &  3096.23 \\
$A_2$ &   3100.17  &  3102.56 \\
$B_2$ &   3101.89  &  3104.44 \\
$B_1$ &   3102.46  &  3105.97 \\
$B_1$ &   3108.24  &  3109.55 \\
$A_1$ &   3110.78  &  3112.92$^b$\\
$A_1$ &   3111.10  &  \textbf{3116.35} \\
$B_1$ &   3116.21  &  3118.38 \\
$B_1$ &   3119.06  &  3122.34 \\
$A_1$ &   3122.94  &  3127.59 \\
$A_2$ &   3124.07  &  3126.44 \\
$A_1$ &   3126.88  &  3129.22 \\
$A_1$ &   \textit{3134.00}  &  \textit{\textbf{3146.79}} \\
$A_2$ &   \textit{3139.32}  &  \textit{3143.09} \\
$B_2$ &   \textit{3139.67}  &  \textit{3143.48} \\
$B_2$ &   \textit{3142.20}  &  \textit{3146.04} \\
$B_1$ &   \textit{3143.98}  &  \textit{\textbf{3156.68}} \\
$A_2$ &   \textit{3149.94}  &  \textit{3153.20} \\
$A_2$ &   3156.84  &  3159.94 \\
$B_2$ &   3157.70  &  3160.91 \\
$B_2$ &   \textit{3159.06} &  \textit{\textbf{3170.99}} \\
$A_1$ &   \textit{3162.45}  &  \textit{3163.64} \\
$A_1$ &   \textit{3164.00}  &  \textit{3168.19} \\
$B_1$ &   \textit{3167.12}  &  \textit{3170.35} \\
$A_1$ &   3169.90  &  3173.84 \\
$A_1$ &   3172.17  &  3174.28 \\
$A_2$ &   3172.61  &  3175.82 \\
$B_2$ &   3176.29  &  3179.58 \\
$A_2$ &   \textit{3180.39}  &  \textit{\textbf{3190.21}} \\
$B_1$ &   \textit{3183.15}  &  \textit{3186.27} \\
$A_2$ &   3191.97  &  3195.37 \\
$B_1$ &   3193.72  &  3197.44 \\
$A_1$ &   3202.65  &  3206.44 \\
$B_1$ &   3202.99  &  3206.78 \\
$A_1$ &   3207.67  &  3210.84 \\
$A_1$ &   \textit{3217.88}  & ??? $^a$ \\        
$B_1$ &   3222.57  &  3223.58 \\
$A_2$ &   3225.77  &  3229.36 \\
$B_2$ &   3232.42  &  3235.64 \\
\hline    
\end{longtable}

\end{center}
Table II. Inverted or permuted wave numbers are italicized, the  VMFCI(11300) numbers which deviate by more than $5~\cm$ are in bold. Some wave numbers are missing in the VMFCI(11300) list and are marked by the symbol $???$. This is  because they appear higher in the spectrum.  As a matter of fact, the quality of the VMFCI(11300) calculation deteriorate beyond $3110~\cm$ where many bold and italicized numbers appear.  Some footnotes in this tables are related to those in Tab.2 of Ref.~\citenum{Thomas2017}.\\

\noindent
$^a$ assigned to $2\nu_5 + 2\nu_{12}$ in the new calculation, missing in the old one.\\
$^b$ this wave number is assigned to $\nu_{12} + \nu_8 + \nu_{15} $ in the new calculation and to $\nu_{12} + \nu_7 + \nu_{14} $  in the old one.\\


\section{Harmonic vibrational analyses\label{appendix1}}

The cubic and quartic force field constants, as well as the first and second electric dipole moment derivatives  that our AnharmoniCaOs code uses were obtained via DFT calculations with the B97-1 exchange-correlation functional with the TZ2P and \mbox{6-31G*} Gaussian basis sets, as described in the text. Since we used numerical differentiation along normal coordinates, we give here the results of the harmonic vibrational analyses for both Pyrene (Table~\ref{tab:harmpyrene}) and Coronene (Table~\ref{tab:harmcoronene}). 

\setlength\LTcapwidth{\hsize}
\begin{longtable}{@{\extracolsep{\fill}}cccccc}
\caption{List of harmonic vibrational normal modes of Pyrene. For each mode numbered according to spectroscopic conventions, we list  the irreducible representation label ("sym." column), the wave number in cm$^{-1}$, and the intensity of the corresponding fundamental IR transition in \km.}\label{tab:harmpyrene} \\[2mm]

\hline \\[.5mm]
& & \multicolumn{2}{c}{B97-1/6-31g$^*$} & \multicolumn{2}{c}{B97-1/TZ2P} \\
mode & sym. & freq. & int. & freq. & int. \\[.5mm]
\hline \\[1mm]
\endfirsthead

\multicolumn{6}{c}{\tablename\ \thetable{} -- continued from previous page} \\[2mm]
\hline \\[.5mm]
& & \multicolumn{2}{c}{B97-1/6-31g$^*$} & \multicolumn{2}{c}{B97-1/TZ2P} \\
mode & sym. & freq. & int. & freq. & int. \\[.5mm]
\hline \\[1mm]
\endhead

\\[1mm] \hline \\[1mm] \multicolumn{6}{r}{{Continued on next page}} \\[1mm] \hline
\endfoot

\\[1mm] \hline
\endlastfoot

  $\nu_{1}$ &	  A$_g$	& 3202.4	& 0.0	&   3182.2 &	    0.0 \\
  $\nu_{2}$ &	  A$_g$	& 3192.4	& 0.0	&   3172.8 &	    0.0 \\
  $\nu_{3}$ &	  A$_g$	& 3175.9	& 0.0	&   3157.5 &	    0.0 \\
  $\nu_{4}$ &	  A$_g$	& 1678.5	& 0.0	&   1661.0 &	    0.0 \\
  $\nu_{5}$ &	  A$_g$	& 1604.9	& 0.0	&   1587.8 &	    0.0 \\
  $\nu_{6}$ &	  A$_g$	& 1435.6	& 0.0	&   1420.8 &	    0.0 \\
  $\nu_{7}$ &	  A$_g$	& 1356.2	& 0.0	&   1347.3 &	    0.0 \\
  $\nu_{8}$ &	  A$_g$	& 1270.5	& 0.0	&   1256.9 &	    0.0 \\
  $\nu_{9}$ &	  A$_g$	& 1172.1	& 0.0	&   1164.9 &	    0.0 \\
 $\nu_{10}$ &	  A$_g$	& 1094.5	& 0.0	&   1085.6 &	    0.0 \\
 $\nu_{11}$ &	  A$_g$	& 809.8		& 0.0	&    811.3 &	    0.0 \\
 $\nu_{12}$ &	  A$_g$	& 594.6		& 0.0	&    593.2 &	    0.0 \\
 $\nu_{13}$ &	  A$_g$	& 409.6		& 0.0	&    409.6 &	    0.0 \\
 $\nu_{14}$ &	  A$_u$	& 969.0		& 0.0 	&    985.1 &	    0.0 \\
 $\nu_{15}$ &	  A$_u$	& 896.2		& 0.0	&    908.2 &	    0.0 \\
 $\nu_{16}$ &	  A$_u$	& 685.5		& 0.0	&    689.7 &	    0.0 \\
 $\nu_{17}$ &	  A$_u$	& 400.9		& 0.0	&    399.9 &	    0.0 \\
 $\nu_{18}$ &	  A$_u$	& 153.1		& 0.0	&    149.9 &	    0.0 \\
 $\nu_{19}$ & B$_{1g}$	& 911.8		& 0.0	&    920.6 &	    0.0 \\
 $\nu_{20}$ & B$_{1g}$	& 816.5		& 0.0	&    815.4 &	    0.0 \\
 $\nu_{21}$ & B$_{1g}$	& 538.5		& 0.0	&	 533.9 &	    0.0 \\
 $\nu_{22}$ & B$_{1g}$	& 250.3		& 0.0	&	 246.9 &	    0.0 \\
 $\nu_{23}$ & B$_{1u}$	& 3202.0	& 71.116 &	3182.0 &	   38.0 \\
 $\nu_{24}$ & B$_{1u}$	& 3176.1	& 0.368	&   3157.8 &	 0.0018 \\
 $\nu_{25}$ & B$_{1u}$	& 3172.9	& 2.711	&   3154.4 &	    1.2 \\
 $\nu_{26}$ & B$_{1u}$	& 1643.2	& 12.622 &  1626.5 &	   13.2 \\
 $\nu_{27}$ & B$_{1u}$	& 1490.5	& 0.320 &   1478.9 &	    1.0 \\
 $\nu_{28}$ & B$_{1u}$	& 1466.4	& 11.194 &  1456.5 &	    8.0 \\
 $\nu_{29}$ & B$_{1u}$	& 1271.2	& 2.776  &  1264.3 &	    2.3 \\
 $\nu_{30}$ & B$_{1u}$	& 1117.8	& 5.675 &   1109.2 &	    7.0 \\
 $\nu_{31}$ & B$_{1u}$	& 1008.1	& 0.830 &   1011.5 &	    1.8 \\
 $\nu_{32}$ & B$_{1u}$	& 830.2		& 3.477 &    830.3 &	    4.6 \\
 $\nu_{33}$ & B$_{1u}$	& 700.6		& 0.151 &    700.8 &	    0.029 \\
 $\nu_{34}$ & B$_{1u}$	& 502.6		& 2.417 &    503.7 &	    3.0 \\
 $\nu_{35}$ & B$_{2g}$	& 980.0		& 0.0	&	 990.7 &	    0.0 \\
 $\nu_{36}$ & B$_{2g}$	& 960.8		& 0.0	&	 977.0 &	    0.0 \\
 $\nu_{37}$ & B$_{2g}$	& 837.2		& 0.0 	&	 859.2 &	    0.0 \\
 $\nu_{38}$ & B$_{2g}$	& 784.5		& 0.0	&	 783.6 &	    0.0 \\
 $\nu_{39}$ & B$_{2g}$	& 583.0		& 0.0	&	 586.5 &	    0.0 \\
 $\nu_{40}$ & B$_{2g}$	& 513.0		& 0.0	&	 511.7 &	    0.0 \\
 $\nu_{41}$ & B$_{2g}$	& 264.2		& 0.0	&	 259.7 &	    0.0 \\
 $\nu_{42}$ & B$_{2u}$	& 3192.1	& 69.930 &	3172.7 &	   42.8 \\
 $\nu_{43}$ & B$_{2u}$	& 3183.0	& 22.516 &  3165.0 &	   12.7 \\
 $\nu_{44}$ & B$_{2u}$	& 1653.6	& 4.393	&   1635.8 &	    2.3 \\
 $\nu_{45}$ & B$_{2u}$	& 1525.0	& 2.782 &   1508.5 &	    3.2 \\
 $\nu_{46}$ & B$_{2u}$	& 1466.2	& 2.046 &   1453.4 &	    3.2 \\
 $\nu_{47}$ & B$_{2u}$	& 1356.3	& 5.519 &   1338.3 &	    4.9 \\
 $\nu_{48}$ & B$_{2u}$	& 1236.1	& 0.026 &   1226.2 &	    0.020 \\
 $\nu_{49}$ & B$_{2u}$	& 1210.9	& 13.525 &  1202.2 &	   14.0 \\
 $\nu_{50}$ & B$_{2u}$	& 1174.7	& 0.418	&   1160.4 &	    0.28 \\
 $\nu_{51}$ & B$_{2u}$	& 979.3		& 0.106 &    975.4 &	    0.0025 \\
 $\nu_{52}$ & B$_{2u}$	& 549.2		& 2.649 &    549.2 &	    2.6 \\
 $\nu_{53}$ & B$_{2u}$	& 355.8		& 1.327	&    356.5 &	    1.7 \\
 $\nu_{54}$ & B$_{3g}$	& 3183.4	& 0.0	&   3165.5 &	    0.0 \\
 $\nu_{55}$ & B$_{3g}$	& 3173.1	& 0.0	&   3154.6 &	    0.0 \\
 $\nu_{56}$ & B$_{3g}$	& 1632.6	& 0.0	&   1618.3 &	    0.0 \\
 $\nu_{57}$ & B$_{3g}$	& 1543.3	& 0.0	&   1528.9 &	    0.0 \\
 $\nu_{58}$ & B$_{3g}$	& 1443.4	& 0.0	&   1433.0 &	    0.0 \\
 $\nu_{59}$ & B$_{3g}$	& 1411.9	& 0.0	&   1393.9 &	    0.0 \\
 $\nu_{60}$ & B$_{3g}$	& 1264.6	& 0.0	&   1260.7 &	    0.0 \\
 $\nu_{61}$ & B$_{3g}$	& 1203.0	& 0.0	&   1194.9 &	    0.0 \\
 $\nu_{62}$ & B$_{3g}$	& 1125.4	& 0.0	&   1122.9 &	    0.0 \\
 $\nu_{63}$ & B$_{3g}$	& 744.4		& 0.0	&    745.2 &	    0.0 \\
 $\nu_{64}$ & B$_{3g}$	& 503.5		& 0.0	&    504.3 &	    0.0 \\
 $\nu_{65}$ & B$_{3g}$	& 459.2		& 0.0	&    459.7 &	    0.0 \\
 $\nu_{66}$ & B$_{3u}$	& 968.6		& 1.753	&    983.8 &	    1.4 \\
 $\nu_{67}$ & B$_{3u}$	& 859.3		& 109.082 &  860.0 &	  111.9 \\
 $\nu_{68}$ & B$_{3u}$	& 759.5		& 11.473 &   755.1 &	   17.5 \\
 $\nu_{69}$ & B$_{3u}$	& 722.5		& 25.863 &   725.6 &	   44.8 \\
 $\nu_{70}$ & B$_{3u}$	& 497.2		& 1.318	&	 496.3 &	    2.1 \\
 $\nu_{71}$ & B$_{3u}$	& 216.2		& 6.968	&    210.2 &	   10.6 \\
 $\nu_{72}$ & B$_{3u}$	& 100.2		& 0.409	&     97.3 &	   0.64 \\
\end{longtable}
\begin{longtable}{@{\extracolsep{\fill}}cccccc}
\caption{List of harmonic vibrational normal modes of coronene. For each mode numbered according to spectroscopic conventions, we list  the irreducible representation label ("sym." column), the wave number in cm$^{-1}$, and the intensity of the corresponding fundamental IR transition in \km}\label{tab:harmcoronene} \\
\hline \\[.5mm]
& & \multicolumn{2}{c}{B97-1/6-31g$^*$} & \multicolumn{2}{c}{B97-1/TZ2P} \\
mode & sym. & freq. & int. & freq. & int. \\[.5mm]
\hline \\[1mm]
\endfirsthead

\multicolumn{6}{c}{\tablename\ \thetable{} -- continued from previous page} \\[2mm]
\hline \\[.5mm]
& & \multicolumn{2}{c}{B97-1/6-31g$^*$} & \multicolumn{2}{c}{B97-1/TZ2P} \\
mode & sym. & freq. & int. & freq. & int. \\[.5mm]
\hline \\[1mm]
\endhead

\\[1mm] \hline \\[1mm] \multicolumn{6}{r}{{Continued on next page}} \\[1mm] \hline
\endfoot

\\[1mm] \hline
\endlastfoot

$\nu_{1}$       & A$_{1g}$      & 3194.3        & 0.0 	& 3174.4	& 0.0 \\
$\nu_{2}$       & A$_{1g}$      & 1646.4        & 0.0 	& 1629.3	& 0.0 \\
$\nu_{3}$       & A$_{1g}$      & 1385.1        & 0.0	& 1366.1	& 0.0 \\
$\nu_{4}$       & A$_{1g}$      & 1250.5        & 0.0	& 1240.7	& 0.0 \\
$\nu_{5}$       & A$_{1g}$      & 1048.4        & 0.0	& 1043.5	& 0.0 \\
$\nu_{6}$       & A$_{1g}$      &  483.9        & 0.0	& 483.0		& 0.0 \\
$\nu_{7}$       & A$_{1u}$      &  946.8        & 0.0	& 964.2		& 0.0 \\
$\nu_{8}$       & A$_{1u}$      &  526.4        & 0.0	& 530.4		& 0.0 \\
$\nu_{9}$       & A$_{2g}$      & 3172.8        & 0.0 	& 3154.1	& 0.0 \\
$\nu_{10}$      & A$_{2g}$      & 1580.0        & 0.0	& 1565.2	& 0.0 \\
$\nu_{11}$      & A$_{2g}$      & 1257.7        & 0.0	& 1254.9	& 0.0 \\
$\nu_{12}$      & A$_{2g}$      &  931.2        & 0.0	& 941.2		& 0.0 \\
$\nu_{13}$      & A$_{2g}$      &  637.5        & 0.0	& 640.1		& 0.0 \\
$\nu_{14}$      & A$_{2u}$      &  873.5        & 158.034	& 873.2	& 167.908 \\
$\nu_{15}$      & A$_{2u}$      &  562.7        & 24.821	& 558.4	& 42.749 \\
$\nu_{16}$      & A$_{2u}$      &  128.0        & 4.784	& 123.4		& 6.911 \\
$\nu_{17}$      & B$_{1g}$      &  775.6        & 0.0	& 773.6		& 0.0 \\
$\nu_{18}$      & B$_{1g}$      &  166.1        & 0.0	& 162.9		& 0.0 \\
$\nu_{19}$      & B$_{1u}$      & 3175.4        & 0.0	& 3156.7	& 0.0 \\
$\nu_{20}$      & B$_{1u}$      & 1585.9        & 0.0	& 1577.7	& 0.0 \\
$\nu_{21}$      & B$_{1u}$      & 1453.2        & 0.0	& 1443.1	& 0.0 \\
$\nu_{22}$      & B$_{1u}$      & 1175.1        & 0.0	& 1181.5	& 0.0 \\
$\nu_{23}$      & B$_{1u}$      &  678.2        & 0.0	& 677.2		& 0.0 \\
$\nu_{24}$      & B$_{1u}$      &  560.0        & 0.0	& 563.3		& 0.0 \\
$\nu_{25}$      & B$_{2g}$      &  972.3        & 0.0	& 985.4		& 0.0 \\
$\nu_{26}$      & B$_{2g}$      &  789.7        & 0.0	& 869.0		& 0.0 \\
$\nu_{27}$      & B$_{2g}$      &  635.7        & 0.0	& 656.5		& 0.0 \\
$\nu_{28}$      & B$_{2g}$      &  227.8        & 0.0	& 224.1		& 0.0 \\
$\nu_{29}$      & B$_{2u}$      & 3190.7        & 0.0	& 3171.4	& 0.0 \\
$\nu_{30}$      & B$_{2u}$      & 1527.8        & 0.0	& 1511.2	& 0.0 \\
$\nu_{31}$      & B$_{2u}$      & 1381.8        & 0.0	& 1367.2	& 0.0 \\
$\nu_{32}$      & B$_{2u}$      & 1218.5        & 0.0	& 1205.0	& 0.0 \\
$\nu_{33}$      & B$_{2u}$      & 1168.5        & 0.0	& 1153.5	& 0.0 \\
$\nu_{34}$      & B$_{2u}$      &  476.9        & 0.0	& 477.0		& 0.0 \\
$\nu_{35}$      & E$_{1g}$      &  954.2        & 0.0	& 971.3		& 0.0 \\
$\nu_{36}$      & E$_{1g}$      &  954.2        & 0.0	& 971.3		& 0.0 \\
$\nu_{37}$      & E$_{1g}$      &  852.9        & 0.0	& 854.8		& 0.0 \\
$\nu_{38}$      & E$_{1g}$      &  852.9        & 0.0	& 854.8		& 0.0 \\
$\nu_{39}$      & E$_{1g}$      &  668.5        & 0.0	& 672.8		& 0.0 \\  
$\nu_{40}$      & E$_{1g}$      &  668.5        & 0.0	& 672.8		& 0.0 \\
$\nu_{41}$      & E$_{1g}$      &  454.3        & 0.0	& 454.3		& 0.0 \\
$\nu_{42}$      & E$_{1g}$      &  454.3        & 0.0	& 454.3		& 0.0 \\
$\nu_{43}$      & E$_{1g}$      &  300.1        & 0.0	& 292.8		& 0.0 \\
$\nu_{44}$      & E$_{1g}$      &  300.1        & 0.0	& 292.8		& 0.0 \\
$\nu_{45}$      & E$_{1u}$      & 3193.0        & 107.828	& 3173.4	& 63.965 \\
$\nu_{46}$      & E$_{1u}$      & 3193.0        & 107.849	& 3173.4	& 63.973 \\
$\nu_{47}$      & E$_{1u}$      & 3173.6        & 6.729	& 3154.9	& 3.395 \\
$\nu_{48}$      & E$_{1u}$      & 3173.6        & 6.728	& 3154.9	& 3.406 \\
$\nu_{49}$      & E$_{1u}$      & 1660.5        & 12.129	& 1643.9	& 7.615 \\
$\nu_{50}$      & E$_{1u}$      & 1660.5        & 12.136	& 1643.9	& 7.623 \\
$\nu_{51}$      & E$_{1u}$      & 1540.8        & 1.630	& 1526.2	& 3.024 \\
$\nu_{52}$      & E$_{1u}$      & 1540.8        & 1.630	& 1526.2	& 3.026 \\
$\nu_{53}$      & E$_{1u}$      & 1435.1        & 0.815	& 1419.0	& 0.387 \\
$\nu_{54}$      & E$_{1u}$      & 1435.1        & 0.815	& 1419.0	& 0.392 \\
$\nu_{55}$      & E$_{1u}$      & 1339.0        & 23.446	& 1333.4	& 21.148\\
$\nu_{56}$      & E$_{1u}$      & 1339.0        & 23.448	& 1333.4	& 21.126 \\
$\nu_{57}$      & E$_{1u}$      & 1241.4        & 1.029	& 1232.7	& 1.255 \\
$\nu_{58}$      & E$_{1u}$      & 1241.4        & 1.028	& 1232.7	& 1.254 \\
$\nu_{59}$      & E$_{1u}$      & 1161.6        & 7.599	& 1155.0	& 8.633 \\
$\nu_{60}$      & E$_{1u}$      & 1161.6        & 7.603	& 1155.0	& 8.628 \\
$\nu_{61}$      & E$_{1u}$      &  821.0        & 0.043	& 819.7	& 0.0014 \\
$\nu_{62}$      & E$_{1u}$      &  821.0        & 0.043	& 819.7	& 0.0012 \\
$\nu_{63}$      & E$_{1u}$      &  779.4        & 6.525	& 780.6	& 6.808 \\
$\nu_{64}$      & E$_{1u}$      &  779.4        & 6.525	& 780.6	& 6.797 \\
$\nu_{65}$      & E$_{1u}$      &  382.3        & 3.153	& 382.6	& 3.872 \\
$\nu_{66}$      & E$_{1u}$      &  382.3        & 3.153	& 382.6	& 3.867 \\
$\nu_{67}$      & E$_{2g}$      & 3191.5        & 0.0	& 3172.1	& 0.0 \\
$\nu_{68}$      & E$_{2g}$      & 3191.5        & 0.0	& 3172.1	& 0.0 \\
$\nu_{69}$      & E$_{2g}$      & 3174.8        & 0.0	& 3156.2	& 0.0 \\
$\nu_{70}$      & E$_{2g}$      & 3174.8        & 0.0	& 3156.2	& 0.0 \\
$\nu_{71}$      & E$_{2g}$      & 1660.0        & 0.0	& 1645.1	& 0.0 \\
$\nu_{72}$      & E$_{2g}$      & 1660.0        & 0.0	& 1645.1	& 0.0 \\
$\nu_{73}$      & E$_{2g}$      & 1487.9        & 0.0	& 1475.9	& 0.0 \\
$\nu_{74}$      & E$_{2g}$      & 1487.9        & 0.0	& 1475.9	& 0.0 \\
$\nu_{75}$      & E$_{2g}$      & 1470.6        & 0.0	& 1457.9	& 0.0 \\
$\nu_{76}$      & E$_{2g}$      & 1470.6        & 0.0	& 1457.9	& 0.0 \\
$\nu_{77}$      & E$_{2g}$      & 1430.8        & 0.0	& 1418.6	& 0.0 \\
$\nu_{78}$      & E$_{2g}$      & 1430.8        & 0.0	& 1418.6	& 0.0 \\
$\nu_{79}$      & E$_{2g}$      & 1254.5        & 0.0	& 1242.3	& 0.0 \\
$\nu_{80}$      & E$_{2g}$      & 1254.5        & 0.0	& 1242.3	& 0.0 \\
$\nu_{81}$      & E$_{2g}$      & 1187.1        & 0.0	& 1177.3	& 0.0 \\
$\nu_{82}$      & E$_{2g}$      & 1187.1        & 0.0	& 1177.3	& 0.0 \\
$\nu_{83}$      & E$_{2g}$      & 1006.6        & 0.0	& 1004.8	& 0.0 \\
$\nu_{84}$      & E$_{2g}$      & 1006.6        & 0.0	& 1004.8	& 0.0 \\
$\nu_{85}$      & E$_{2g}$      &  687.2        & 0.0	& 686.7		& 0.0 \\
$\nu_{86}$      & E$_{2g}$      &  687.2        & 0.0	& 686.7		& 0.0 \\
$\nu_{87}$      & E$_{2g}$      &  492.3        & 0.0	& 493.3		& 0.0 \\
$\nu_{88}$      & E$_{2g}$      &  492.3        & 0.0	& 493.3		& 0.0 \\
$\nu_{89}$      & E$_{2g}$      &  367.9        & 0.0	& 368.0		& 0.0 \\
$\nu_{90}$      & E$_{2g}$      &  367.9        & 0.0	& 368.0		& 0.0 \\
$\nu_{91}$      & E$_{2u}$      &  965.3        & 0.0	& 980.7		& 0.0 \\
$\nu_{92}$      & E$_{2u}$      &  965.3        & 0.0	& 980.7		& 0.0 \\
$\nu_{93}$      & E$_{2u}$      &  814.8        & 0.0	& 829.2		& 0.0 \\
$\nu_{94}$      & E$_{2u}$      &  814.8        & 0.0	& 829.2		& 0.0 \\
$\nu_{95}$      & E$_{2u}$      &  770.2        & 0.0	& 780.6		& 0.0 \\
$\nu_{96}$      & E$_{2u}$      &  770.2        & 0.0	& 780.6		& 0.0 \\
$\nu_{97}$      & E$_{2u}$      &  552.9        & 0.0	& 551.6		& 0.0 \\
$\nu_{98}$      & E$_{2u}$      &  552.9        & 0.0	& 551.6		& 0.0 \\
$\nu_{99}$      & E$_{2u}$      &  302.9        & 0.0	& 299.2		& 0.0 \\
$\nu_{100}$     & E$_{2u}$      &  302.9        & 0.0	& 299.2		& 0.0 \\
$\nu_{101}$     & E$_{2u}$      &   89.1        & 0.0	& 86.7		& 0.0 \\
$\nu_{102}$     & E$_{2u}$      &   89.1        & 0.0	& 86.7		& 0.0 \\
\end{longtable}
It is well known that vibrational frequencies computed in the harmonic approximation can be brought in fairly good agreement with experimental values by scaling them with empirical scaling factors (see e.~g. Ref.~\citenum{Irikura2005}). The scaling factor(s) are typically obtained by fitting a large number of identified vibrational transitions, depend on the level of theory and basis set used, and may be different for different types of vibrations (i.e. stretches, bends, twists, the kinds of bonds involved etc.).\cite{Pauzat1995,Pauzat1997} Scaling factors for the B97-1 functional in combination with the TZ2P basis set were obtained by Can\'e, Miani, and Trombetti\cite{Cane2007} and confirmed by Maltseva \emph{et al.}\cite{Maltseva2015,Maltseva2016} with minimal variation. For the sake of comparison with anharmonic calculations we here used the original ones taken from Ref.~\citenum{Cane2007}, namely 0.9660 for C-H stretches and 0.982 for all the other bands. Given the very small sample of bands these factors were calibrated on, their general validity is not adequately tested, and they should be used with caution.

\section{Anharmonic bands of pyrene as a function of $r$\label{appendix2}}

The large set of bands listed in Table~\ref{tab:pyrtestthresnocor} results from anharmonic calculations with different $r$ values,  keeping fixed $h=8$. State assignment was done, as far as this table is concerned, on the calculation with $r=0.3$, where it is the least ambiguous. In most eigenstates a single harmonic basis function is dominant. In the few cases where several harmonic basis functions have weights in the same range, the latter are all listed. Only transitions with non-negligible intensity are included in the table. When decreasing the $r$-value, some bands split in several ones due to resonances. Since this splitting may involve significant mixing with bands which are distinct at the $r=0.3$ level, sometimes the same wave number appears several times in columns corresponding to smaller $r$ values. This happens, for example, for the bands which are computed at 3034, 3037, and 3042 cm$^{-1}$ at the $r=0.3$ level. These give rise to a multitude of bands at smaller $r$ values, of which only the strongest are listed. Note also that, in some cases, distinct bands are almost coincident in position. However, they are listed on different lines in the  table. 
\setlength{\LTcapwidth}{2.0\hsize}
\begin{longtable*}{@{\extracolsep{\fill}}p{2cm}p{1cm}p{4.5cm}p{4.5cm}p{4.5cm}}
\caption{Calculated anharmonic band positions for pyrene as a function of the threshold $r$ and for fixed $h=8$.}\label{tab:pyrtestthresnocor}\\[2mm]
\hline\\
mode & $r$=0.3 & $r$=0.2 & $r$=0.1 & $r$=0.05 \\[.5mm]
\hline\\[.5mm]
\endfirsthead

\multicolumn{5}{c}{\tablename\ \thetable{} -- continued from previous page} \\[2mm]
\hline\\[.5mm]
mode & $r$=0.3 & $r$=0.2 & $r$=0.1 & $r$=0.05 \\[.5mm]
\hline\\[.5mm]
\endhead

\\[.5mm] \hline \\[1mm] \multicolumn{5}{r}{{Continued on next page}} \\[.5mm] \hline
\endfoot

\\[.5mm] \hline
\endlastfoot

$\nu_{44} + \nu_{48} + \nu_{53}$, $\nu_5 + \nu_{44}$	& 3144	& 3138, 3144, 3153	& 3134, 3144, 3146, 3154	& 3145, 3156 \\
$\nu_{44} + \nu_{57}$	& 3107	& 3105, 3106	& 3103, 3104	& 3098, 3102 \\
$\nu_{10} + \nu_{52} + \nu_{57}$	& 3102	& 3100	& 3099	& 3098, 3102 \\
$\nu_{4} + \nu_{45}$	& 3097	& 3097	& 3089, 3098	& 3092, 3098, 3102 \\
$\nu_{34} + \nu_{57} + \nu_{62}$	& 3096	& 3095	& 3092, 3093	& 3083, 3092, 3095 \\
$\nu_{26} + \nu_{57}$	& 3080	& 3079	& 3080	& 3062, 3073, 3078 \\
$\nu_{23}$	& 3055	& 3039, 3062, 3064, 3069	& 3010, 3029, 3035, 3069, 3104	& 3019, 3026, 3030, 3067, 3076 \\
$\nu_{4} + \nu_{28}$	& 3051	& 3023, 3030, 3036	& 3029, 3045, 3055, 3058	& 3056 \\
$\nu_{45} + \nu_{56}$	& 3043	& 3029, 3039, 3062, 3064, 3069	& 3029, 3035, 3045, 3055, 3069	& 3030, 3043, 3067, 3076 \\
$\nu_{42}$, $\nu_{43}$	& 3042	& 3042	& 3019, 3029, 3037, 3053	& 3023, 3040  \\
$\nu_{4} + \nu_{46}$	& 3037	& 3038	& 3029, 3037	& 3032, 3033, 3040, 3042, 3044, 3045 \\
$\nu_{43}$, $\nu_{42}$	& 3034	& 3035	& 2946, 3029, 3053	& 2940, 3032, 3033, 3044, 3045 \\
$\nu_{24}$	& 3031	& 3030	& 2954, 2969, 3010, 3045, 3078, 3079	& 2956, 3018, 3019, 3026, 3076 \\
$\nu_{49} + \nu_{61}$	& 2359	& 2359	& 2359	& 2359 \\
$\nu_{14} + \nu_{35}$	& 1936	& 1936	& 1937	& 1937 \\
$\nu_{35} + \nu_{66}$	& 1932	& 1932	& 1932	& 1934 \\
$\nu_{14} + \nu_{36}$	& 1922	& 1923	& 1919, 1937	& 1921, 1937 \\
$\nu_{36} + \nu_{66}$	& 1919	& 1919	& 1918	& 1919 \\
$\nu_{14} + \nu_{19}$	& 1866	& 1866	& 1865	& 1866 \\
$\nu_{15} + \nu_{35}$	& 1852	& 1552	& 1850	& 1848 \\
$\nu_{15} + \nu_{36}$	& 1839	& 1839	& 1838	& 1842 \\
$\nu_{35} + \nu_{67}$	& 1824	& 1825	& 1825	& 1823 \\
$\nu_{36} + \nu_{67}$	& 1809	& 1810	& 1810	& 1766, 1801, 1823 \\
$\nu_{37} + \nu_{66}$	& 1803	& 1803	& 1801, 1803	& 1801, 1809, 1810 \\
$\nu_{15} + \nu_{19}$	& 1790	& 1788	& 1790	& 1792 \\
$\nu_{14} + \nu_{20}$	& 1758	& 1761	& 1768	& 1766 \\
$\nu_{19} + \nu_{67}$	& 1745	& 1745	& 1745	& 1745 \\
$\nu_{17} + \nu_{18} + \nu_{35} + \nu_{71}$	& 1742	& 1742	& 1746	& 1741 \\
$\nu_{38} + \nu_{66}$	& 1740	& 1745	& 1746	& 1743 \\
$\nu_{15} + \nu_{37}$	& 1727	& 1727	& 1726	& 1726 \\
$\nu_{36} + \nu_{68}$, $\nu_{35} + \nu_{68}$	& 1705	& 1700	& 1701	& 1659, 1669, 1697 \\
$\nu_{15} + \nu_{20}$	& 1692	& 1692	& 1694	& 1659, 1669, 1697 \\
$\nu_{35} + \nu_{68}$, $\nu_{35} + \nu_{69}$, $\nu_{38} + \nu_{66}$	& 1687	& 1692	& 1684	& 1659, 1697, 1766 \\
$\nu_{36} + \nu_{69}$	& 1669	& 1665, 1668	& 1661, 1701	& 1659, 1669, 1697 \\
$\nu_{16} + \nu_{35}$	& 1655	& 1656	& 1655	& 1651 \\
$\nu_{20} + \nu_{67}$	& 1646	& 1646	& 1645	& 1645 \\
$\nu_{16} + \nu_{36}$	& 1636	& 1637	& 1639	& 1636 \\
$\nu_{19} + \nu_{68}$	& 1617	& 1607, 1619	& 1606, 1618	& 1617 \\
$\nu_{38} + \nu_{67}$	& 1613	& 1613	& 1612	& 1612 \\
$\nu_{44}$	& 1597	& 1591, 1599	& 1592, 1598	& 1588, 1596, 1600 \\
$\nu_{53} + \nu_{60}$	& 1592	& 1592	& 1581, 1591	& 1590 \\
$\nu_{26}$	& 1585	& 1585	& 1586	& 1585 \\
$\nu_{16} + \nu_{19}$	& \multirow{2}{*}{1578}	& \multirow{2}{*}{1578}	& \multirow{2}{*}{1576, 1578}	& \multirow{2}{*}{1575, 1578, 1581, 1585, 1590} \\
$\nu_{17} + \nu_{53} + \nu_{67}$, $\nu_{53} + \nu_{60}$ \\
$\nu_{20} + \nu_{68}$	& 1544	& 1544	& 1544	& 1544 \\
$\nu_{39} + \nu_{66}$	& 1537	& 1540	& 1540	& 1537 \\
$\nu_{16} + \nu_{37}$	& 1521	& 1521	& 1521	& 1521 \\
$\nu_{20} + \nu_{69}$	& 1514	& 1514	& 1513	& 1510 \\
$\nu_{38} + \nu_{68}$	& 1509	& 1506	& 1509	& 1505 \\
$\nu_{21} + \nu_{66}$	& 1489	& 1489	& 1486	& 1485 \\
$\nu_{45}$	& 1475	& 1465, 1475	& 1461, 1463, 1474, 1486	& 1473 \\
$\nu_{14} + \nu_{40}$	& \multirow{2}{*}{1466}	& \multirow{2}{*}{1464, 1465}	& \multirow{2}{*}{1462, 1463}	& \multirow{2}{*}{1447, 1453, 1459, 1460} \\
$\nu_{51} + \nu_{64}$ \\
$\nu_{27}$	& 1444	& 1444	& 1442	& 1438, 1442, 1447 \\
$\nu_{28}$	& 1429	& 1417, 1425, 1429	& 1404, 1412, 1419, 1425, 1442	& 1424 \\
$\nu_{46}$	& \multirow{2}{*}{1419}	& \multirow{2}{*}{1420}	& \multirow{2}{*}{1414, 1418}	& \multirow{2}{*}{1412, 1417} \\
$\nu_{39} + \nu_{67}$ \\
$\nu_{51} + \nu_{65}$	& 1417	& 1411, 1417, 1429	& 1412, 1419, 1425, 1429	& 1409, 1418, 1424 \\
$\nu_{15} + \nu_{21}$	& 1412	& 1412	& 1410	& 1410 \\
$\nu_{47}$	& 1311	& 1311	& 1310	& 1311 \\
$\nu_{20} + \nu_{70}$	& 1286	& 1286	& 1286	& 1286 \\
$\nu_{21} + \nu_{68}$	& 1266	& 1266	& 1264	& 1264 \\
$\nu_{40} + \nu_{68}$	& 1243	& 1243	& 1242	& 1241 \\
$\nu_{29}$	& 1240	& 1240	& 1238	& 1236 \\
$\nu_{3} + \nu_{64}$	& 1190	& 1189	& 1189	& 1189 \\
$\nu_{49}$	& 1184	& 1184	& 1183	& 1179, 1180, 1181 \\
$\nu_{16} + \nu_{40}$	& 1181	& 1180	& 1180	& 1180, 1181 \\
$\nu_{17} + \nu_{38}$	& 1157	& 1157	& 1155	& 1153 \\
$\nu_{13} + \nu_{33}$	& 1096	& 1096	& 1097	& 1096 \\
$\nu_{30}$	& 1093	& 1093	& 1091	& 1089 \\
$\nu_{12} + \nu_{34}$	& 1084	& 1084	& 1081, 1082	& 1080, 1081 \\
$\nu_{53} + \nu_{63}$	& 1083	& 1083	& 1081, 1082	& 1080, 1081 \\
$\nu_{39} + \nu_{70}$	& 1059	& 1059	& 1058	& 1058 \\
$\nu_{52} + \nu_{65}$, $\nu_{31}$	& 999	& 997, 999	& 994, 995, 997	& 993, 996 \\
$\nu_{31}$, $\nu_{52} + \nu_{65}$	& 997	& 995, 997, 999	& 986, 994, 997	& 985, 993, 996 \\
$\nu_{66}$	& 962	& 962	& 962	& 962 \\
$\nu_{67}$	& 845	& 845	& 844	& 843 \\
$\nu_{32}$	& 820	& 820	& 817	& 816 \\
$\nu_{68}$	& 742	& 742	& 741	& 739 \\
$\nu_{69}$	& 712	& 712	& 711	& 710 \\
$\nu_{52}$	& 542	& 542	& 540	& 538 \\
$\nu_{34}$	& 497	& 497	& 495	& 493 \\
$\nu_{70}$	& 486	& 486	& 484	& 483 \\
$\nu_{53}$	& 349	& 349	& 349	& 342, 347 \\
$\nu_{71}$	& 202	& 202	& 200	& 199 \\
$\nu_{72}$	& 95	& 95	& 94	& 92 \\
\end{longtable*}

\section{Anharmonic spectra of Pyrene and Coronene} \label{sec:anharmspectra}

In this section we provide, for reference and easier comparison with experimental and/or other theoretical spectra, complete plots of our best computed anharmonic spectra for Pyrene (from Fig.~\ref{plotpyr1} to \ref{plotpyr13}) and Coronene (from Fig.~\ref{plotcor1} to \ref{plotcor14}). Each figure covers a range of 300~cm$^{-1}$, and consecutive plots have an overlap of 50~cm$^{-1}$. The overall spectrum, in red, was obtained by convolving individual bands with a Lorentzian profile with a 1 cm$^{-1}$ width. Intensities of the convolved spectra are absolute and expressed in km~mol$^{-1}$cm, the standard units of quantum chemistry calculations. The precise positions of individual bands are marked in the spectra by black bars. The complete, anharmonic, tabulated spectra of Pyrene and Coronene can be found in the online PAH database by Malloci, Joblin, and Mulas.\cite{Malloci2007b}


\begin{figure}
\includegraphics[width=\hsize]{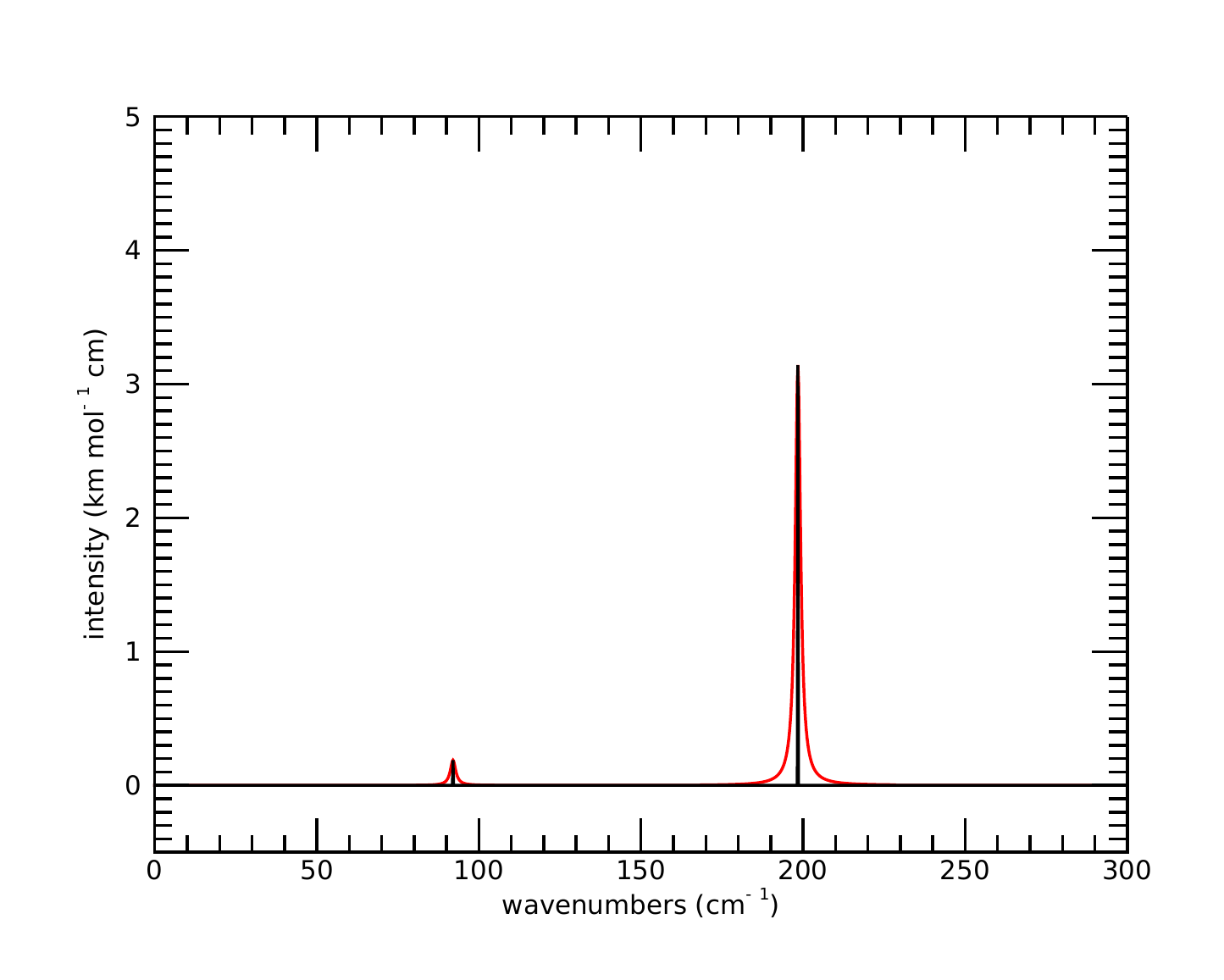}
\caption{Anharmonic spectrum of Pyrene computed with $r=0.05$ and $h=8$ in the range from 0 to 300 cm$^{-1}$. Black bars indicate the precise position of individual bands, whereas red envelopes are convolved with Lorentzian profiles with a 1 cm$^{-1}$ width.}\label{plotpyr1}
\end{figure}
\begin{figure}
\includegraphics[width=\hsize]{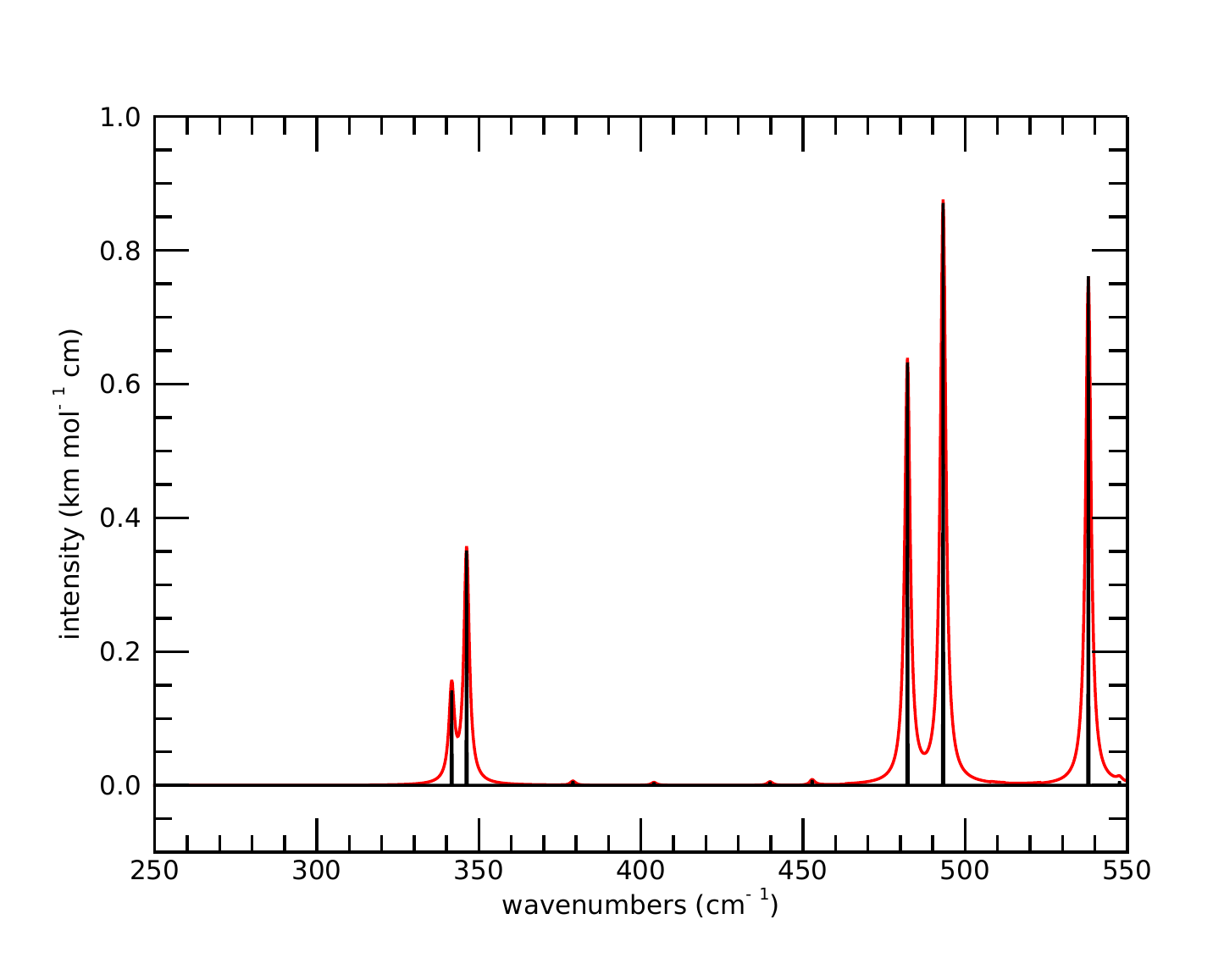}
\caption{Same as Fig.~\ref{plotpyr1} in the range from 250 to 550 cm$^{-1}$.}\label{plotpyr2}
\end{figure}
\begin{figure}
\includegraphics[width=\hsize]{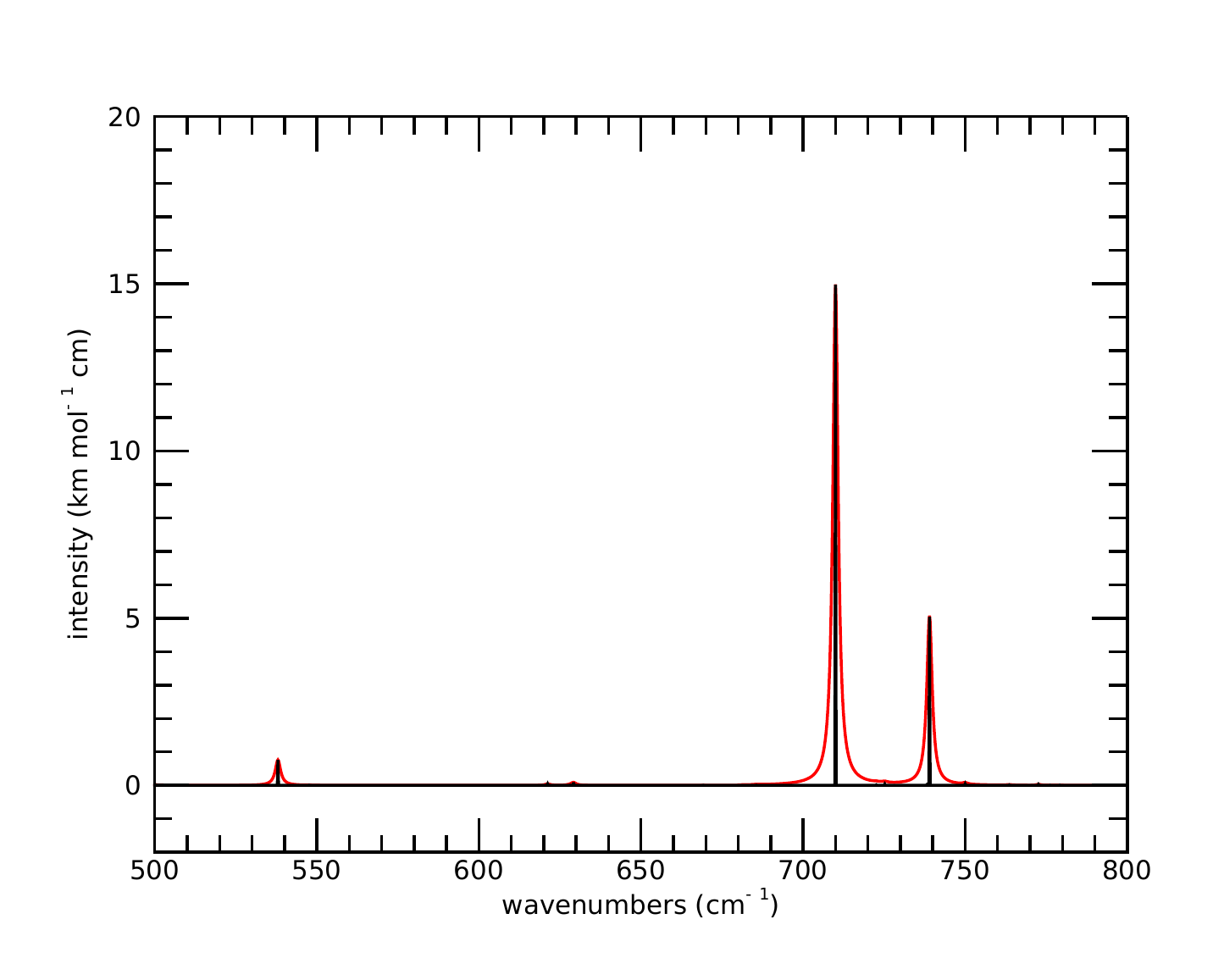}
\caption{Same as Fig.~\ref{plotpyr1} in the range from 500 to 800 cm$^{-1}$.}\label{plotpyr3}
\end{figure}
\begin{figure}
\includegraphics[width=\hsize]{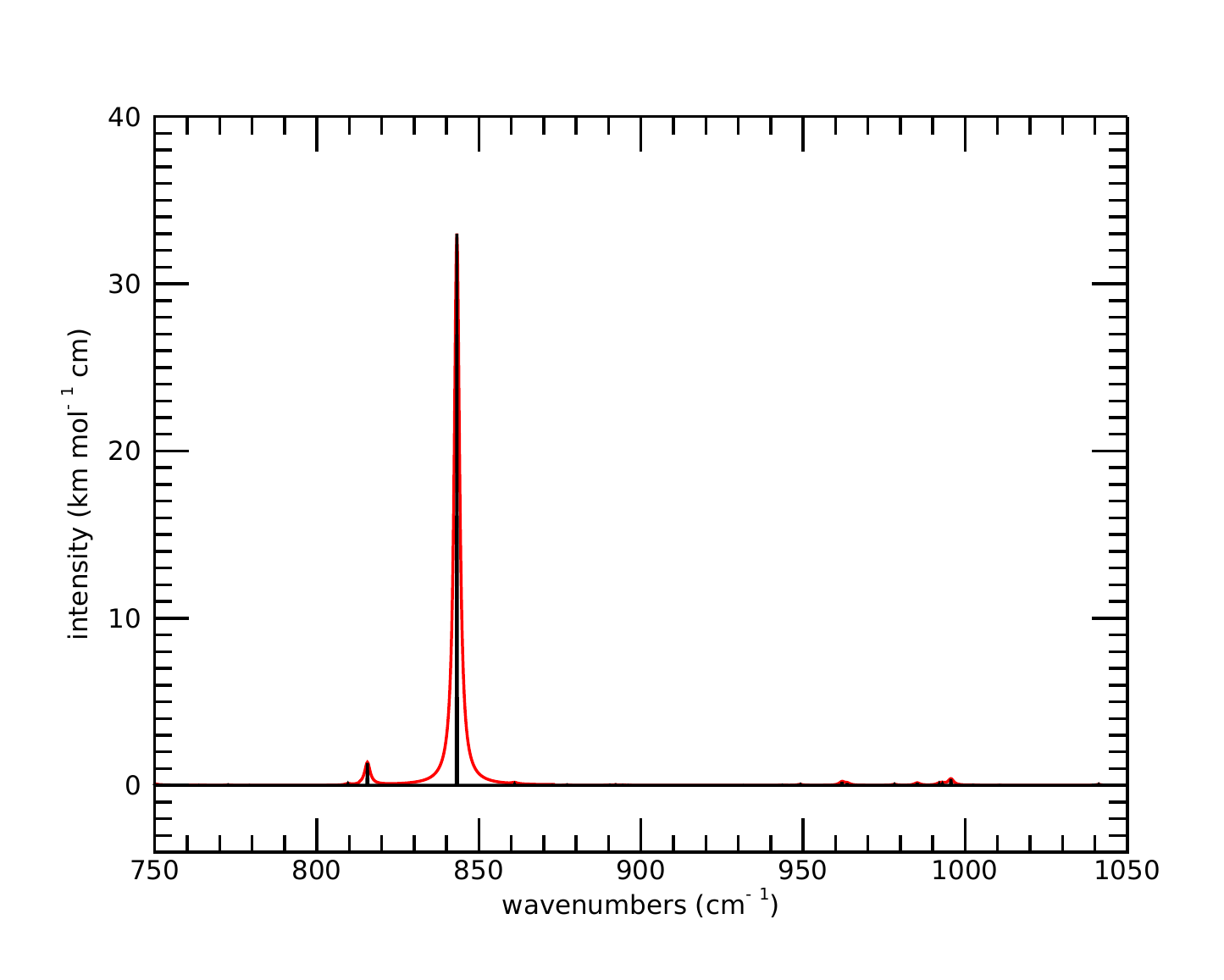}
\caption{Same as Fig.~\ref{plotpyr1} in the range from 750 to 1050 cm$^{-1}$.}\label{plotpyr4}
\end{figure}
\begin{figure}
\includegraphics[width=\hsize]{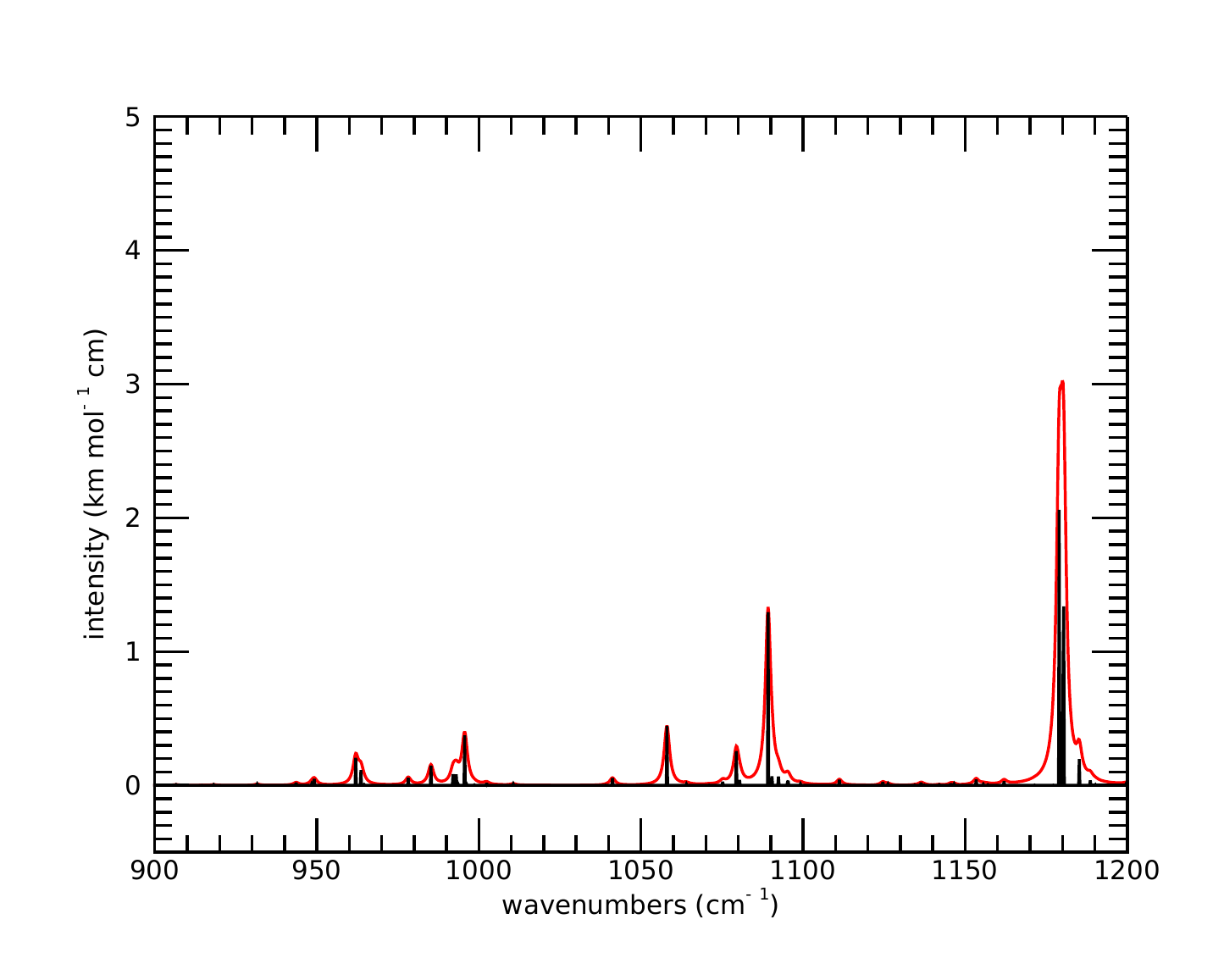}
\caption{Same as Fig.~\ref{plotpyr1} in the range from 900 to 1200 cm$^{-1}$.}\label{plotpyr5}
\end{figure}
\begin{figure}
\includegraphics[width=\hsize]{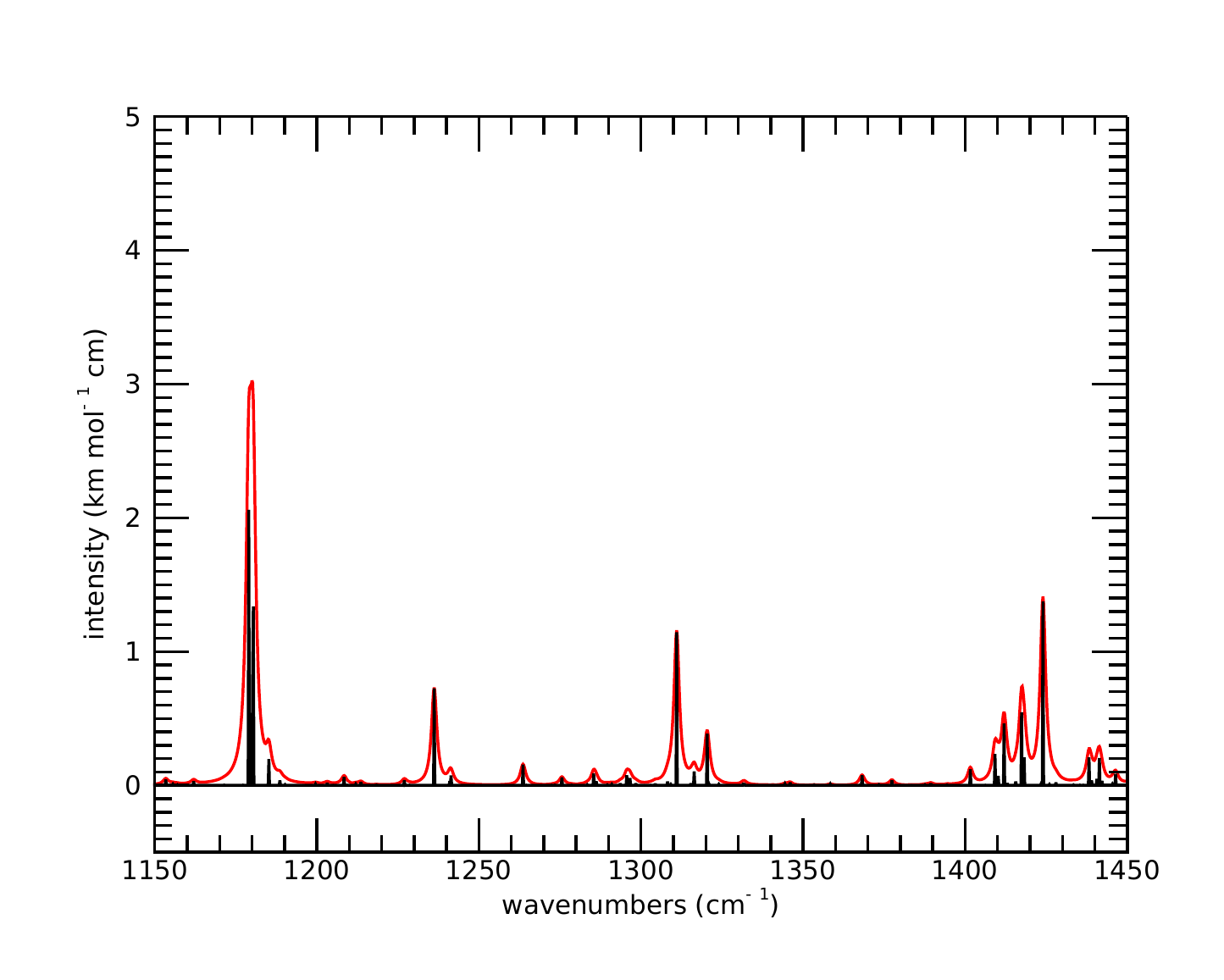}
\caption{Same as Fig.~\ref{plotpyr1} in the range from 1150 to 1450 cm$^{-1}$.}\label{plotpyr6}
\end{figure}
\begin{figure}
\includegraphics[width=\hsize]{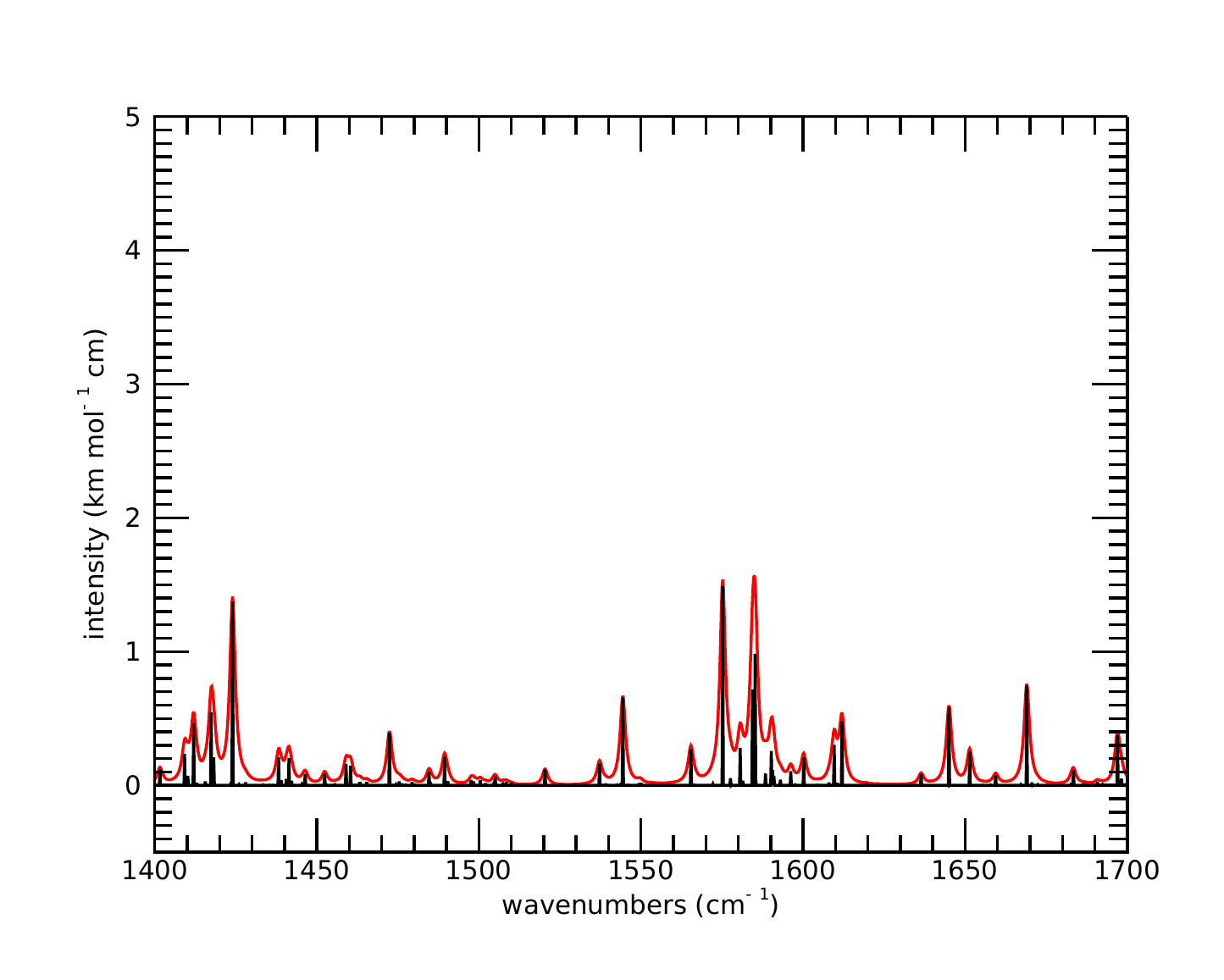}
\caption{Same as Fig.~\ref{plotpyr1} in the range from 1400 to 1700 cm$^{-1}$.}\label{plotpyr7}
\end{figure}
\begin{figure}
\includegraphics[width=\hsize]{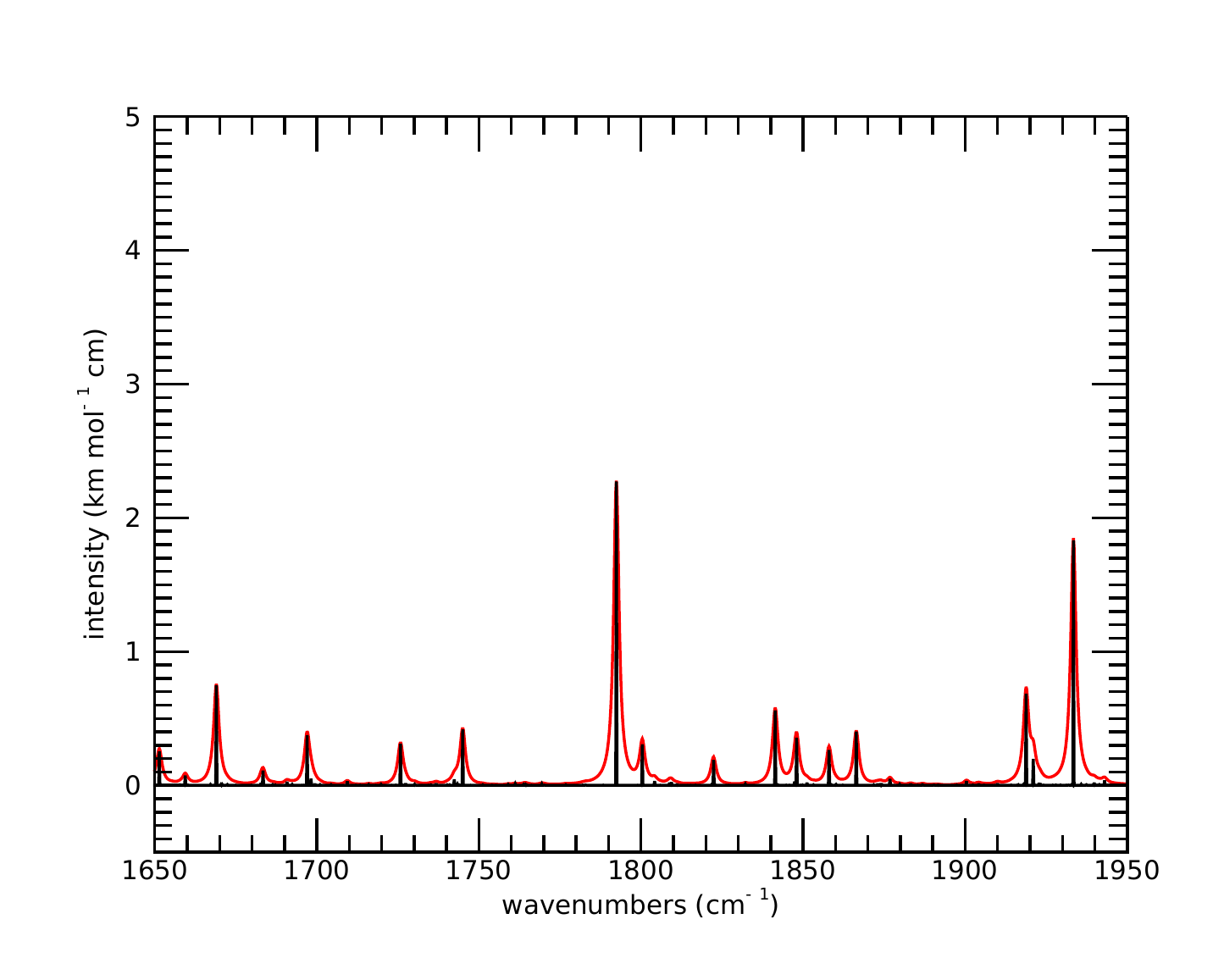}
\caption{Same as Fig.~\ref{plotpyr1} in the range from 1650 to 1950 cm$^{-1}$.}\label{plotpyr8}
\end{figure}
\begin{figure}
\includegraphics[width=\hsize]{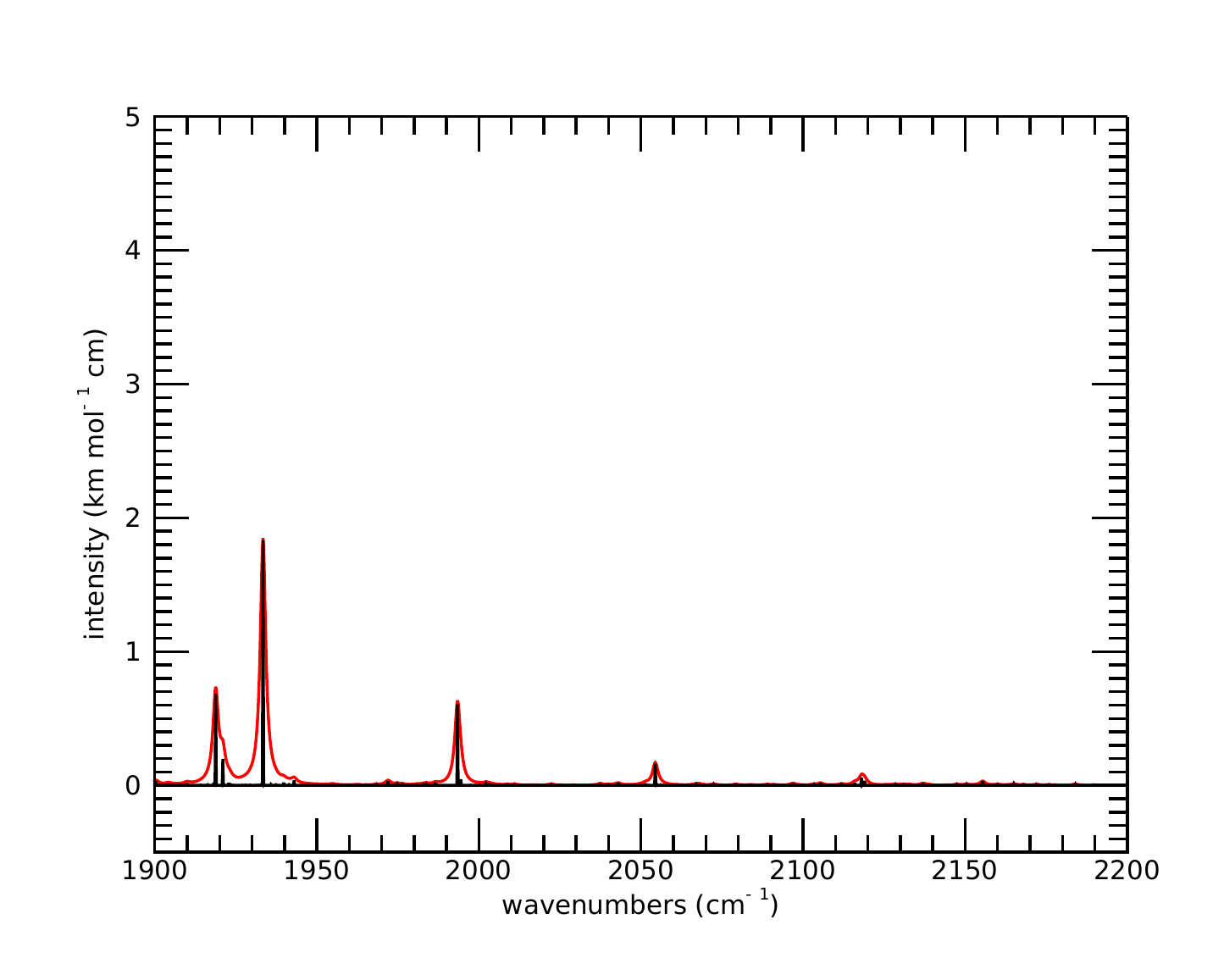}
\caption{Same as Fig.~\ref{plotpyr1} in the range from 1900 to 2200 cm$^{-1}$.}\label{plotpyr9}
\end{figure}
\begin{figure}
\includegraphics[width=\hsize]{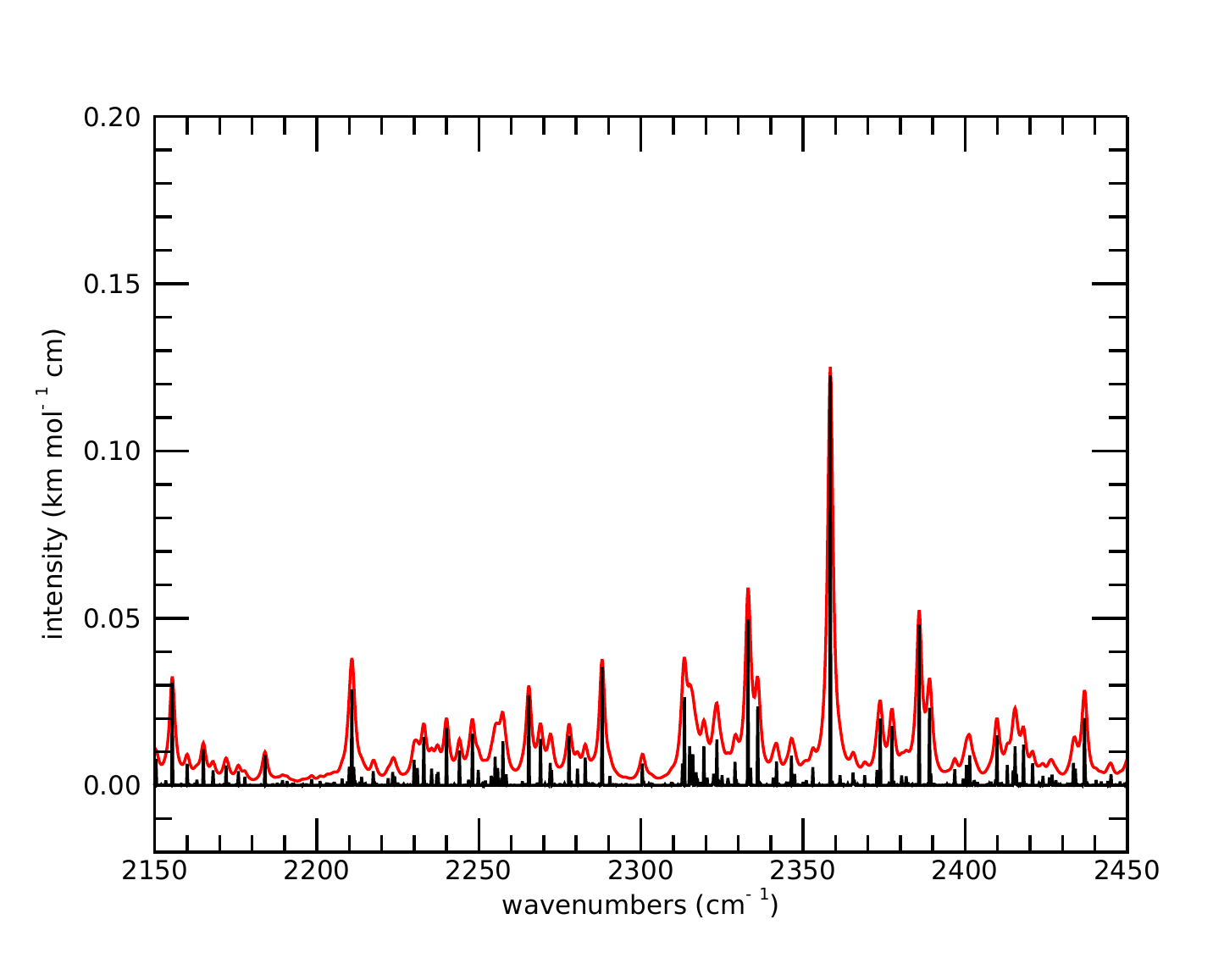}
\caption{Same as Fig.~\ref{plotpyr1} in the range from 2150 to 2450 cm$^{-1}$.}\label{plotpyr10}
\end{figure}
\begin{figure}
\includegraphics[width=\hsize]{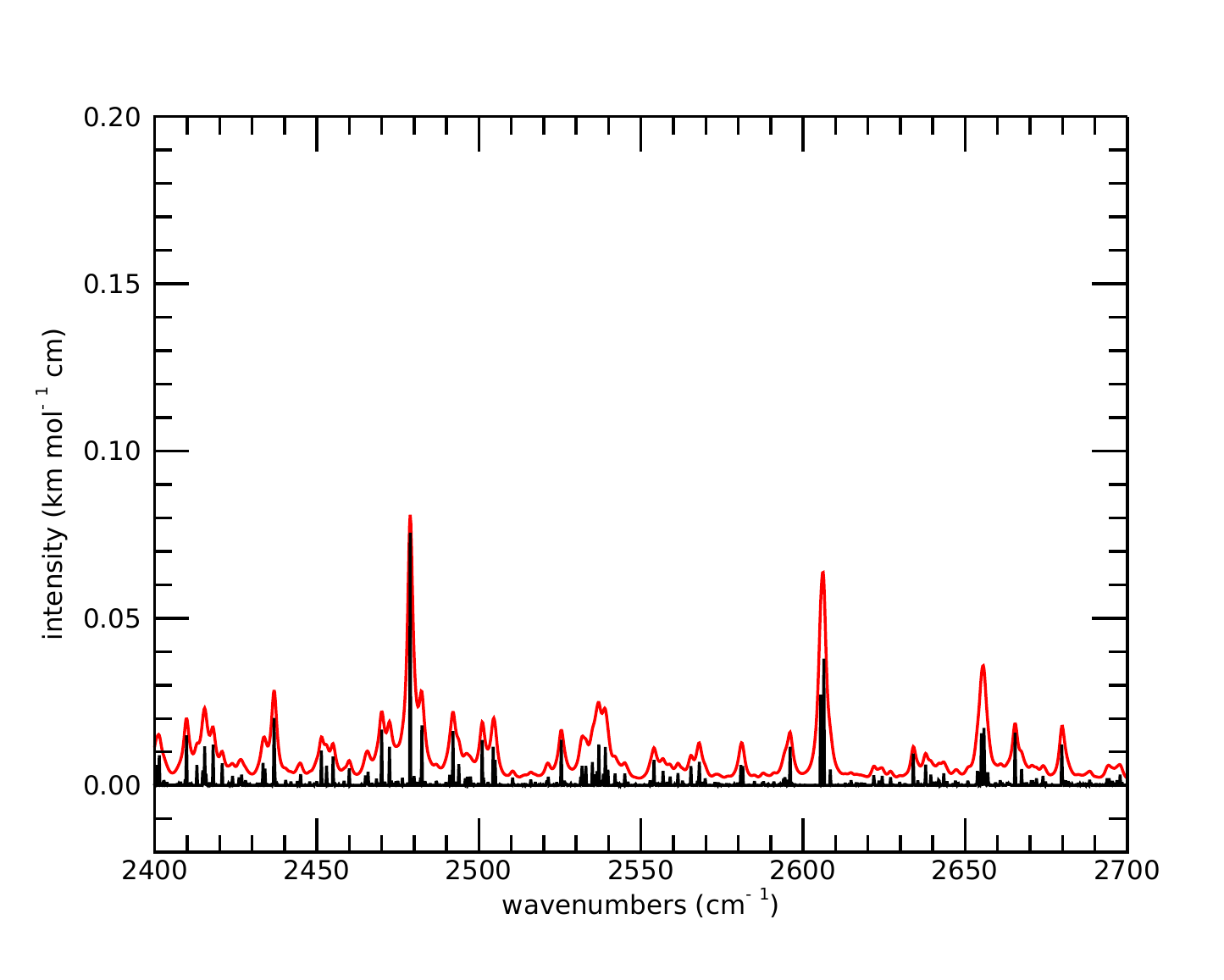}
\caption{Same as Fig.~\ref{plotpyr1} in the range from 2400 to 2700 cm$^{-1}$.}\label{plotpyr11}
\end{figure}
\begin{figure}
\includegraphics[width=\hsize]{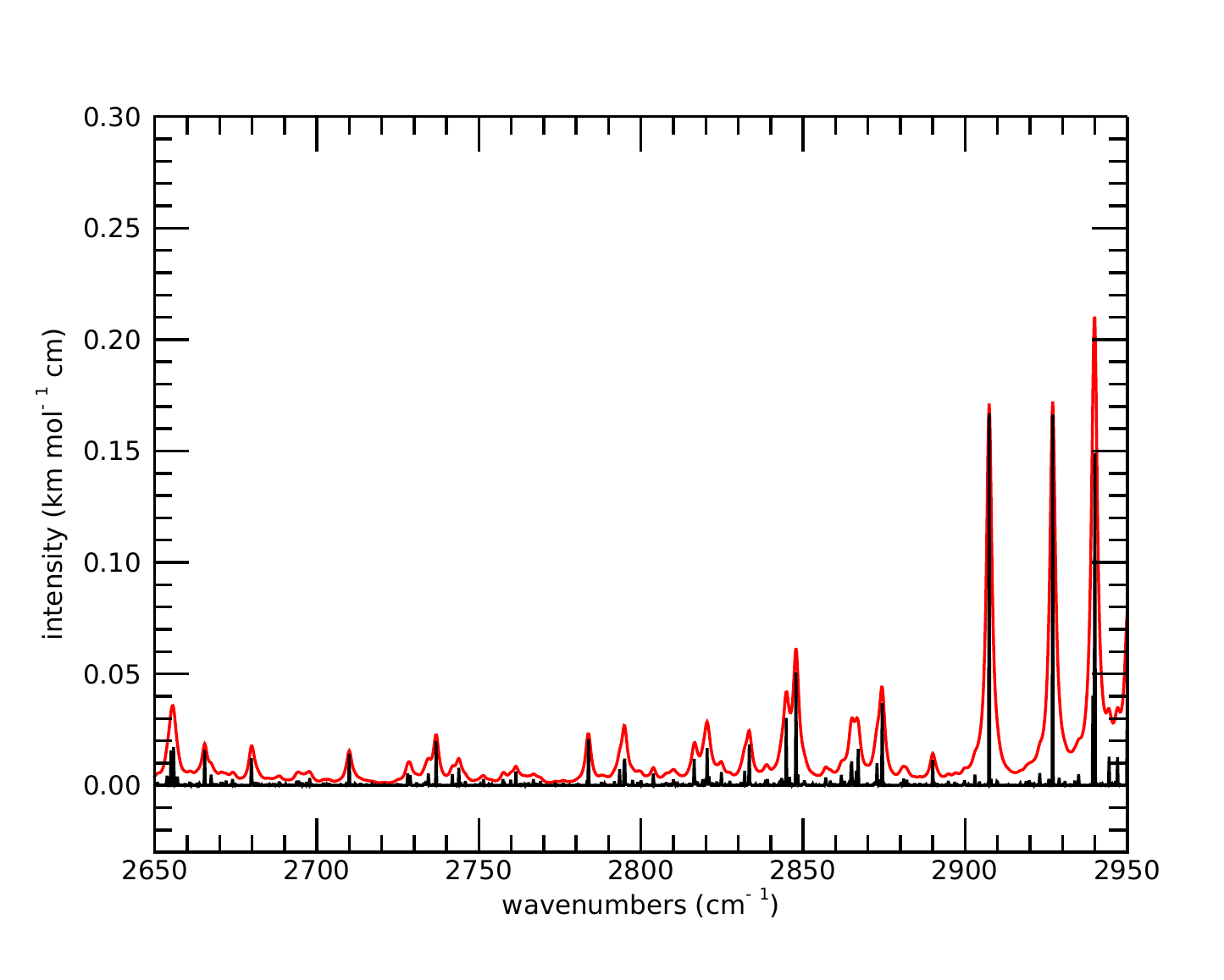}
\caption{Same as Fig.~\ref{plotpyr1} in the range from 2650 to 2950 cm$^{-1}$.}\label{plotpyr12}
\end{figure}
\begin{figure}
\includegraphics[width=\hsize]{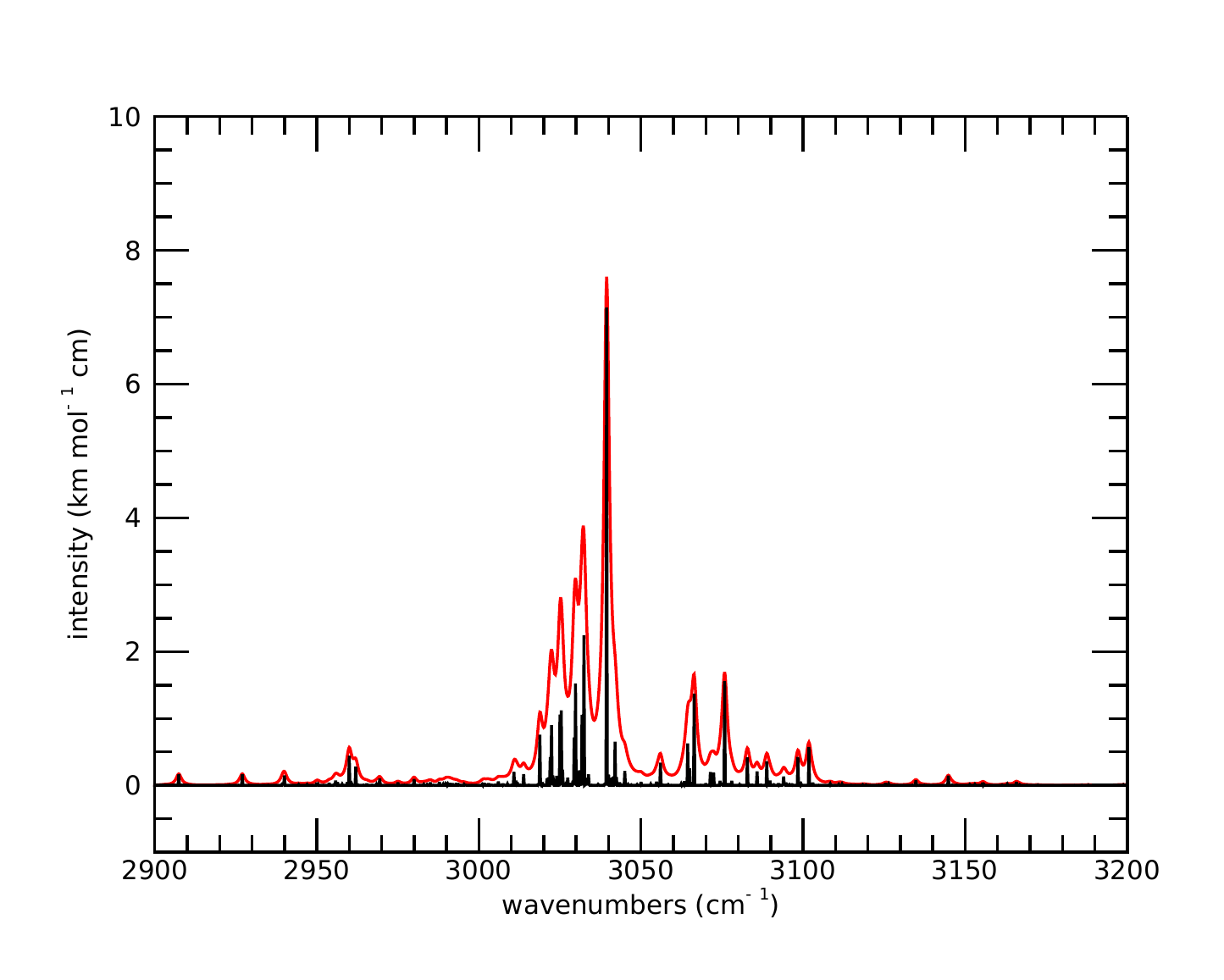}
\caption{Same as Fig.~\ref{plotpyr1} in the range from 2900 to 3200 cm$^{-1}$.}\label{plotpyr13}
\end{figure}
\begin{figure}
\includegraphics[width=\hsize]{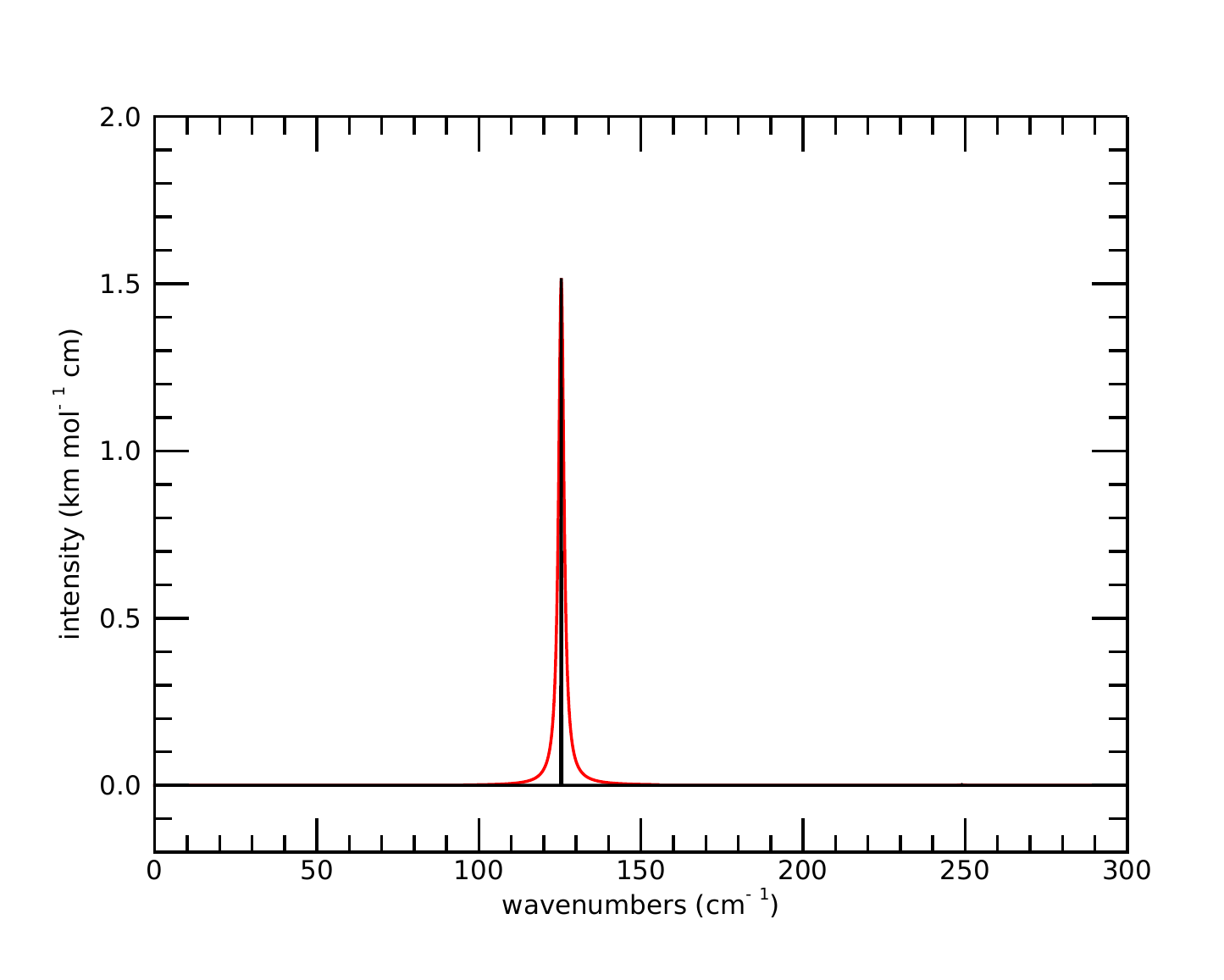}
\caption{Anharmonic spectrum of Coronene computed with $r=0.1$ and $h=6$ in the range from 0 to 300 cm$^{-1}$. Black bars indicate the precise position of individual bands, whereas red envelopes are convolved with Lorentzian profiles with a 1 cm$^{-1}$ width.}\label{plotcor1}
\end{figure}
\begin{figure}
\includegraphics[width=\hsize]{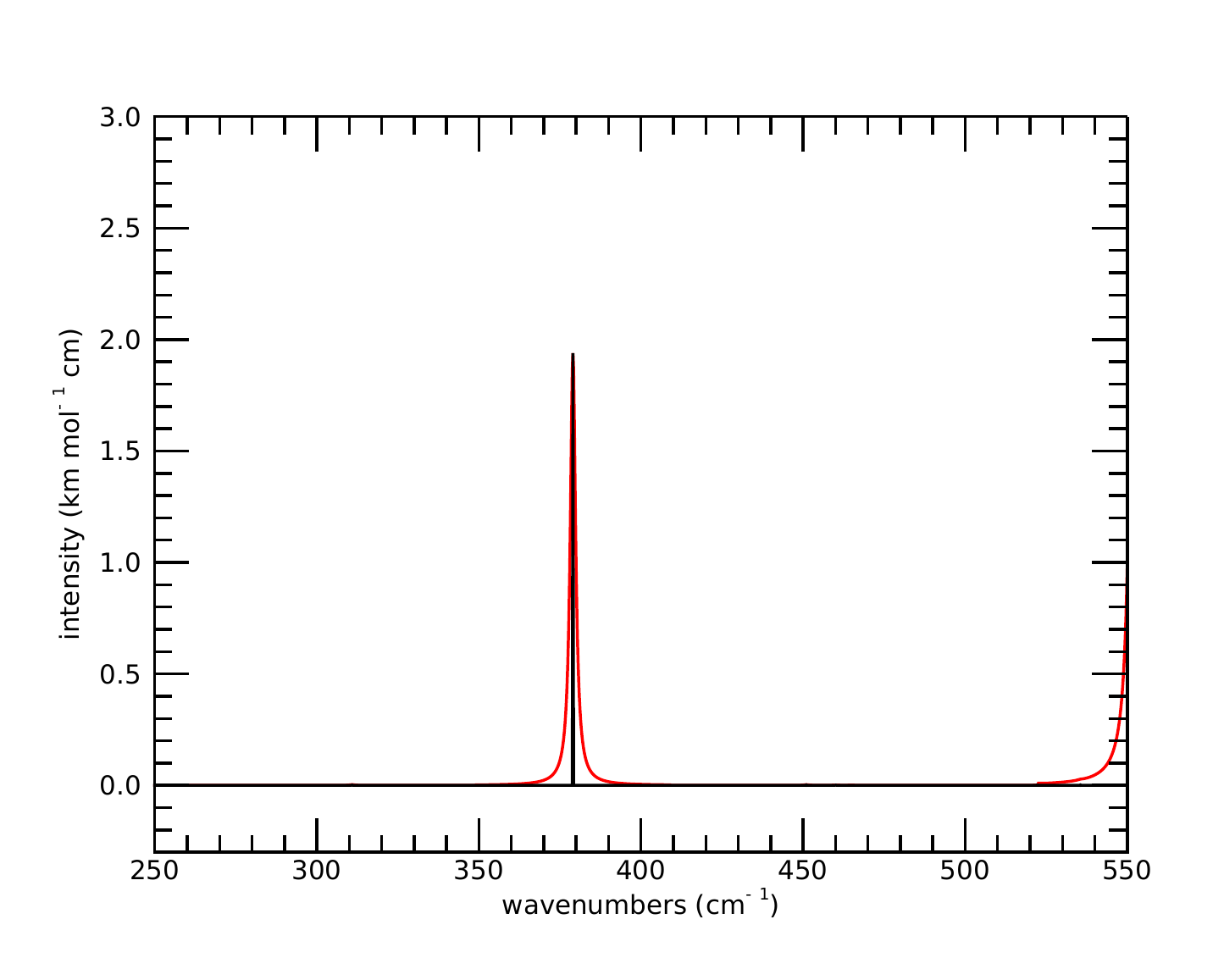}
\caption{Same as Fig.~\ref{plotcor1} in the range from 250 to 550 cm$^{-1}$.}\label{plotcor2}
\end{figure}
\begin{figure}
\includegraphics[width=\hsize]{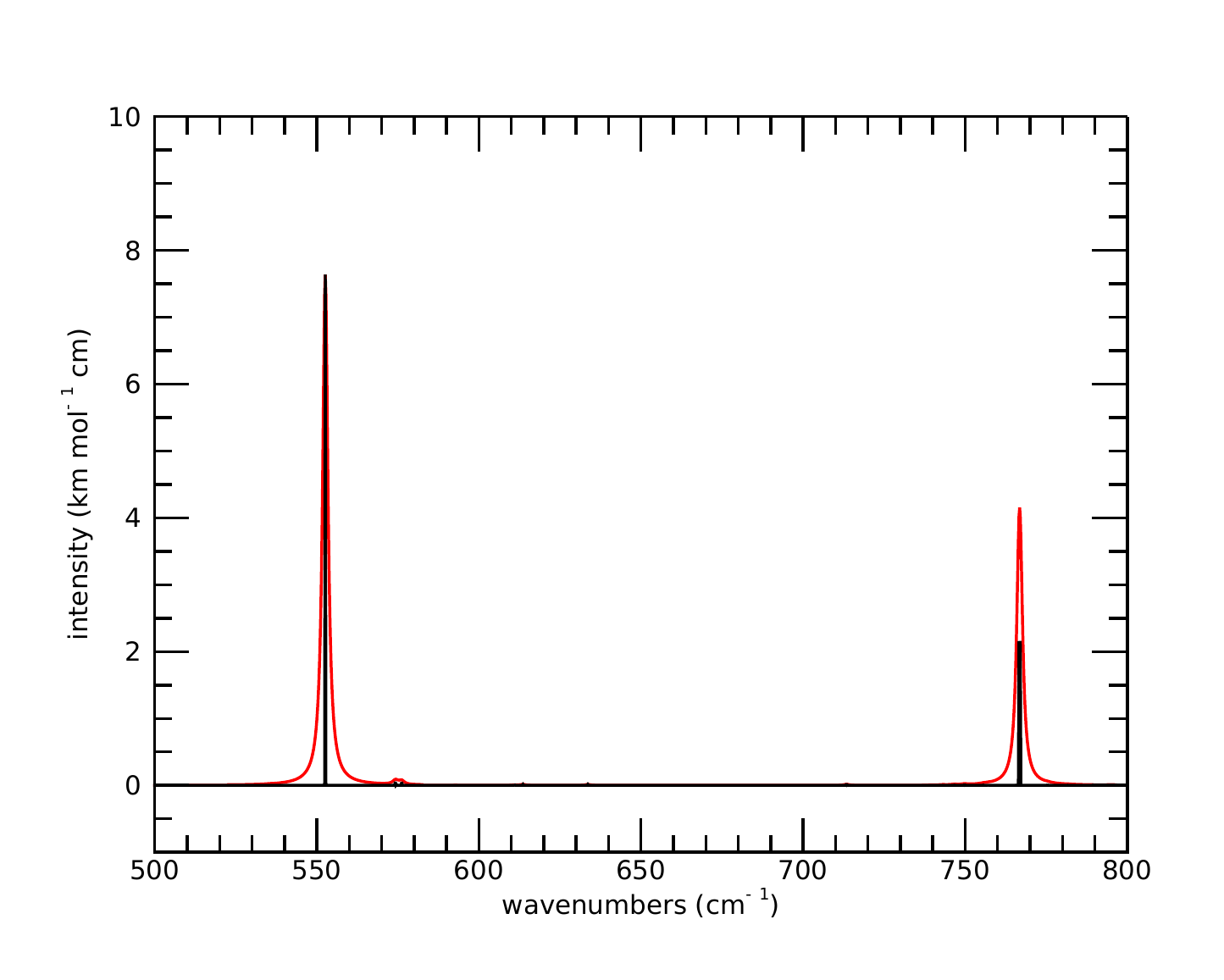}
\caption{Same as Fig.~\ref{plotcor1} in the range from 500 to 800 cm$^{-1}$.}\label{plotcor3}
\end{figure}
\begin{figure}
\includegraphics[width=\hsize]{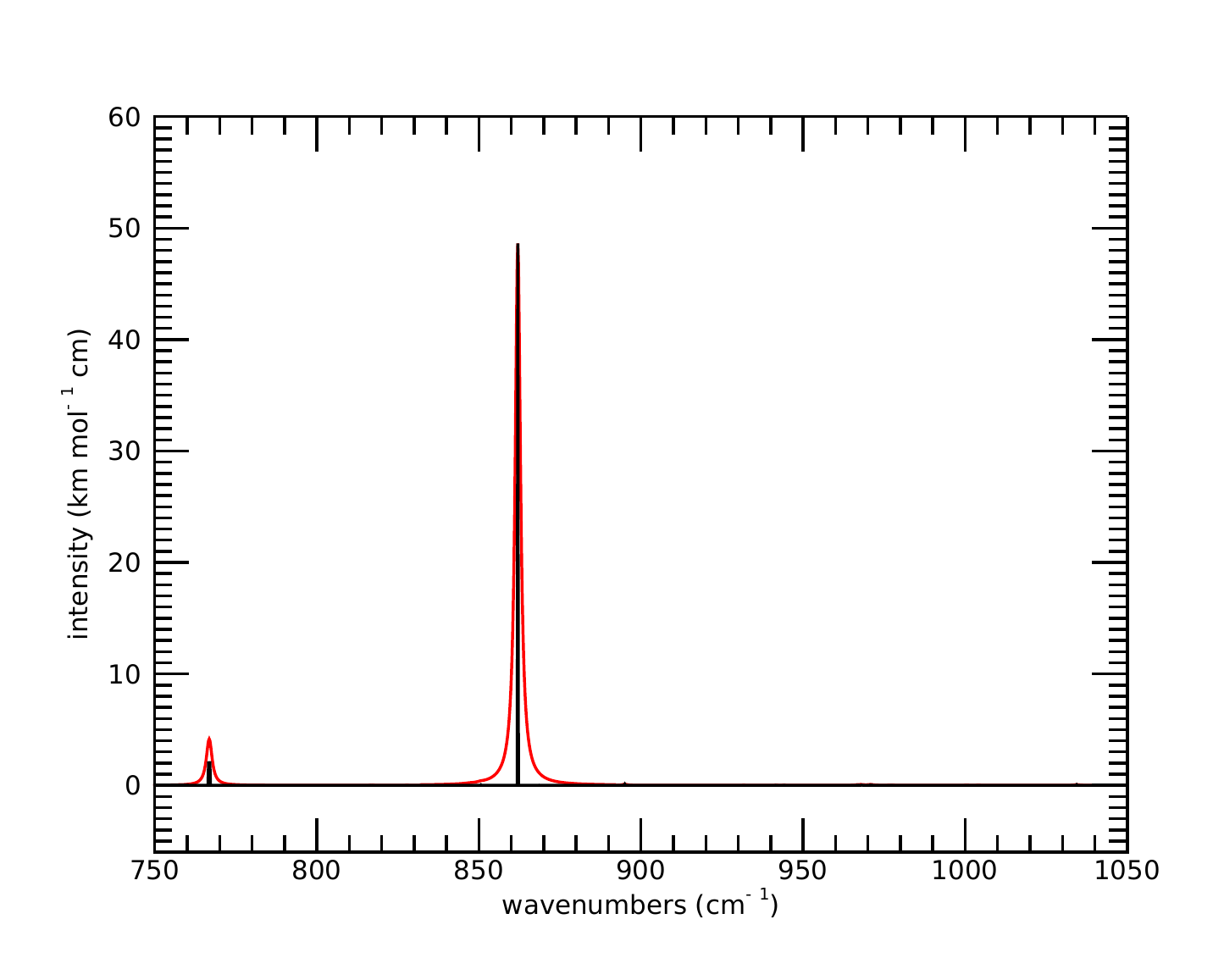}
\caption{Same as Fig.~\ref{plotcor1} in the range from 750 to 1050 cm$^{-1}$.}\label{plotcor4}
\end{figure}
\begin{figure}
\includegraphics[width=\hsize]{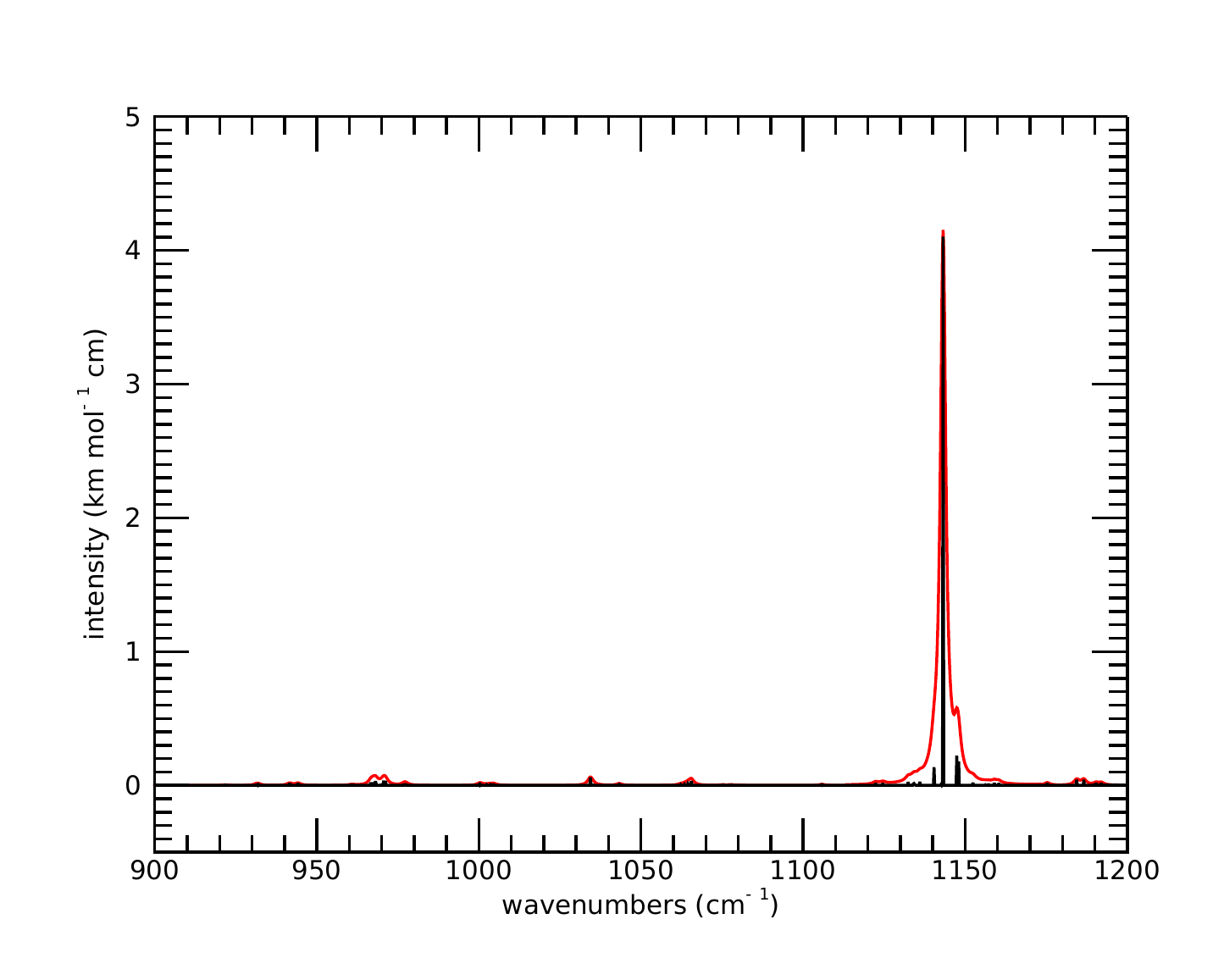}
\caption{Same as Fig.~\ref{plotcor1} in the range from 900 to 1200 cm$^{-1}$.}\label{plotcor5}
\end{figure}
\begin{figure}
\includegraphics[width=\hsize]{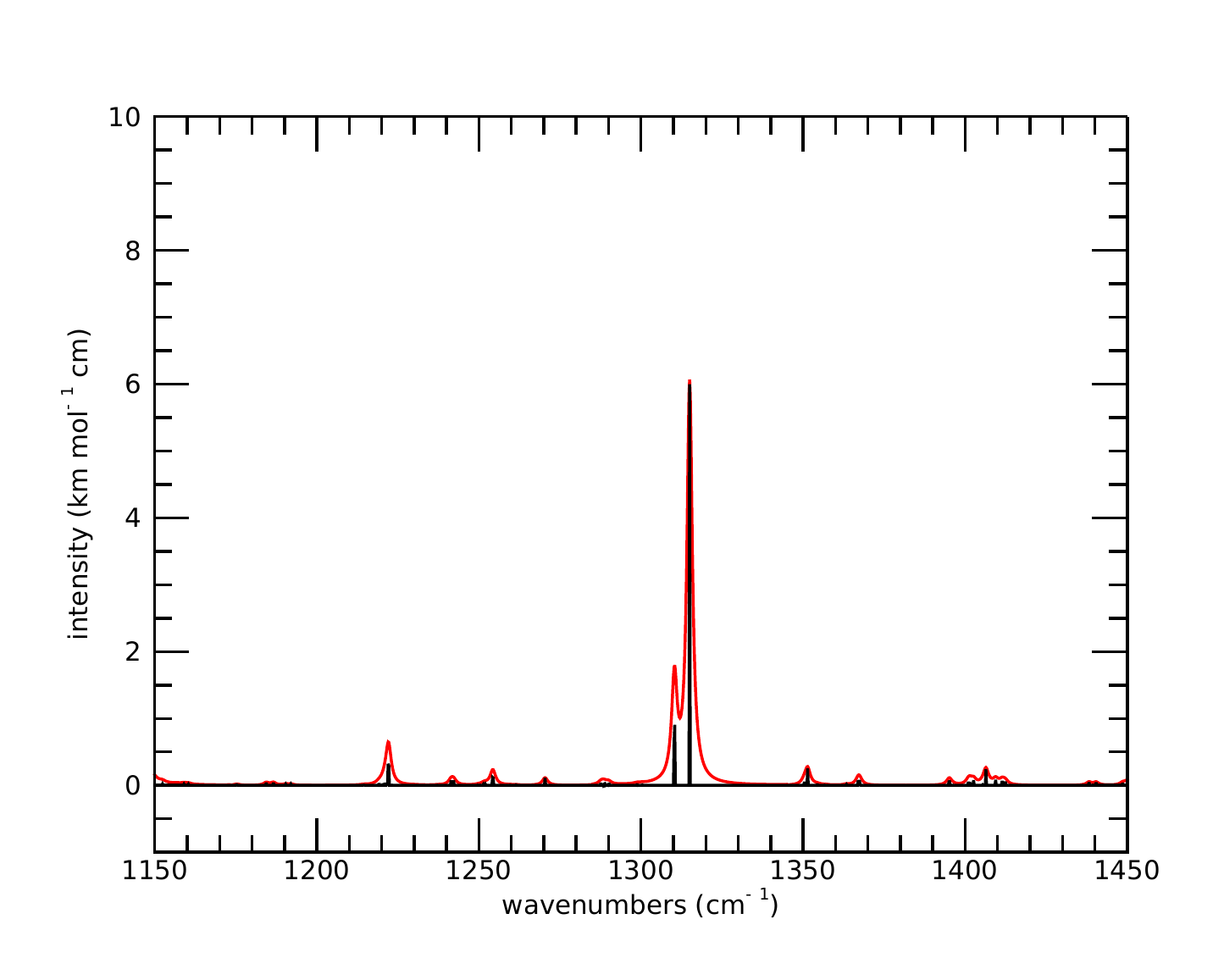}
\caption{Same as Fig.~\ref{plotcor1} in the range from 1150 to 1450 cm$^{-1}$.}\label{plotcor6}
\end{figure}
\begin{figure}
\includegraphics[width=\hsize]{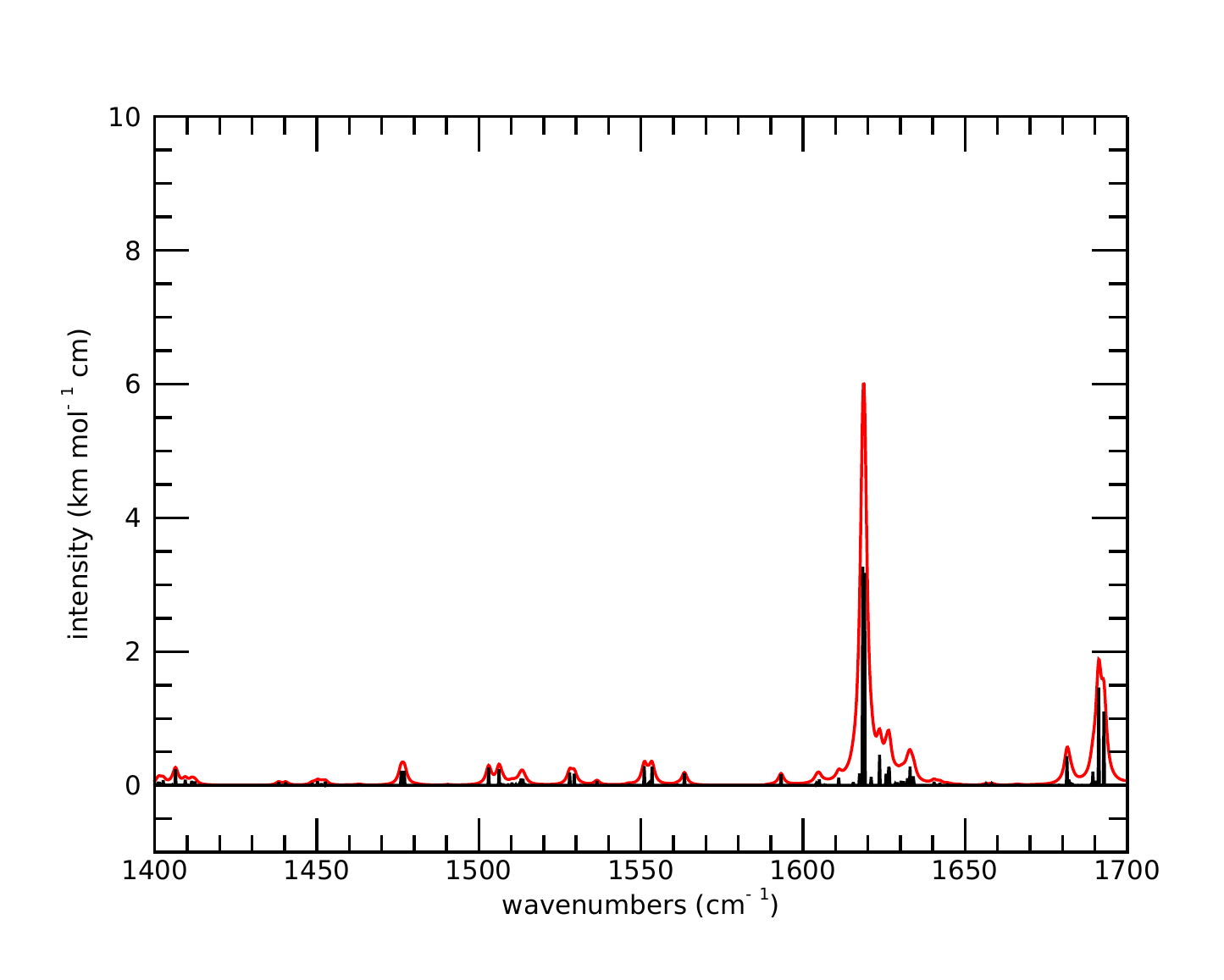}
\caption{Same as Fig.~\ref{plotcor1} in the range from 1400 to 1700 cm$^{-1}$.}\label{plotcor7}
\end{figure}
\begin{figure}
\includegraphics[width=\hsize]{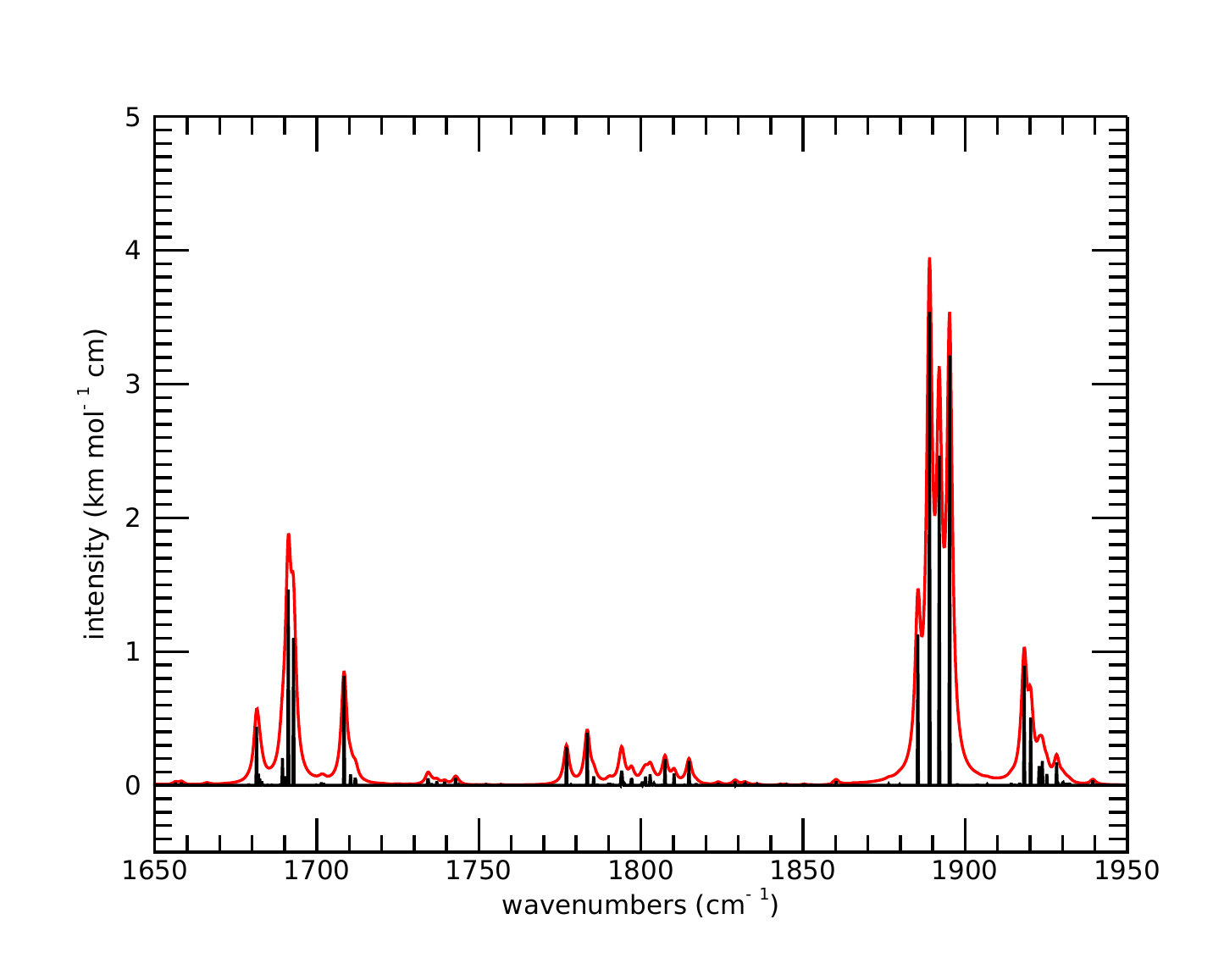}
\caption{Same as Fig.~\ref{plotcor1} in the range from 1650 to 1950 cm$^{-1}$.}\label{plotcor8}
\end{figure}
\begin{figure}
\includegraphics[width=\hsize]{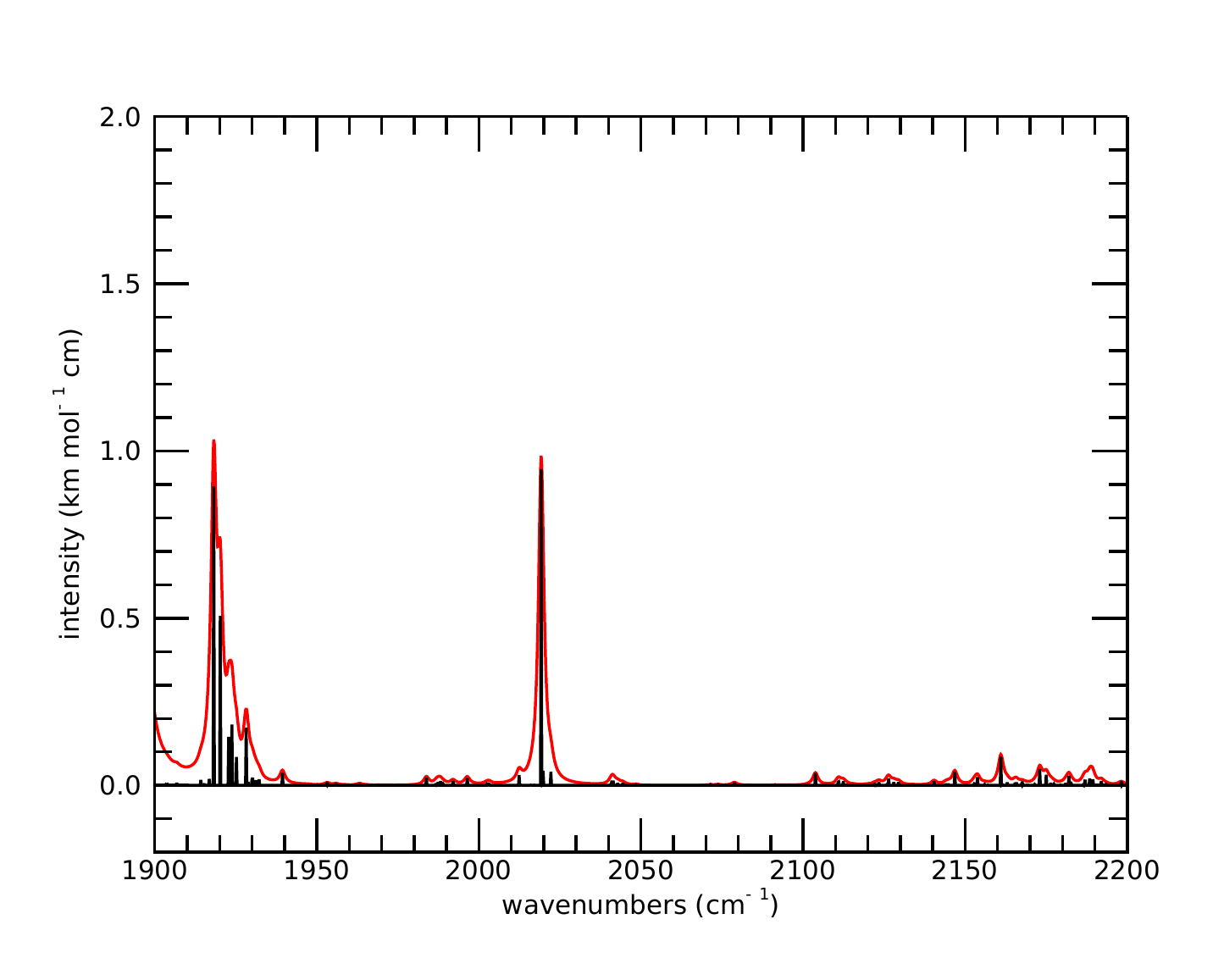}
\caption{Same as Fig.~\ref{plotcor1} in the range from 1900 to 2200 cm$^{-1}$.}\label{plotcor9}
\end{figure}
\begin{figure}
\includegraphics[width=\hsize]{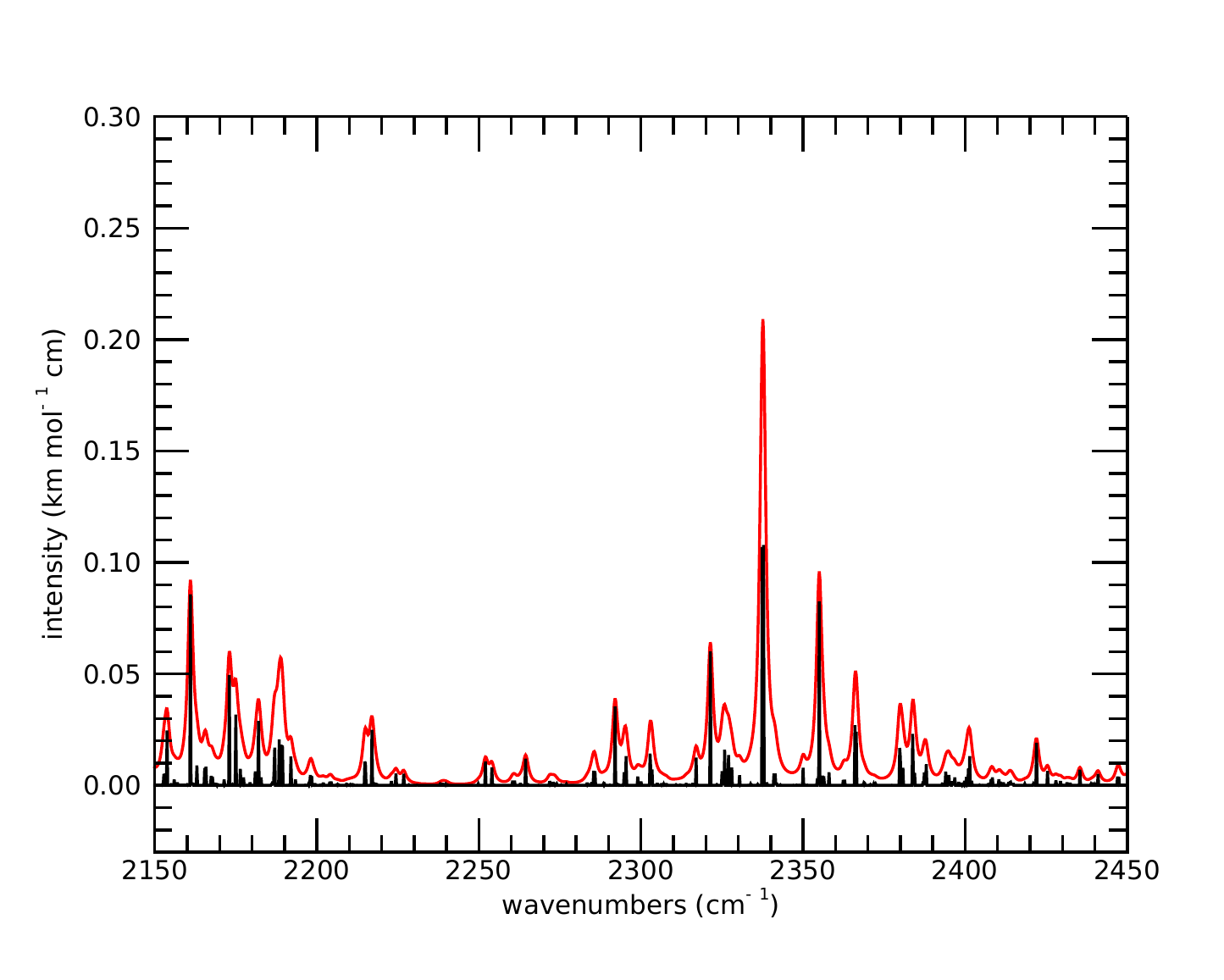}
\caption{Same as Fig.~\ref{plotcor1} in the range from 2150 to 2450 cm$^{-1}$.}\label{plotcor10}
\end{figure}
\begin{figure}
\includegraphics[width=\hsize]{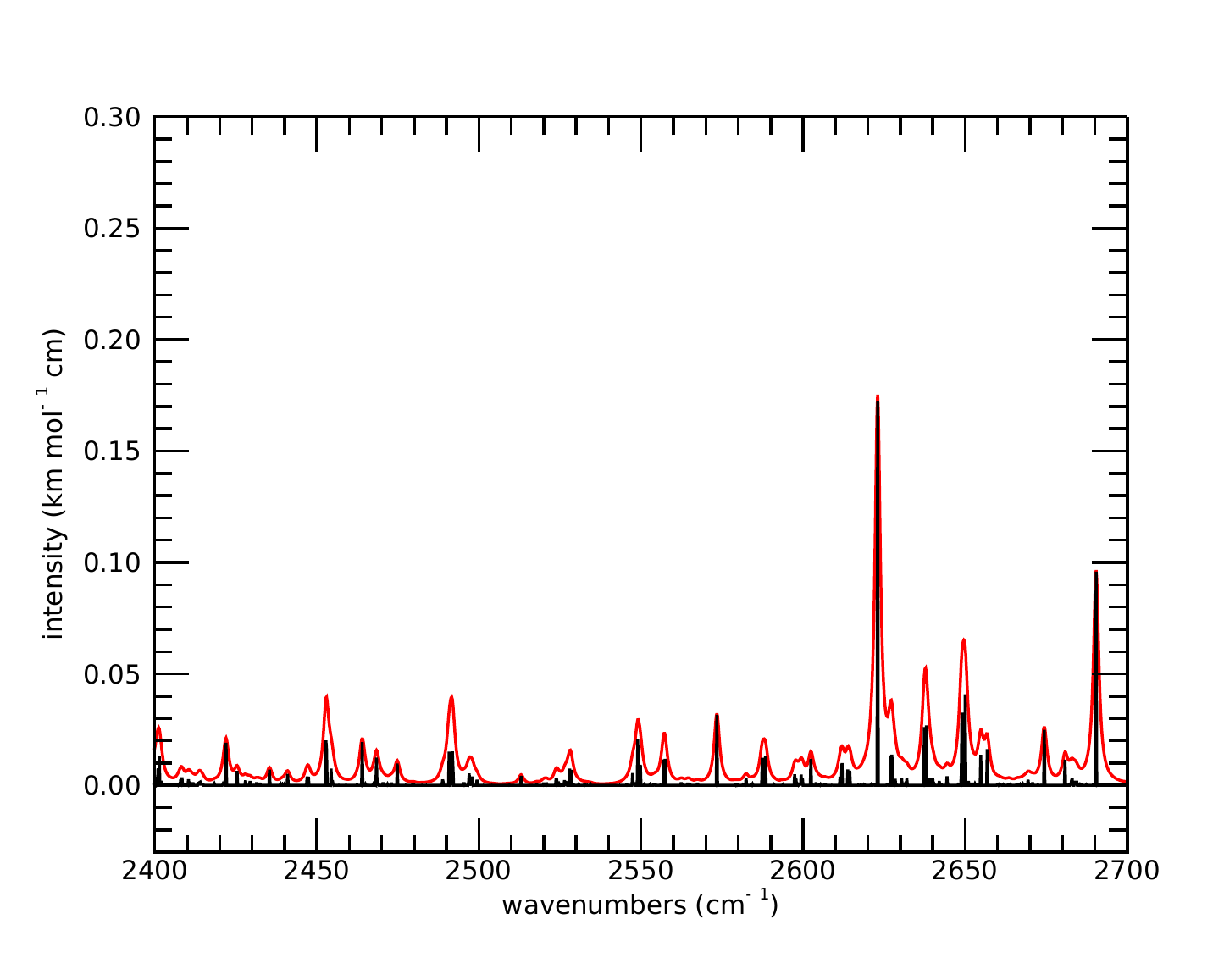}
\caption{Same as Fig.~\ref{plotcor1} in the range from 2400 to 2700 cm$^{-1}$.}\label{plotcor11}
\end{figure}
\begin{figure}
\includegraphics[width=\hsize]{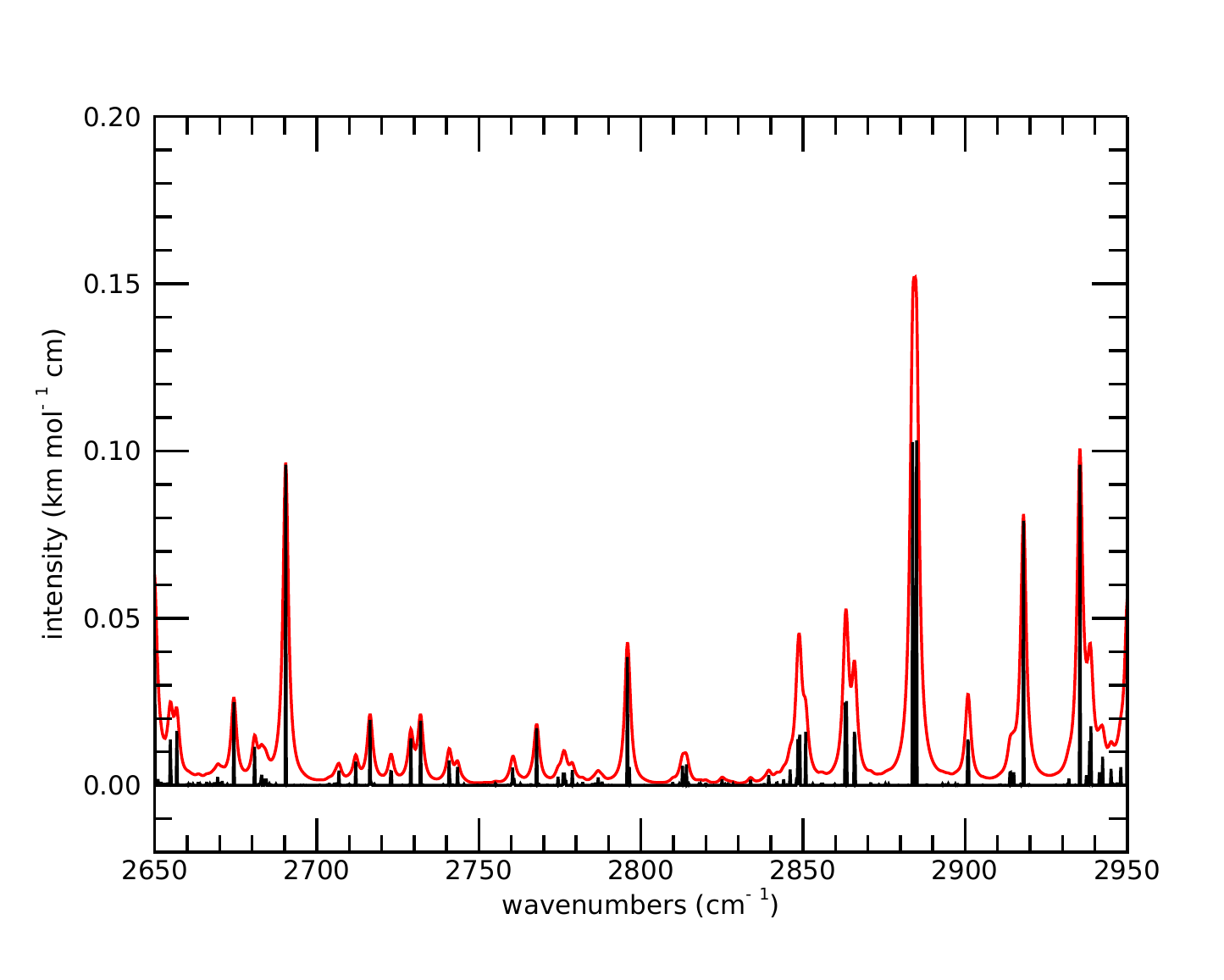}
\caption{Same as Fig.~\ref{plotcor1} in the range from 2650 to 2950 cm$^{-1}$.}\label{plotcor12}
\end{figure}
\begin{figure}
\includegraphics[width=\hsize]{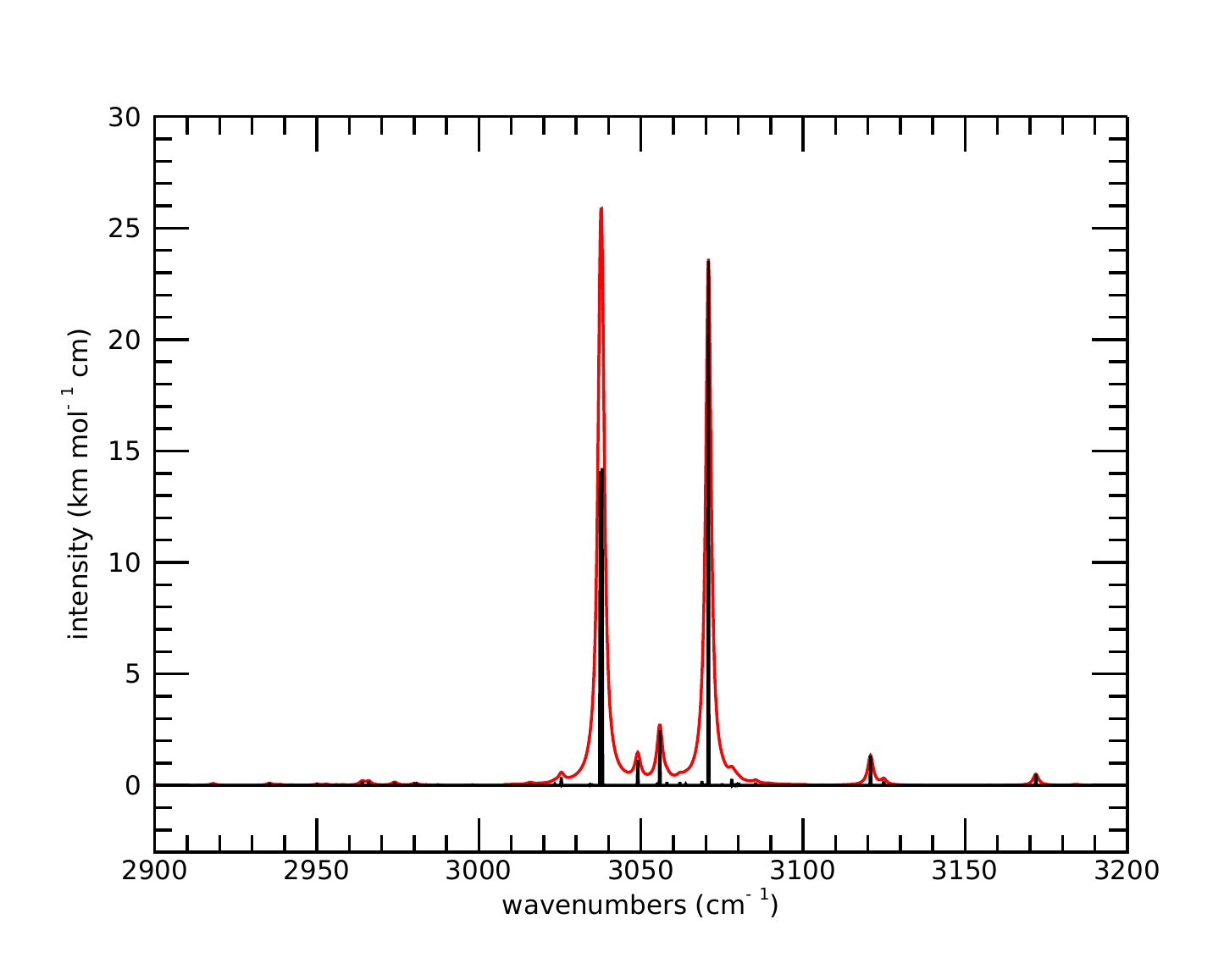}
\caption{Same as Fig.~\ref{plotcor1} in the range from 2900 to 3200 cm$^{-1}$.}\label{plotcor13}
\end{figure}
\begin{figure}
\includegraphics[width=\hsize]{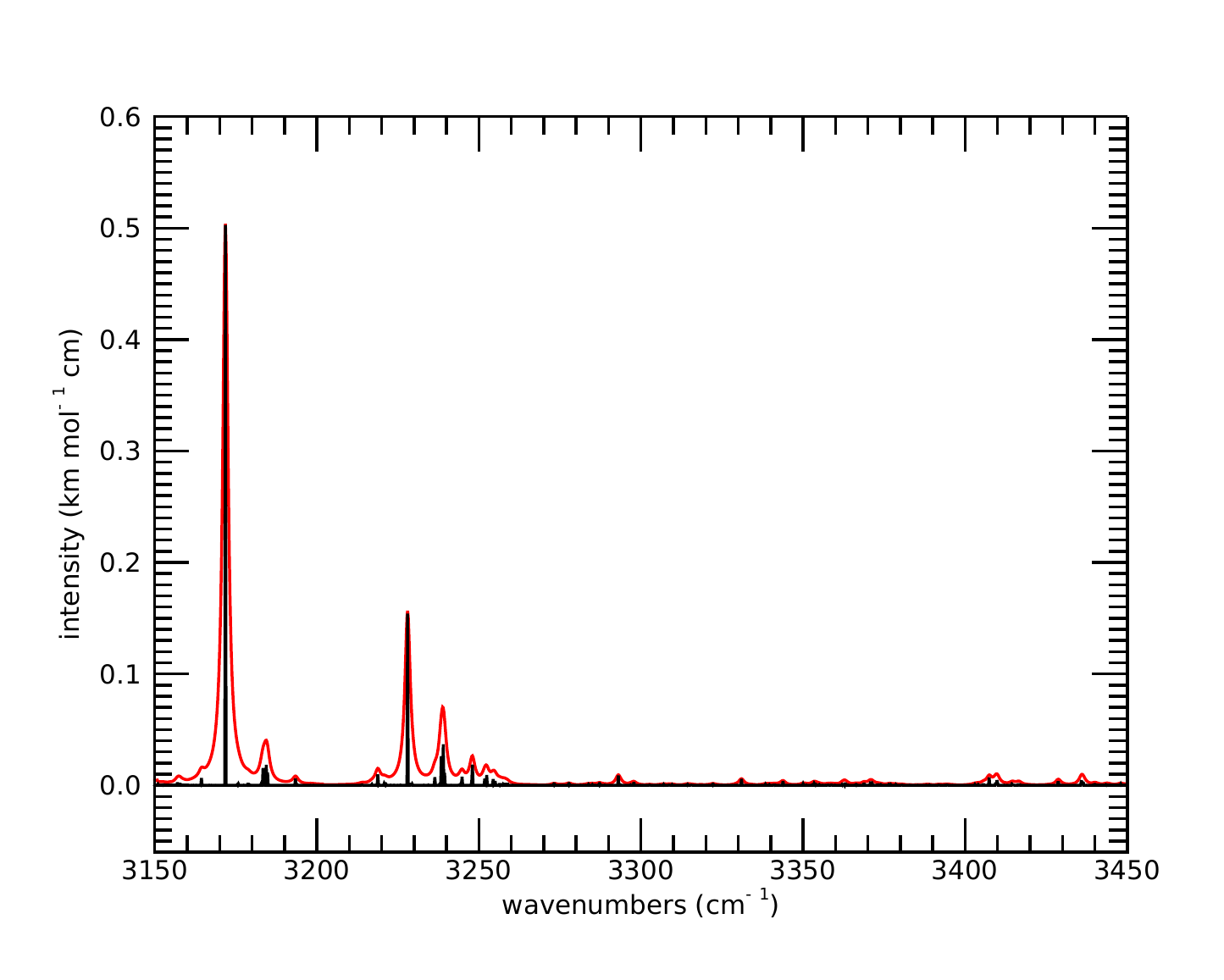}
\caption{Same as Fig.~\ref{plotcor1} in the range from 3150 to 3450 cm$^{-1}$.}\label{plotcor14}
\end{figure}

%
%
\bibliography{biblio}